\documentclass[oneside,12pt]{thesis}

\ifpdf
    \pdfcatalog { /PageMode (/UseOutlines)
                  /OpenAction (fitbh)  }
\fi

\title{Study of the Structure and Dynamics of Complex Biological Networks}

\ifpdf
  \author{Areejit Samal}
  \collegeordept{Department of Physics and Astrophysics}
  \university{University of Delhi}
\else
  \author{Areejit Samal}
  \collegeordept{Department of Physics and Astrophysics}
  \university{University of Delhi}
\fi

\degree{Doctor of Philosophy}

\degreedate{May 2008}

\usepackage{amssymb}

\def\signature#1#2{\parbox[b]{1in}{\smash{#1}\vskip12pt}
\hfill \parbox[t]{3in}{\shortstack{\vrule width 3in height
0.4pt\\\small#2}}}

\begin{document}

\renewcommand\baselinestretch{1.2}
\baselineskip=18pt plus1pt

\maketitle

\frontmatter

\newpage
\pagestyle{empty} \mbox{}
\newpage

\chapter*{}
\thispagestyle{empty}
\begin{center}
{\Huge \bf Declaration}
\end{center}
\vspace*{5ex} \noindent {\large This thesis describes work done by
the candidate during his tenure as Ph.D. student at the Department
of Physics and Astrophysics, University of Delhi, Delhi, India under
the supervision of Prof. Sanjay Jain and Prof. Shobhit Mahajan. The
work reported in this thesis is original and it has not been
submitted earlier for any degree to any university.}

\par
\vspace*{8ex} \signature{Candidate:}{Areejit Samal}
\par
\vspace*{8ex} \signature{Supervisor:}{Sanjay Jain}
\par
\vspace*{8ex} \signature{Supervisor:}{Shobhit Mahajan}
\par
\vspace*{8ex} \signature{Head of the Department:}{D. S. Kulshreshtha}

\newpage
\pagestyle{empty} \mbox{}
\newpage

\thispagestyle{empty}

\section*{List of Publications}

\subsubsection*{In International Refereed Journals}
\begin{enumerate}
\item Low degree metabolites explain essential reactions and enhance modularity in biological
networks,\\
{\bf Areejit Samal}, Shalini Singh, Varun Giri, Sandeep Krishna, N.
Raghuram and Sanjay Jain,\\
{\it BMC Bioinformatics}, {\bf 7:118} (2006).

\item The regulatory network of {\it E. coli} metabolism as a Boolean dynamical system exhibits
both homeostasis and flexibility of response,\\
{\bf Areejit Samal} and Sanjay Jain,\\
{\it BMC Systems Biology}, {\bf 2:21} (2008).

\item A universal power law and proportionate change process characterize the evolution of metabolic networks,\\
Shalini Singh, {\bf Areejit Samal}, Varun Giri, Sandeep Krishna, N.
Raghuram and Sanjay Jain,\\
{\it Eur. Phys. J. B}, {\bf 57:75-80} (2007).

\end{enumerate}

\noindent The work reported in this thesis is based on publications 1 and 2 mentioned above.

\subsubsection*{In Conference Proceedings}

\begin{enumerate}

\item Modelling Stock Market: Expectation Bubbles and Crashes,\\
Varun Giri and {\bf Areejit Samal},\\
Student paper for IMSc Complex Systems School, January 2-27 2006,
held at the Institute of Mathematical Sciences, Chennai,
India in association with Santa Fe Institute, New Mexico, USA.

\item India's Input Output System: an Enquiry,\\
Bino Paul, Ramachandran Venkat and {\bf Areejit Samal},\\
Proceedings of the 43$^{rd}$ Annual Conference of The Indian Econometric
Society, January 5-7 2007, held at the Indian Institute of Technology,
Bombay, India.

\end{enumerate}

\section*{Important Talks and Oral Presentations}

\begin{enumerate}

\item Presented talk titled ``System level dynamics and robustness
of the genetic network regulating {\it E. coli} metabolism'' at Second European
PhD Complexity School: Stochastic Effects in Differential Nonlinear Models
- From Neutrality in Evolution to Efficiency in Markets,
November 22-27, 2007, held at ISI, Torino, Italy.

\item Invited talk with hands-on-session titled ``Some computational
systems biology techniques to study gene regulation of metabolism''
at Heraeus International Summer School - Statistical Physics of
Gene Regulation, July 16-27, 2007, held at Jacobs University, Bremen,
Germany.

\item Invited talk with hands-on-session titled ``A practical guide to
graph-theoretic analysis of large scale biological networks'' at Workshop
on knowledge discovery in Life Sciences: Tools \& Techniques in Bioinformatics,
January 29 to February 2, 2007, held at Bioinformatics Centre,
University of Pune, India. This talk was presented with Varun Giri.

\item Oral presentation titled ``The Genetic Network controlling {\it E. coli}
metabolism as a Dynamical System'' at Computational Insights into Biological
Systems, December 26-28, 2006, held at Indian Institute of Science, Bangalore,
India.

\item Invited talk titled ``Computational Systems Biology: An Overview'' at
Systems Biology: A New Era in Bioinformatics, May 4, 2006, held at
Bioinformatics Centre, Department of Biotechnology, Himachal Pradesh University,
Shimla, India.

\item Invited talk titled ``Essentiality and Modularity in large scale metabolic
networks'' at Systems Biology: A New Era in Bioinformatics, May 4, 2006, held at
Bioinformatics Centre, Department of Biotechnology, Himachal Pradesh University,
Shimla, India.

\end{enumerate}

\section*{Poster Presentations}

\begin{enumerate}

\item Presented poster titled ``The {\it Escherichia coli} transcriptional regulatory
network exhibits both homeostasis and flexibility of response'' at International
Conference on Bioinformatics, December 18-20, 2006, held at Hotel Ashok, New Delhi,
India.

\item Presented poster titled ``Low degree metabolites enhance modularity in metabolic
and regulatory networks'' at 22$^{nd}$ Jerusalem Winter School in Theoretical Physics
on Biological Networks and Evolution, December 27, 2004 to January 7, 2005, held at
Institute for Advanced Studies of the Hebrew University, Jerusalem, Israel.

\item Presented poster titled ``Statistics and roles of linear pathways in metabolic
networks'' at STATPHYS - KOLKATA V Complex Networks: Structure, Function and Processes,
June 27 to July 1, 2004, held at Satyendra Nath Bose National Centre for Basic Sciences,
Kolkata, India.

\end{enumerate}
\newpage
\pagestyle{empty} \mbox{}
\newpage

\setcounter{secnumdepth}{3} \setcounter{tocdepth}{3}

\tableofcontents

\printglossary

\mainmatter
\pagestyle{plain}


\chapter{Introduction}

\label{introduction}

\section{Motivation}

The study of networks is important from the point of view
of understanding complex systems in nature
\cite{WS1998,JK1998,BA1999,S2001,AB2002,Handbookofgraphs}.
A living cell can be viewed as a complex dynamical system
consisting of several thousand different types of molecules.
These molecules are all connected to each other by a
complex web of interactions.
This web can be thought of as an overlay of different
networks including the metabolic network, protein-protein
interaction network and genetic regulatory network.
Much of the work on living systems in the twentieth century
was focused towards understanding the behaviour of individual
molecules inside cells.
However, most systemic properties of living systems are a
result of complex interactions between various microscopic
constituents such as genes, proteins and metabolites.
Hence, it is important to study the large scale structure
and system level dynamics of complex biological networks
\cite{HHLM1999,K2002,A2003,BO2004,Palssonbook,Alonbook}.

Technical advances in data collection techniques and the
availability of complete genome sequences has led to a
reconstruction of many cellular networks.
Structural studies of large scale metabolic, protein-protein
interaction and genetic regulatory networks have uncovered
some unexpected patterns in these networks which are in
common with complex social and technological networks
(for reviews see
\cite{AB2002,Handbookofgraphs,A2003,AA2003,B2003,BO2004,BLGT2004}).
However, there is limited understanding of how the observed
structural properties of biological networks are related to
cellular functions.
Further, at present, much less is understood about how the
observed structural regularities in biological networks
arose in the course of evolution.

Although structural studies of complex biological networks
have discovered some interesting patterns in these networks,
they have an inherent limitation.
For example, when more than one link converges at a single node
in the network, an input function needs to be specified for the
node.
The input functions of the nodes in the network can have
important dynamical consequences.
It was shown that the coherent feed-forward loop can act as a
sign sensitive delay circuit while the incoherent feed-forward
loop as a sign sensitive accelerator \cite{MA2003}.
Note that from a pure topological perspective, both coherent
and incoherent feed-forward loops have the same triangle
architecture, but incorporating the knowledge of the nature of
different regulatory links (positive or negative) leads to
different dynamical consequences.
Guet {\it et al} synthetically engineered three gene networks
employing a library of promoters with varying strengths and
three genes to show experimentally that networks with same
topology but different input functions can lead to different
behaviours \cite{GEHL2002}.
On the other hand, they also found that networks with different
topology can have the same logical behaviour.
The above mentioned results  show that pure topological
similarity of two circuits may not imply similar behaviour for
both circuits.
Thus, the network topology alone cannot determine the
network behaviour.

Over the years, several dynamical models describing various
subsystems inside the cell have been proposed and studied
extensively (see, e.g., \cite{BL1997,BI1999,VMMO2000}).
Barkai and Leibler proposed a theoretical model of
{\it E. coli} chemotactic pathway which reproduced
the observed property of `adaptation' of the chemotactic
response, and moreover showed that this property is
robust to parameter variation in the model \cite{BL1997}.
von Dassow {\it et al} studied a model of segment polarity
network in {\it Drosophila} to show that the spatial pattern
of gene expression was robust to changes in certain kinetic
parameters \cite{VMMO2000}.
Kacser and Burns showed that perturbation of individual
enzyme concentrations within a metabolic pathway rarely
affects the molecular flux through the pathway provided
the enzymes follow Michaelis-Menten kinetics and are not
saturated with substrate \cite{KB1973}.
These dynamical studies of biological systems have
provided understanding of their functional robustness.
However, one expects that new insights on the whole system
level will be obtained by studying the dynamics of large
scale networks which incorporates information about most or
a large fraction of interacting molecules constituting the
network.
It is for the whole that distinctive properties unique to life
are most dramatically visible.
Thus, it is important to study the system level dynamics of
large scale biological networks.

In this thesis, we have studied the large scale structure
and system level dynamics of certain biological networks
using tools from graph theory, computational biology and
dynamical systems.
In chapter 2, we study the structure and dynamics of
large scale metabolic networks inside three organisms,
{\it Escherichia coli}, {\it Saccharomyces cerevisiae} and
{\it Staphylococcus aureus}.
In chapters 3 and 4, we study the dynamics of the
large scale genetic network controlling {\it E. coli}
metabolism.
We have tried to explain the observed system level
dynamical properties of these networks in terms of their
underlying structure.
Our studies of the system level dynamics of these large
scale biological networks provide a different perspective
on their functioning compared to that obtained from purely
structural studies.
Our study also leads to some new insights on features such as
robustness, fragility and modularity of these large scale
biological networks.
We also shed light on how different networks inside the
cell such as metabolic networks and genetic networks
influence each other.


\section{Biochemical networks in cells}

Biological networks are abstract representations of the
molecular components of living systems and their interactions.
The molecular constituents inside the cell include DNA, RNA,
proteins, metabolites and small molecules.
The interactions between the various types of molecular
constituents inside the cell, e.g., protein-DNA,
protein-protein, protein-metabolite, etc., have distinctive
features that lead to various kinds of networks inside the
cell.
These include the metabolic network, transcriptional
regulatory network and protein-protein interaction network.


\subsection{Metabolic network}

The metabolic network represents the set of biochemical
reactions that are responsible for the uptake of food
molecules or nutrients from the external environment and
converting them into other molecules that are building
blocks required for the growth and maintenance of the cell.
The latter include ATP, the energy currency of the cell,
other nucleotides, amino acids, lipid molecules and other
molecules.
These are sometimes called `biomass' metabolites and
constitute the output of the metabolic network.
The inputs are the food molecules such as sugars as well as
other organic molecules varying from organism to organism
and inorganic molecules such as water, hydrogen ions, oxygen
and sources of nitrogen, phosphorus, sulphur, iron, sodium,
potassium, etc.
These typically enter the cell through its membrane.
The bulk of the metabolic network are the chemical reactions
(ranging from several hundred to more than a thousand reactions
in different organisms) which transform the input molecules
into the output molecules.
The various reactions in the metabolic network are catalyzed
by enzymes.
Enzymes are proteins which are coded by genes.
Metabolites are the reactants or products of various
reactions.

\begin{figure*}
\centering
\includegraphics[height=7cm]{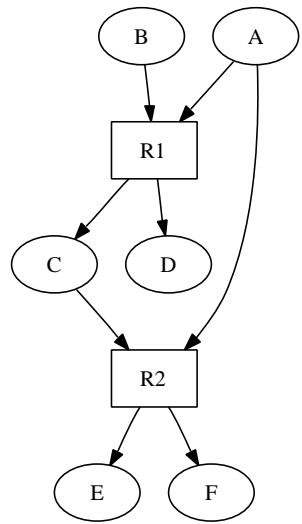}
\caption{
Example of a directed bipartite graph for a hypothetical
metabolic network with two reactions.
In this figure, rectangles represent reactions and ovals
metabolites.
In reaction R1, metabolites A and B are reactants and
metabolites C and D are products.
In reaction R2, metabolites A and C are reactants and
metabolites E and F are products.
Arrows to (from) metabolites represent their production
(consumption) in reactions.
Notice that there are no direct links between two metabolites
or two reactions in the graph.
}
\label{bipartite}
\end{figure*}

The metabolic network can be represented as a bipartite
graph consisting of two types of nodes:
metabolites and reactions.
An example of a directed bipartite graph is shown in
Fig. \ref{bipartite}.
In the directed bipartite metabolic graph, there is a link
from a metabolite pointing to a reaction node if the metabolite
is a reactant of the reaction, and a link from a reaction
pointing to a metabolite node if the metabolite is a product
of the reaction.
In the bipartite metabolic graph, there are no direct links
between two metabolites or two reactions.
In chapter 2 of this thesis, we have studied the metabolic
networks inside three organisms as a directed bipartite
graph.

The metabolic networks are also sometimes represented as
unipartite graphs (which could be directed or undirected)
in which there is only one type of node (metabolite nodes
or reaction nodes).
In the undirected unipartite metabolite graph, for example,
the nodes of the network represent metabolites, with edges
between two metabolites if they participate in a single
reaction.
Similarly, in the undirected unipartite reaction graph, the
nodes of the network represent reactions, with edges between
two reactions if they share the same metabolite.
In the directed unipartite reaction graph, there is a link
pointing from one reaction node to another if the former
produces a metabolite that is consumed by the latter.

The bipartite representation of the metabolic network has
more information compared to the above unipartite
representations.
A more detailed description of the metabolic network would
require three types of nodes:
metabolites, reactions and enzymes.
Such a tripartite network could additionally account for
the correspondence between enzymes and reactions and
regulatory interactions between metabolites and enzymes
inside the cell.


\subsection{Transcriptional regulatory network}

Genes are segments of DNA that code for proteins inside the
cell.
Transcription is the process by which an enzyme, RNA polymerase,
reads the sequence of bases on a gene and constructs an mRNA
molecule from that sequence.
Translation is the process in which a ribosome, a macromolecular
assembly, reads the information contained in the mRNA molecule
and synthesizes a protein molecule from the sequence on the mRNA
molecule.
Thus, each protein molecule is a product of the gene that codes
for it.
In turn, proteins are responsible for carrying out various
functions inside the cell, including catalyzing reactions of the
metabolic network as enzymes, building various cellular structures,
etc.

The transcription of a gene to an mRNA molecule is also regulated
by proteins.
The proteins that regulate the expression of genes inside
cells are referred to as transcription factors.
A transcription factor may activate or inhibit the
expression of a gene inside the cell by binding to regions
upstream or downstream of the gene on the DNA molecule.
This process may in turn facilitate or prevent RNA
polymerase and rest of the transcription machinery
from binding and initiating the transcription of the gene.
Thus, the genes inside cells interact amongst each other
via intermediate transcription factors to influence each
other's expression.
This network of interacting genes inside the cell is
referred to as the transcriptional regulatory network.
This network may be represented by a graph in which every
gene is represented by a node, and where an arrow from
one gene node to another means that the former codes for a
transcription factor that regulates the transcription of
the latter gene.

A more detailed description of the transcriptional
regulatory network would require us to characterize the
regulatory links in the network as activating or
repressing.
Further, an input function needs to be specified for each
gene node when more than one transcription factor
regulates the expression of that gene in the network.
In chapters 3 and 4 of this thesis, we have studied the
part of the transcriptional regulatory network in
{\it E.coli} that controls metabolism.

\subsection{Other interaction networks}

Inside the cell, the proteins interact with each other
to influence each other's activity.
Extracellular signals are mediated to the inside of a cell
by protein-protein interactions of signaling molecules.
Proteins that interact for a long time with each other
structurally can form part of a protein complex.
A protein may also act as a carrier for another protein.
A protein may interact briefly with another protein leading
to a transfer of a phosphate group.
These protein-protein interactions are crucial for living
systems.
We can represent various protein-protein interactions inside
the cell by a network with nodes as proteins and links between
two nodes representing a physical interaction
between two proteins.
Such networks are referred to as protein-protein interaction
networks.
The links between two nodes can be of various different types
corresponding to different types of interactions between proteins.

\subsection{Cell as a network of networks}

It is important to emphasize that the metabolic,
transcriptional regulatory and protein-protein interaction
networks mentioned above are not independent of each other
inside the cell.
The state of the genes in the transcriptional regulatory
network determines the activity of the metabolic network.
The concentration of metabolites in the metabolic
network determines the activity of transcription factors or
proteins which regulate the expression of genes in the
regulatory network.
The protein-protein interactions determine the activity of
various proteins inside the cell.
Thus, the above mentioned and other biochemical networks
together with the interactions at the interface of these
networks form a `network of networks' inside the cell that
determines the overall behaviour of the organism.

\section{Architectural features of large scale biological
networks}

Recent advances in the development of high-throughput
data collection techniques coupled with the systematic analysis
of fully sequenced genomes has generated detailed lists of
molecular components inside various organisms.
The available information has led to the mapping of
different cellular networks inside many organisms.
In this section, we review some of the known features of
large scale metabolic and transcriptional regulatory
networks.
This also enables us to introduce more concretely the work
done in this thesis.

\subsection{Metabolic networks}

The knowledge of enzymes along with their functional
assignments have led to a reconstruction of nearly
complete lists of organism specific metabolic reactions
\cite{GCRG,KEGG,Ecocyc}.
Jeong {\it et al} \cite{JTAOB2000} studied the structure
of the metabolic networks inside 43 different organisms.
They represented the metabolic network as a bipartite graph
with two types of nodes: metabolites and reactions.
The degree of a node is defined as the number of links
attached to that node in the graph
\cite{Hararybook,Bollobasbook}.
Since the metabolic network is a directed graph, each
metabolite in the network has an in-degree and an
out-degree.
The in-degree of a node denotes the number of links that
the node has to other nodes in the directed graph.
The out-degree of a node denotes the number of links that
start from the node to other nodes in the directed graph.
The degree distribution of a graph, $P(k)$, gives the
probability that a randomly selected node has exactly
$k$ links in the graph \cite{Hararybook,Bollobasbook}.
Jeong {\it et al} found both the in-degree and out-degree
distribution of metabolites to approximate a power law
form $P(k) \sim k^{-\gamma}$ for the metabolic networks
inside 43 organisms.
$\gamma$ is the degree exponent which they found to be
universal and close to 2.2 for all the organisms studied,
both for the in-degree and out-degree distribution.

Independently, Wagner and Fell \cite{WF2001} studied the
large scale metabolic network of {\it E. coli}.
They represented the {\it E. coli} metabolic network as
two different unipartite graphs: metabolite graph and
reaction graph \cite{WF2001}.
Wagner and Fell also found the connectivity of the
metabolites in the network to follow a power law
distribution.
The two independent studies by Jeong {\it et al}
\cite{JTAOB2000} and Wagner and Fell \cite{WF2001} show
that most metabolites participate in a few reactions
while there are a few metabolites which participate
in many reactions in metabolite networks.
The ubiquitous metabolites such as ATP that participate in
several reactions and have a high degree are also referred
to as hubs of the network.

The distance or shortest path between nodes $i$ and $j$ in
a graph is defined as the minimum number of links that have
to be traversed to reach from node $i$ to $j$.
The average path length of a graph is defined as the average
over the shortest paths between all pairs of nodes in the
network \cite{Hararybook,Bollobasbook}.
The diameter of a graph is defined as the supremum of the
shortest paths between all pairs of nodes in the network
\cite{Hararybook,Bollobasbook}.
The studies by Jeong {\it et al} \cite{JTAOB2000} and Wagner
and Fell \cite{WF2001} found the metabolic networks inside
organisms to have the small-world \cite{WS1998} property,
i.e., any two nodes in the system can be connected by relatively
short paths along existing links.
Jeong {\it et al} showed that the sequential removal of the
high degree nodes or hubs from metabolic networks results
in a sharp rise of network diameter as the network
disintegrates into small isolated clusters.
On the other hand, when they removed a set of randomly chosen
metabolite nodes from the network, the average path length
between the remaining nodes was not affected \cite{JTAOB2000}.
This observation led them to conclude that the hubs of the
metabolic network are crucial for maintaining functionality
of metabolic networks.

Ma and Zeng \cite{MZ2003} have further explored the global
connectivity structure of metabolic networks by classifying
nodes in the metabolite graph into four subsets based on
their mutual connectivity properties and location in the
network: a giant strong component, in-component,
out-component and an isolated subset.
Two nodes $i$ and $j$ are said to strongly connected in a
directed graph, if there is a path from $i$ to $j$ and from
$j$ to $i$ in the graph.
A strong component is a subset of nodes in the directed graph
such that for any pair of nodes $i$ and $j$ in the subset there
is a path from $i$ to $j$ and $j$ to $i$ in the directed graph.
The largest strong component of a directed network is referred
to as the giant strong component.
The set of nodes that are not in the giant strong component
but from which the nodes in the giant component can be reached
forms the in-component.
The set of nodes that are not in the giant strong component
but that can be reached from the nodes in the giant component
forms the out-component.
The set of nodes that have no path to the nodes in the giant
strong component forms the isolated subset.
The decomposition of the metabolite nodes into the above
mentioned four connected components revealed a `bow-tie'
macroscopic structure of the metabolic network \cite{MZ2003}.
The bow-tie structure of the metabolic network was similar
to that observed by Broder {\it et al} for the World Wide
Web \cite{BKMRRSTW2000}.
In uncovering the bow-tie structure, Ma and Zeng removed the
connections through the ubiquitous currency metabolites in the
metabolic network.
Csete and Doyle have argued that the bow tie architecture
of the metabolic network with a conserved core and
plug-and-play modularity around core can contribute toward
robustness and evolvability of the system \cite{CD2004}.

Ma and Zeng \cite{MZ2003} also tried to account for the
preferred directionality of reactions in the graph for the
metabolic network, and found the average path length between
metabolites to be almost double of that observed by Jeong
{\it et al}.
The average path length between nodes in the giant strong
component was found to determine the average path length of the
whole network.
An alternative study by Arita \cite{A2004} tried to account
for the actual structural changes in connecting metabolites in
the graph of the {\it E. coli} metabolic network, and found
again the average path length to be almost double of that
observed by Jeong {\it et al}.
Thus, accounting for the directionality of reactions, activity
of reactions and functional transfer of biochemical groups in
a more biologically meaningful graph representation of the
metabolic network gives an average path length larger than that
obtained by Jeong {\it et al}.

The above mentioned studies of the structure of the metabolic
networks have revealed a large variation in the metabolite
connectivity inside these networks.
It has been suggested that one of the important consequences
of power law degree distribution is the vulnerability of the
network to selective attack on hubs while being robust to random
deletion of nodes from the network as most nodes are of low
degree and their deletion does not change the average path length
between remaining nodes in the network \cite{AJB2000}.
For the protein-protein interaction network of
{\it S. cerevisiae}, it was shown that the essentiality of a
protein is correlated with its degree in the network
\cite{JMBO2001}.
This observation has been suggested as evidence for the
importance of the hubs in maintaining the overall structure
and function of cellular networks.
Although the role of high degree metabolites or hubs in
maintaining the overall structure of the metabolic networks
has been well emphasized in the literature, the role of low
degree metabolites has attracted little or no attention.

In chapter 2 of this thesis, we show that certain low degree
metabolites introduce fragility for flows in metabolic network.
There we have used a computational method to determine essential
reactions for growth in the metabolic networks inside three
organisms.
A reaction is designated as `essential' if its knockout from
the metabolic network renders the organism unviable.
We show that the low degree metabolites as opposed to high
degree metabolites explain essential reactions in
metabolic networks \cite{SSGKRJ2006}.
It is the low degree metabolites that are critical from the
point of view of functional robustness of the system.

In chapter 2, we also show that certain low degree
metabolites lead to clusters of reactions with highly
correlated reaction fluxes in the metabolic network.
We then show that genes corresponding to reactions of
such clusters predict regulatory modules in {\it E. coli}.
Thus, the modularity observed by us at the metabolic level
is also reflected at the genetic level.
Our work therefore shows that low degree metabolites play a role in
two hitherto unconnected properties of biological networks:
on the one hand they cause certain reactions to become
essential for the viability of the organism, and on the other
hand they contribute to the modularity of biological networks.

\subsection{Transcriptional regulatory networks}

The available information regarding the target genes of
transcription factors has led to a reconstruction of
transcriptional regulatory networks inside model
organisms like {\it E. coli} and {\it S. cerevisiae}
\cite{RegulonDB,SMMA2002,LRROBGHHTSZJMGRWTVFGY2002,GBBK2002,LBYSTG2004}.
The presently available transcriptional regulatory maps
are highly incomplete due to ongoing annotation of the
fully sequenced genomes.
It has been estimated \cite{CKRHP2004} that the coverage
of the known transcriptional regulatory network of
{\it E. coli} is only about $25\%$ of the actual network
inside the organism.

The out-degree distribution for the known transcriptional
regulatory network of {\it E. coli} and {\it S. cerevisiae}
was shown to approximate a power law.
However, the in-degree
distribution for the two networks followed a restricted
exponential function \cite{SMMA2002,LRROBGHHTSZJMGRWTVFGY2002}.
This example shows that not all biological networks are
characterized by a power law degree distribution of nodes.
The out-degree of a node in the transcriptional regulatory
network represents the number of target genes regulated by
a transcription factor.
The in-degree of a node in the transcriptional regulatory
network represents the number of transcription factors
regulating a target gene.
The exponential in-degree distribution for the transcriptional
regulatory network suggests that a very large promoter region
required for the combinatorial regulation of a target gene by
many transcription factors is highly unlikely inside the cell.

`Network motifs' have been defined as patterns of
interconnections or subgraphs that are over-represented in
a real network compared to randomized versions of the same
network with similar local connectivity
\cite{SMMA2002,MSIKCA2002}.
Network motifs can be detected by algorithms that compare
the patterns found in the real network to those found in
suitably randomized networks.
The method is analogous to detection of sequence motifs in
genomes as recurring sequences that are very rare in random
sequences.
By studying the {\it E. coli} transcriptional regulatory
network, Alon and colleagues found that the `feed-forward
loop' (FFL) is a motif in the regulatory network
\cite{SMMA2002}.
The structure of a feed-forward loop (FFL) motif is defined
by a transcription factor X that regulates a second
transcription factor Y, such that both X and Y jointly
regulate a gene or operon Z.
Other motifs found in the {\it E. coli} transcriptional
regulatory network include the `single-input module' (SIM)
and `dense overlapping regulons' (DOR) \cite{SMMA2002}.
The structure of single-input module (SIM) is defined by
a set of genes or operons that are controlled by a single
transcription factor.
The structure of dense overlapping regulons (DOR) is
defined by a layer of overlapping interactions between genes
or operons and a group of input transcription factors.
Later, the motifs found in the {\it E. coli} transcriptional
regulatory network were also found in the {\it S. cerevisiae}
transcriptional regulatory network
\cite{LRROBGHHTSZJMGRWTVFGY2002,MSIKCA2002}.

By studying the dynamics of various motifs found in the
regulatory networks, it has been shown that these motifs may
perform important information processing tasks.
The coherent feed-forward loop motif has been shown to
filter out noise or spurious signals in the network
\cite{SMMA2002,MA2003}.
The single-input module motif was shown to help generate
temporal programs of gene expression
\cite{SMMA2002,RRSA2002}.
Recent studies have shown that the appearance of same motifs
in regulatory networks of {\it E. coli} and {\it S. cerevisiae}
does not necessarily imply that these motifs are
evolutionarily conserved \cite{CW2003,W2003b}.
Evolution seems to have converged on the same patterns of
interconnections in different organisms after a lot of
tinkering perhaps due to the specific information processing
tasks these motifs perform inside a cell \cite{A2003,W2003b}.
An objective of the exercise of detecting motifs in different
biological networks is to create a library of motifs and their
possible functions.

Ma {\it et al} \cite{MBZ2004} tried to decompose the
{\it E. coli} transcriptional regulatory network into
various connected components.
They found that there were no strong components in the
{\it E. coli} transcriptional regulatory network.
This was consistent with earlier observations by Shen-Orr
{\it et al} \cite{SMMA2002} that the network had no cycles
of length $\ge$ 2.
There were only autoregulatory loops in the presently known
transcriptional regulatory network of {\it E. coli}
\cite{SMMA2002}.
Ma {\it et al} \cite{MBZ2004} found the structure of the
{\it E. coli} transcriptional regulatory network to be
hierarchical.
A similar lack of strong components or cycles was also
observed for the transcriptional regulatory network of
{\it S. cerevisiae} \cite{GBBK2002}.
The transcriptional regulatory network of {\it E. coli} had
a five layered hierarchical architecture with genes at the
top layer having no incoming links.
Balaszi {\it et al} found that the genes at the top layer
are regulated by distinct environmental signals in the
transcriptional regulatory network of {\it E. coli}
\cite{BBO2005}.

In chapters 3 and 4 of this thesis, we have studied the
large scale structure and system level dynamics of the
transcriptional regulatory network controlling metabolism
in {\it E. coli}.
Our study reinforces the hierarchical, essentially acyclic
structure with environmental control of the genes belonging
to the top layer of the regulatory network of {\it E. coli},
also found by previous studies mentioned above.
Further, our dynamical study of regulatory network of
{\it E. coli} metabolism elucidates the functional
consequences of this observed architecture of the network
\cite{SJ2008}.
We show that the regulatory network of {\it E. coli}
metabolism exhibits two types of robustness.
One, the regulatory network of {\it E. coli} metabolism
exhibits an insensitivity to perturbations of gene
configurations for a fixed environment, i.e., the system
returns to the same attractor when gene configurations are
perturbed for a fixed environment.
Two, the regulatory network of {\it E. coli} metabolism
exhibits a flexible response to changed environments,
i.e., the system moves to a new attractor that enables
it to maintain its key functionality when it encounters a
changed environment in a sustained manner.
The hierarchical acyclic architecture of the regulatory
network of {\it E. coli} metabolism with control
variables as external metabolites explains the observed
robust dynamics of the system.
Further, we observe a highly disconnected and modular
architecture at the intermediate level of the hierarchical
graph of the regulatory network.
We find that the modules at the intermediate level of this
hierarchical graph are regulated by different sets of
environmental signals, and the modules interact only at
the lowest level of the graph contributing to the robust
response of the system to changed environments.
This modular architecture of the regulatory network may
also contribute towards the evolvability of the system.
Thus, our study sheds new light on how structural design
features of the regulatory network of {\it E. coli}
contribute towards robustness and modularity of the
system.

\section{Methods for studying system level dynamics of large
scale biological networks}

Our work mentioned in the previous section employs graph
theoretic and statistical methods to describe the structure
of large scale metabolic and genetic regulatory network (like
the other works reviewed in that section), but it also goes
beyond structure to investigate flows and other dynamical
phenomena.
In this section, we mention some dynamical methods used for
studying biological networks and discuss the kind of methods
that are appropriate for a systems level study.

Many differential equations based models have been proposed
and studied extensively to understand the dynamics of
subsystems or pathways inside organisms.
Two of the best examples are the {\it E. coli} chemotactic
pathway and the segment polarity network of {\it Drosophila
melanogaster}.
Barkai and Leibler have studied extensively a kinetic model for
the {\it E. coli} chemotactic pathway.
They showed that the property of chemotactic adaptation
(whereby a cell resets its tumbling frequency to the
same basal value after a change of chemoattractant
concentration) is robust to variations in a specific set of kinetic
parameters.
In their model, the robust behaviour of the system was
a consequence of negative feedback, and this was later
confirmed experimentally \cite{ASBL1999}.
Yi {\it et al} showed that the robust behaviour of the
{\it E. coli} chemotactic pathway is a consequence of a specific
type of negative feedback control strategy, namely, integral
feedback control \cite{YHSD2000} which is a commonly used
strategy in engineering.
von Dassow {\it et al} modelled the segment polarity network of
{\it Drosophila melanogaster} using differential equations and
showed that the spatial distribution of gene expression patterns
was robust to changes in a set of initial conditions, rate
constants or genetic perturbations \cite{VMMO2000}.
They showed that positive feedback contributes to robustness
in the model for the segment polarity network by amplifying
the stimuli and enhancing the sensitivity of the system.
Since then Ingolia has analyzed the model by von Dassow
{\it et al} and showed that the bistability caused by positive
feedback loops is responsible for the robust patten formation
\cite{I2004}.

The above mentioned examples of differential equations based
models for subsystems inside cells containing a few nodes have
provided important insights about the robust behaviour of
these subsystems.
However, we expect to gain qualitatively different insights
regarding the dynamical behaviour at the whole cell level by
studying the dynamics of large scale biological networks that
incorporate the collective functioning of a substantial fraction
of the nodes in the system, compared to those obtained by
studying the dynamics of smaller subsystems.
At present, there is limited knowledge of kinetic parameters
such as rate constants, enzyme concentrations, etc., for large
scale biological networks.
Further, the measured kinetic parameters for a given cell may
show wide variation across the population of cells.

Due to paucity of kinetic data, a differential equation based
simulation of large scale biological networks is not feasible
at present and the large number of unknown parameters would
also render the results of such a simulation difficult to
interpret \cite{B2005}.
While deciding on a modelling approach to a large scale biological
system, we need to account for the available knowledge
about the system being studied.
The choice of the method and the level of abstraction would also
depend upon the questions we wish to address for the biological
system at hand.
In the absence of large scale kinetic data, alternative modelling
approaches such as flux balance analysis (FBA) and Boolean
networks can be used to simulate the dynamics of metabolic networks
and genetic regulatory networks, respectively.

At present, the list of reactions along with the stoichiometric
coefficients of the involved metabolites is largely known for
metabolic networks inside many single celled organisms.
However, we currently lack the knowledge of kinetic rate constants
for most reactions that can occur inside the cell.
Due to lack of kinetic data, constraint based modelling approaches
such as flux balance analysis (FBA)
\cite{VP1994a,EP2000,EIP2001,Palssonbook} can be used to perform a
steady state analysis of the large scale metabolic networks.
FBA is a computational technique that can be used to determine the
steady state fluxes of all reactions in the metabolic network and
predict the growth rate of the cell for a given nutrient medium.
The key requirement for FBA technique is the knowledge of network
structure along with stoichiometric coefficients of the involved
metabolites which is largely known for many organisms.
The predictions of FBA for few reaction fluxes and growth rate of
{\it E. coli} under few minimal media have been shown to have good
agreement with experimentally measured values
\cite{EIP2001,IEP2002}.
In chapter 2 of this thesis, we have used FBA to determine
essential reactions for growth in the metabolic networks for
{\it E. coli}, {\it S. cerevisiae} and {\it S. aureus}.

In the case of genetic regulatory networks, the current
availability of biological data is limited to network
structure and the information regarding the nature of
regulatory links, i.e., activating or repressing.
When more than one regulatory link converges at a
single gene in the network, an input function needs to be
specified for the gene.
In the absence of quantitative data on genetic regulatory
networks, the Boolean network approach may be used to
perform qualitative simulations.
Kauffman proposed the framework of Boolean networks to
study the dynamics of genetic regulatory networks
\cite{K1969a,K1969b,Kauffmanbook}.
The Boolean approach provides a coarse grained description
of the dynamics of genetic regulatory networks where each
gene in the network is in one of the two states:
active or inactive.
In this approach, the state of each gene at a given time
instant is determined by the state of its input genes at
the previous time instant based on a Boolean input function.
The input function may be written in terms of the AND, OR
and NOT Boolean operators.
This is a discrete dynamical system;
the genes' states may be updated synchronously or
asynchronously.
Boolean network models of small cellular subsystems have
also provided useful biological insights
\cite{ST2001,AO2003,LLLOT2004,EPA2004,LAA2006}.
Recently, two databases for large scale transcriptional
regulatory networks inside model organisms, {\it E. coli}
and {\it S. cerevisiae}, have been reconstructed using
empirical data that contain both the network structure
and Boolean input functions \cite{CKRHP2004,HLPP2006}.
In chapter 3 of this thesis, we have used the Boolean
approach to study the dynamics of the large scale genetic
network controlling {\it E. coli} metabolism as
represented in the database iMC1010$^{v1}$ \cite{CKRHP2004}.

\section{Thesis organization}

The subsequent chapters in this thesis are organized as
follows:

\begin{itemize}

\item {\bf Chapter 2} studies the structure and dynamics of
the metabolic networks inside three organisms.
We determine metabolites based on their low degree of
connectivity in the metabolic network.
We show that certain low degree metabolites lead to clusters
of highly correlated reactions in the metabolic network.
We find that these clusters at the metabolic level correspond
to regulatory modules at the genetic level.
The computational technique of flux balance analysis (FBA)
is then used to determine `essential' reactions for growth
in the metabolic networks of {\it E. coli}, {\it S. cerevisiae}
and {\it S. aureus}.
We show that most essential reactions in metabolic networks
are explained by their association with a low degree
metabolite.
In this chapter, we show that low degree metabolites are
implicated in two seemingly unrelated properties in
metabolic networks: modularity and essentiality.

\item {\bf Chapter 3} studies in detail the dynamics of the
large scale genetic network controlling {\it E. coli}
metabolism.
Using the information contained in a previously published
database \cite{CKRHP2004} representing the genetic network
controlling {\it E. coli} metabolism, we construct an
effective Boolean dynamical system of genes and external
metabolites describing the network.
We study the dependence of the attractors of this Boolean
dynamical system on the initial conditions of genes and
state of the external environments.
We find that the attractors of the Boolean dynamical system
are fixed points or low period cycles for any fixed
environment.
We show that the system exhibits the property of homeostasis
in that the attractor is highly insensitive to initial
conditions or perturbation of genes for a fixed environment.
However, we find that the attractors corresponding to
different environments have a wide variation.
We also show that for most environmental conditions, the
attractors of the genetic network allow close to optimal
metabolic growth.
In this chapter, we show that the genetic network controlling
{\it E. coli} metabolism simultaneously exhibits the twin
dynamical properties of homeostasis and flexibility of
response.

\item {\bf Chapter 4} studies the design features of the
genetic network controlling {\it E. coli} metabolism in order
to understand the origin of observed dynamical properties
of homeostasis and response flexibility.
We find that the genetic network controlling {\it E. coli}
metabolism is an essentially acyclic graph.
The root nodes of this acyclic graph are external metabolites
that act as control variables of the dynamical system.
The leaf nodes of the acyclic graph are the genes coding for
enzymes while the genes coding for transcription factors are
at the intermediate level.
We shown that deleting the leaf nodes corresponding to the enzyme
coding genes along with their links from the full graph leads to
a subgraph with many disconnected components that may be regarded
as modules of the genetic network.
The localization and dynamical autonomy of the disconnected
components or modules may contribute towards evolvability of
the network.
In this chapter, it is shown that the architecture of the genetic
network endows the system with the twin properties of homeostasis
and flexibility of response.

\item {\bf Chapter 5} is a perspective of the work reported
in this thesis in relation to the overall subject.
It discusses some of the limitations associated with
this work.
It also suggests some future directions of research based on
work reported here.

\item {\bf Appendix A} lists the 85 UP-UC clusters in the
{\it E. coli} metabolic network.

\item {\bf Appendix B} reviews the computational technique
of flux balance analysis (FBA).

\item {\bf Appendix C} describes various computer programs
used to obtain results reported in this thesis.
These programs can be downloaded from the associated
website: {\it http://areejit.samal.googlepages.com/programs}.

\end{itemize}


\chapter{Low degree metabolites enhance modularity and explain essential reactions in metabolic networks}
\label{lowdeg}

In this chapter, we have studied the metabolic networks of {\it Escherichia coli},
{\it Saccharomyces cerevisiae} and {\it Staphylococcus aureus}.
We first locate metabolites based purely on their low degree in the metabolic
network.
We then show that certain low degree metabolites contribute to a rigidity or coherence
of reaction fluxes in the metabolic network resulting in clusters of highly correlated
reactions.
We find that these clusters of metabolic reactions in the {\it E. coli} metabolic
network predict genetic regulatory modules, as captured in the structure of operons,
with a high probability.
We then use a computational method to determine the essential reactions for growth in
the metabolic networks of {\it E. coli}, {\it S. cerevisiae} and {\it S. aureus}.
We show that most essential metabolic reactions in {\it E. coli}, {\it S. cerevisiae}
and {\it S. aureus} can be explained by the fact that they are associated with a low
degree metabolite.


\section{`Uniquely Produced' (`Uniquely Consumed') metabolites and their
associated reactions}

It is convenient to represent the metabolic network as a bipartite graph
consisting of two types of nodes: metabolites and reactions.
In a directed bipartite graph, there are two types of links:
(a) from metabolite nodes to reaction nodes and
(b) from reaction nodes to metabolite nodes.
The first type of links defines the reactants.
The second type of links defines the products.
In a bipartite graph, there are no links between two similar
types of nodes.

We have designated a metabolite as `uniquely produced' or `UP'
(`uniquely consumed' or `UC'), if there is only a single
reaction in the metabolic network that produces (consumes)
the metabolite \cite{SSGKRJ2006}.
A UP(UC) metabolite has in-degree (out-degree) equal to
unity in the bipartite graph.
A metabolite that is both UP and UC may be designated as a
`UP-UC metabolite' \cite{SSGKRJ2006}.
A UP-UC metabolite has both in-degree and out-degree equal to unity.
Such a metabolite has degree two in the network.
In general, a metabolite that is either UP or UC or both has a
low degree in the metabolic network as it participates in very
few reactions.

We have designated a reaction as `uniquely producing' or `UP'
(`uniquely consuming' or `UC'), if it produced (consumed) a
UP(UC) metabolite in the bipartite metabolic network
\cite{SSGKRJ2006}.
A reaction is UP(UC), if it is the only process by which some
metabolite can be produced (consumed) in the complete metabolic
network.
We designate reactions in the metabolic network that
are either UP or UC or both as `UP/UC reactions'.

The metabolic network is an input-output network which takes in
nutrients from the external environment as inputs and produces
key molecules contributing towards growth and maintenance of the
cell as outputs.
In this chapter, we have studied the metabolic networks inside
three organisms:  {\it E. coli} (version iJR904 \cite{RVSP2003}),
{\it S. cerevisiae} (version iND750 \cite{DHP2004}) and
{\it S. aureus} (version iSB619 \cite{BP2005}).
The databases iJR904, iND750 and iSB619 for {\it E. coli},
{\it S. cerevisiae} and {\it S. aureus}, respectively, have been
reconstructed using the annotation of fully sequenced genomes for
these organisms and biochemical literature sources.
The databases were downloaded from the website \cite{GCRG}.
The reactions inside these metabolic network databases can be
broadly classified into internal and transport reactions.
The transport reactions in the metabolic network represent
transport processes of metabolites across the cell boundary.
The internal reactions in the metabolic network are confined
to the cell boundary.
In addition to internal and transport reactions, the metabolic
network databases considered here contain a fictitious reaction
referred to as the biomass reaction representing the ratios of
various metabolic precursors that are required for unit biomass
production or growth of the organism.

For convenience, the metabolites that can be transported across
the cell boundary are represented by two nodes in the metabolic
network databases considered here.
One of the nodes represents the external version of the metabolite
and the other node represents the internal version of the
metabolite, and the transport of the metabolite across the cell
boundary is treated as a dynamical reaction in the above mentioned
databases converting one type of node into another.
In the databases considered here, most external metabolites are
usually involved in only two unidirectional transport reactions in
the metabolic network representing their transport process across
the cell boundary.
One of the two reactions transports a external metabolite into
the cell while the other transports it outside the cell.
Thus, most external metabolites in these metabolic network databases
satisfy the property of UP or UC or both.
We do not consider the external metabolites while determining the set
of UP(UC) metabolites in the network as the external metabolite nodes
are a matter of convention in the databases.
We consider only the internal metabolites while determining the set
of UP(UC) metabolites in the network.


\subsection{Detection of UP(UC) metabolites and reactions}
\label{bipartitematrix}

The list of reactions in a reconstructed metabolic network
for an organism includes both reversible and irreversible
reactions.
Starting from the reconstructed metabolic network of an organism,
we prepare a list of metabolic reactions that has each reversible
reaction in the original network replaced by two unidirectional
reactions (one reaction each for the forward and the backward
direction).
From this list of unidirectional metabolic reactions, construct a matrix
${\bf A}$ = $(A_{ij})$ of dimensions $m \times n$, where $m$ is the
number of internal metabolites and $n$ is the number of reactions
in the network.
The rows of matrix ${\bf A}$ correspond to metabolites and the columns
correspond to reactions in the metabolic network.
The matrix element $A_{ij}$ is set equal to -1, if metabolite $i$
is consumed in reaction $j$, +1 if metabolite $i$ is produced in
reaction $j$, and 0 if metabolite $i$ does not participate in
reaction $j$.
The matrix ${\bf A}$ is a compact representation of the bipartite
metabolic graph.

A UP(UC) metabolite has in-degree (out-degree) equal to unity in the
bipartite metabolic graph.
A metabolite $i$ in the network is UP(UC), if the $i^{th}$ row of
matrix ${\bf A}$ has exactly one entry that equals -1 (+1).
A UP-UC metabolite has in-degree and out-degree equal to unity in
the bipartite metabolic graph.
A metabolite $i$ in the network is UP-UC, if the $i^{th}$ row of matrix
${\bf A}$ has exactly one entry that equals -1, one entry that equals
+1 and has all other entries 0.
A reaction $j$ is UP(UC), if the $j^{th}$ column of matrix ${\bf A}$
has at least one entry +1 (-1) such that the row corresponding to the
entry +1 (-1) in column $j$ has no other entry equal to +1 (-1).

As mentioned earlier, we do not consider the external metabolites while
determining the set of UP(UC) metabolites.
This amounts to excluding the rows corresponding to external metabolites
from the matrix ${\bf A}$ for the computation of UP(UC) metabolites.
Thus, the matrix ${\bf A}$ does not contain any rows corresponding to
external metabolites and the rows of matrix ${\bf A}$ correspond
only to internal metabolites in the network.
The biomass reaction which defines the ratios of various metabolic
precursors that are required for the unit biomass production of the
organism is also included in the list of reactions while determining
the UP(UC) metabolites.
Usually, the last column of the bipartite matrix ${\bf A}$ corresponds
to the biomass reaction.


\subsection{UP(UC) statistics for the metabolic networks of
{\it E. coli}, {\it S. cerevisiae} and {\it S. aureus}}

The databases were downloaded from the website \cite{GCRG}.
The {\it E. coli} metabolic network iJR904 accounts
for 761 metabolites participating in 931 reactions
(686 irreversible and 245 reversible reactions).
The {\it S. cerevisiae} metabolic network iND750
accounts for 1061 metabolites participating in 1149 reactions
(719 irreversible and 430 reversible reactions).
The {\it S. aureus} metabolic network iSB619 accounts
for 645 metabolites participating in 644 reactions
(423 irreversible and 221 reversible reactions)..
Following the steps outlined in section \ref{bipartitematrix},
the bipartite matrix ${\bf A}$ was constructed for the three
metabolic networks.
We found the dimensions ($m$,$n$) of matrix ${\bf A}$ to be
(618,1177), (945,1580) and (561,866) for the metabolic networks
of {\it E. coli}, {\it S. cerevisiae} and {\it S. aureus},
respectively.
In obtaining the bipartite matrix ${\bf A}$ for the three metabolic
networks, each reversible reaction was converted into two one sided
reactions.
Further, the biomass reaction was added to the list of metabolic
reactions.
The number of external metabolites in the metabolic networks of
{\it E. coli}, {\it S. cerevisiae} and {\it S. aureus} was 143,
116 and 84, respectively.
The rows corresponding to these external metabolites were not
included in the matrix ${\bf A}$ for determining UP or UC
metabolites.

\begin{sidewaysfigure}
\centering
\includegraphics[width=20cm]{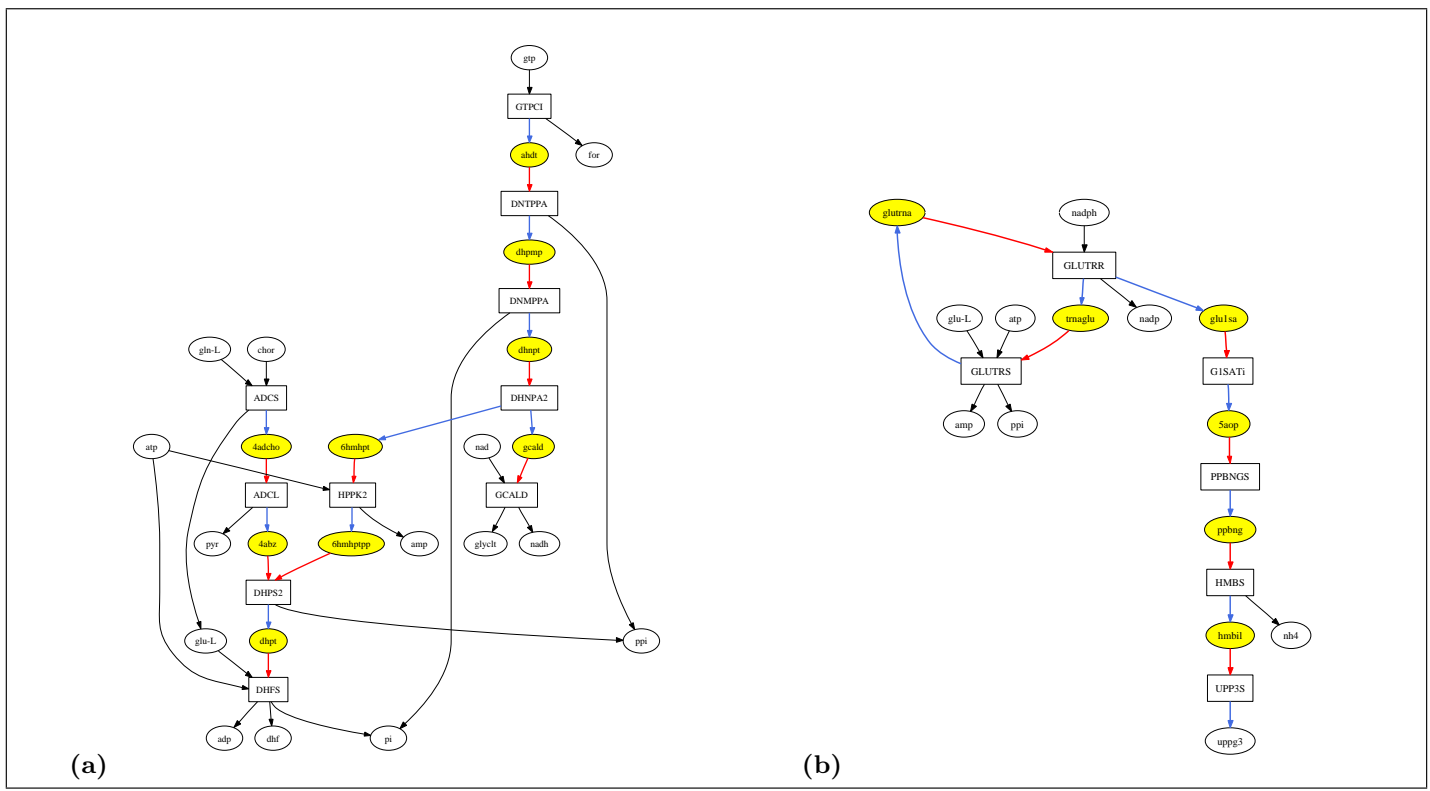}
\renewcommand{\captionfont}{\scriptsize}
\caption{
(a) UP-UC metabolites in the {\it E. coli} metabolic
network forming a UP-UC cluster of 10 reactions.
(b) UP-UC metabolites in the {\it S. aureus} metabolic network
forming a UP-UC cluster of 6 reactions.
In this figure, rectangles represent reactions and ovals
metabolites.
Yellow ovals represent UP-UC metabolites.
Arrows to (from) metabolites represent their production (consumption)
in reactions.
A blue (red) link represents the production (consumption) of a
UP(UC) metabolite.
Notice that UP-UC clusters are not strictly linear pathways.
For example, in part (a) the reactions in the cluster are not all in
a single chain and in part (b) there is a cycle inside the UP-UC
cluster.
Nevertheless fixing the flux of any one reaction in a UP-UC cluster
fixes the fluxes of all other reactions in the cluster in any steady
state, since the production rate of every UP-UC metabolite must be
the same as its consumption rate.
Hence, in part (a), fixing the flux of reaction GCALD fixes the flux
of reaction DHNPA2 (because of the intermediate UP-UC metabolite gcald),
which in turn fixes the fluxes of reactions HPPK2 and DNMPPA, and so
on.
To reduce clutter, nodes corresponding to $h$ (proton) and $h_2o$ have
been omitted.
Abbreviation of metabolite and reaction names in part (a) are as
in \cite{RVSP2003} and in part (b) as in \cite{BP2005}.
The figure has been drawn using Graphviz software
\cite{Graphviz}.}
\label{upuccluster}
\end{sidewaysfigure}

Using the bipartite matrix ${\bf A}$ for each one of the three
organisms, we determined UP(UC) metabolites and reactions in the
metabolic networks of {\it E. coli}, {\it S. cerevisiae} and
{\it S. aureus}.
We found the number of UP(UC) metabolites in the metabolic
networks of {\it E. coli}, {\it S. cerevisiae} and {\it S. aureus}
to be 291 (285), 395 (376) and 282 (237), respectively.
We found the number of UP-UC metabolites in the metabolic
networks of {\it E. coli}, {\it S. cerevisiae} and {\it S. aureus}
to be 185, 178 and 145, respectively.
Examples of UP-UC metabolites in {\it E. coli} and {\it S. aureus}
networks are shown in Fig. \ref{upuccluster}.
We found the number of UP(UC) reactions in the metabolic networks of
{\it E. coli}, {\it S. cerevisiae} and {\it S. aureus} to be 289 (272),
391 (370) and 277 (218), respectively.
The number of reactions that were either UP or UC or both (the
`UP/UC reactions') in the metabolic networks of {\it E. coli},
{\it S. cerevisiae} and {\it S. aureus} were found to be 418,
583 and 376, respectively.


\section{`UP-UC cluster' of reactions}

A UP-UC metabolite has one reaction that produces it and one reaction
that consumes it in the metabolic network.
A steady state is defined as one where all metabolite concentrations and
reaction velocities are constant.
In any steady state, the flux of the reaction producing a UP-UC metabolite
is always proportional to the flux of the reaction consuming the metabolite,
with the proportionality constant determined by the stoichiometric
coefficients of the metabolite in the two reactions.
Then, maintaining the steady state requires the enzymes of the two reactions
associated with a UP-UC metabolite to be simultaneously active.
This raises the question as to whether the genes coding for enzymes of the two
reactions associated with a UP-UC metabolite are coexpressed.
We have defined a `UP-UC cluster' of reactions as a set of reactions connected
by UP-UC metabolites \cite{SSGKRJ2006}.
Examples of UP-UC clusters of reactions in the metabolic networks of {\it E. coli}
and {\it S. aureus} are shown in Fig. \ref{upuccluster}.
In steady state, fluxes of all reactions that are part of a single UP-UC cluster
are proportional to each other.
Fixing the flux of any reaction in a UP-UC cluster fixes the fluxes of all other
reactions in the cluster under steady state.
Further, for any steady state analysis, each UP-UC cluster can be replaced by
a single effective reaction and this can be used to coarse-grain metabolic
networks \cite{PSNMS1999,SKWMP2002}.
Notice that UP-UC clusters include linear pathways but can lead to branched
or cyclic structures as shown in Fig. \ref{upuccluster}.
UP-UC clusters of reactions are special cases of reaction/enzyme subsets
\cite{PSNMS1999,SKWMP2002,SKBSG2002}, co-sets \cite{PPP2002,RP2004}
and fully coupled reactions \cite{BNSM2004} that have been discussed
earlier in the literature.


\subsection{Algorithm to determine UP-UC clusters}
\label{clusteralgorithm}

We now describe in detail the algorithm to determine UP-UC clusters in any
metabolic network.

\begin{enumerate}

\item Starting from the reconstructed metabolic network of an organism,
construct the bipartite matrix ${\bf A}$ of dimensions $m \times n$,
where $m$ is the number of internal metabolites and $n$ is the number of reactions
in the network as described in section \ref{bipartitematrix}.

\item In the matrix ${\bf A}$, determine the rows corresponding to UP-UC
metabolites as described in section \ref{bipartitematrix}.

\item  Obtain a matrix ${\bf B}$ from matrix ${\bf A}$ by setting every
entry of each row in ${\bf A}$ that corresponds to a non UP-UC metabolite
equal to zero, i.e., delete all links in the graph except those going
into or out of UP-UC metabolites.

\item From the matrix ${\bf B}$, construct a reaction-reaction graph
in which each node corresponds to a reaction.
The $n \times n$ adjacency matrix ${\bf C}$ = $C_{jk}$ of this graph is
defined as $C_{jk}$ = 1 if $B_{ij}$ = 1 and $B_{ik}$ = -1, else $C_{jk}$ = 0.
The matrix ${\bf C}$ represents a directed graph.

\item The weak components of size $\ge$ 2 of the graph ${\bf C}$ are
the various UP-UC clusters.
These are obtained as follows:
First convert the directed graph ${\bf C}$ into the associated undirected graph
${\bf \tilde{C}}$ by dropping all the directions of the arrows, i.e.,
$\tilde{C}_{jk}$ = 1 if $C_{jk}$ = 1 or $C_{kj}$ = 1 or both, else
$\tilde{C}_{jk}$ = 0.
Two nodes $j$ and $k$ in ${\bf C}$ are said to be weakly connected if there
exists a path between them in the associated undirected graph ${\bf \tilde{C}}$.
A weak component is a maximal set of nodes that are weakly connected to each other.

By construction, if two reaction nodes $j$ and $k$ are adjacent (i.e., connected
by a link) in ${\bf \tilde{C}}$, there exists a UP-UC metabolite that is produced
in one of those reactions and consumed in the other.
Thus, the fluxes of those two reactions will have a constant ratio in all steady
states.
This logic extends to entire connected cluster in ${\bf \tilde{C}}$ to which those
reactions belong.
\end{enumerate}

\noindent Note that a choice has to be made as to whether to include or
exclude the biomass reaction from the list of reactions in matrix
${\bf A}$ and ${\bf B}$.
Its inclusion/exclusion gives slightly different results for the set of
UP-UC metabolites and their clusters.
The biomass reaction should be included in the matrix ${\bf A}$ in
steps 1-2 above (identification of UP-UC metabolites), for if it is
not, then those biomass metabolites which are consumed by only one
reaction other than the biomass reaction get identified as UC metabolites
resulting in some spurious UP-UC clusters.
However, in the matrix ${\bf B}$ (steps 3-5) it is a matter of convention
whether the biomass reaction is included or not; results in the two cases
are different but each is valid in its own right.
The results reported in this chapter correspond to the following choice:
In steps 1-2 above, the matrix  ${\bf A}$ includes the biomass reaction
and in steps 3-5 the matrix  ${\bf B}$ excludes it.
When the biomass reaction is included in the matrix ${\bf B}$ the size of
the largest UP-UC cluster increases.
A program to determine UP-UC clusters in the {\it E. coli} metabolic network
is contained in Appendix \ref{program}.


\section{UP-UC clusters predict regulatory modules in {\it E. coli}}
\label{operon}

\begin{figure*}
\centering
\includegraphics[width=14cm]{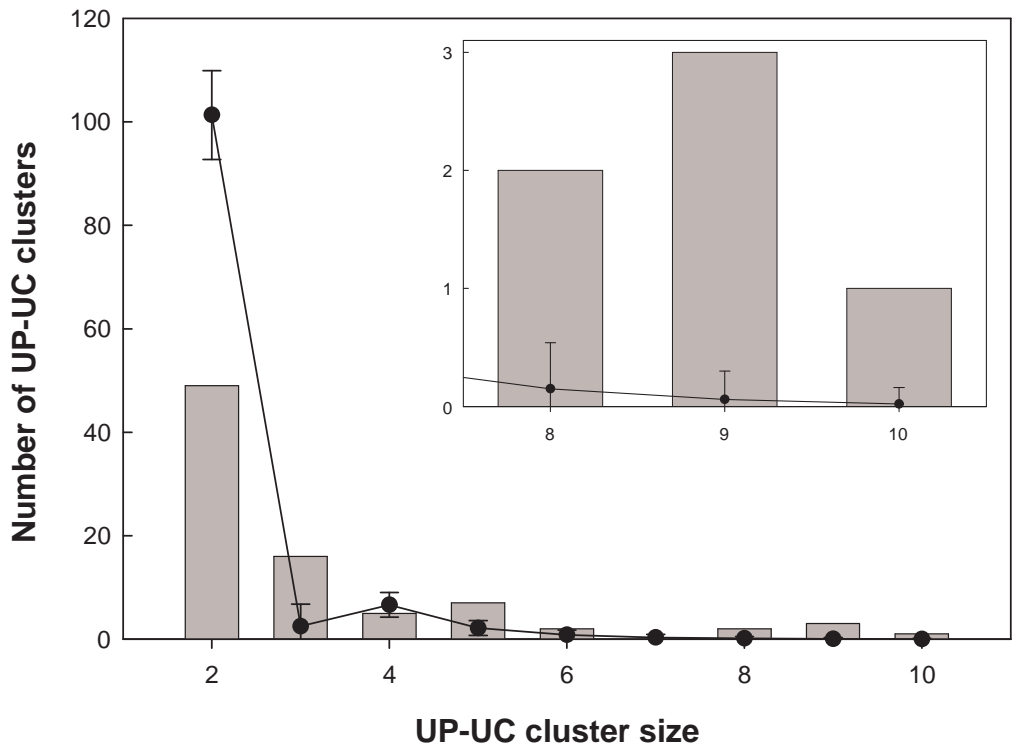}
\caption{
Frequency histogram of UP-UC cluster sizes in the
{\it E. coli} metabolic network (grey bars).
The data is shown in Table \ref{motiftable}.
The black line is the frequency distribution for the
randomized versions of the network (averaged over 1000
realizations) that preserve the in-degree and out-degree
of all nodes.
Error bars show one standard deviation of the randomized
ensemble.
{\bf Inset:} Enlargement of the graph for the larger sized
clusters.
In the real network, larger UP-UC clusters (size $\geq 8$)
occur much more often than in the randomized version
($p < 0.001$).
On the other hand, smaller UP-UC clusters (size $\leq 3$)
occur much less often than in the randomized version
($p < 0.001$).}
\label{motif}
\end{figure*}

We used the algorithm mentioned in section \ref{clusteralgorithm} to
determine UP-UC clusters in the {\it E. coli} metabolic network
iJR904.
The total number of UP-UC clusters in the {\it E. coli} metabolic network
was found to be 85.
The list of 85 UP-UC clusters in the {\it E. coli} metabolic network is
contained in Appendix \ref{upuclist}.
The size distribution of UP-UC clusters in the {\it E. coli} metabolic network
is shown by grey bars in Fig. \ref{motif} and listed in Table \ref{motiftable}.

Since the fluxes of reactions forming a UP-UC cluster have fixed ratios with
respect to each other for all steady states, the set of genes that code for
enzymes catalyzing various reactions of the cluster may be expected to be
coregulated forming a transcriptional module.
The bacteria {\it E. coli} is a prokaryote.
In prokaryotes, the genes are grouped into transcriptional modules called
operons.
An operon is a set of genes which are transcribed into a single mRNA molecule
that may code for more than one protein.
The set of genes that form a single operon are therefore guaranteed to be
coexpressed.
We investigated whether the genes coding for enzymes catalyzing reactions of
a UP-UC cluster are part of the same operon in {\it E. coli}.
At present, the genes corresponding to enzymes of reactions that constitute the
{\it E. coli} metabolic network iJR904 have been identified for only part of the
network.
Of the 85 UP-UC clusters in {\it E. coli}, only 69 UP-UC clusters had two or more
reactions with known corresponding genes.
The regulation of these 69 UP-UC clusters was investigated using the known operon
information from RegulonDB \cite{RegulonDB} and Ecocyc \cite{Ecocyc} databases for
{\it E. coli}.
Genes of reactions within 85 UP-UC clusters for {\it E. coli} that belong to the
same operon are indicated in the table listed in Appendix \ref{upuclist}.
For 42 of the 69 UP-UC clusters, two or more genes of the cluster were found to be
part of the same operon.
Furthermore, we found that 36 of these 42 UP-UC clusters had at least half of their
genes belonging to the same operon.
For 21 UP-UC clusters, we found all reactions in a cluster to be covered by the
same operon in the sense that at least one gene catalyzing each reaction in the
set belonged to the same operon.

The following test was performed to show that two genes belonging to a UP-UC cluster
in {\it E. coli} have greater probability of lying on the same operon than otherwise
expected.
We found that there were 251 unique genes catalyzing various reactions in the 69
UP-UC clusters.
If we randomly pick any two of these 251 genes, the probability that the two genes
lie on the same operon is 0.0057.
If we randomly pick a pair of genes that belong to the same UP-UC cluster from this
set of 251 genes, the probability that the two genes lie on the same operon is 0.29.
Thus, genes belonging to a UP-UC cluster have a much greater probability of being
coregulated than otherwise.
This shows that the set of genes that correspond to a UP-UC cluster in the {\it E. coli}
metabolic network are strongly correlated with regulatory modules at the genetic level.
Our analysis here rests only on the available operon data for {\it E. coli}.
However, it is possible that two genes which do not belong to the same operon are
coexpressed inside the cell.
For example, a set of genes that are regulated by the same transcription factor inside
the cell may be coexpressed.
So, it is possible that UP-UC clusters may find even greater correspondence with
regulatory modules when expression data for {\it E. coli} is analyzed.


\begin{table*}
\centering
\begin{tabular}{|c|c|c|}

\hline
{\small Size of UP-UC}&{\small Number of clusters}&{\small Number of clusters in}\\
{\small cluster}&{\small in real network} &{\small randomized networks}\\
{\small }&{\small }&{\small Mean $\pm$ S.D.}\\
\hline

\small 2 &\small  49 &\small  101.32 $\pm$ 8.60  \\

\small 3 &\small  16 &\small  22.49 $\pm$ 4.28  \\

\small 4 &\small  5 &\small  6.62 $\pm$ 2.38  \\

\small 5 &\small  7 &\small  2.15 $\pm$ 1.44 \\

\small 6 &\small  2 &\small  0.84 $\pm$ 0.89 \\

\small 7 &\small  0 &\small  0.34 $\pm$ 0.57 \\

\small 8 &\small  2 &\small  0.15 $\pm$ 0.39 \\

\small 9 &\small  3 &\small  0.06 $\pm$ 0.24 \\

\small 10 &\small  1 &\small  0.02 $\pm$ 0.14 \\
\hline

\end{tabular}

\caption{
Size distribution of UP-UC clusters in {\it E. coli}
network and its randomized versions.
85 UP-UC clusters of size ranging from 2 to 10 reactions
were found in the real network.
The number of clusters of each size is given in the
second column of the table.
The third column gives the UP-UC cluster size distribution
for randomized networks with same local connectivity as
the real network, averaged over 1000 realizations of the
randomized network.}
\label{motiftable}
\end{table*}


\section{Large UP-UC clusters are over-represented in the real
metabolic network}

The bunching up of UP-UC metabolites next to each other in the metabolic
network results in formation of UP-UC clusters with more than two reactions.
We asked the question: Is it expected that a network like the {\it E. coli}
metabolic network of 618 internal metabolites and 1177 reactions with
185 UP-UC metabolites will have a size distribution of UP-UC clusters as
given in Table \ref{motiftable}?
To answer this question, we compared the distribution of UP-UC clusters in
the real {\it E. coli} metabolic network with a suitably randomized version
of the original network.
The randomized network has the same number of metabolite nodes and reaction
nodes and the same number of incoming and outgoing links at each node as the
real {\it E. coli} metabolic network.

The randomized networks with same local connectivity as the real {\it E. coli}
metabolic network were generated using the following algorithm.
Starting from the reconstructed metabolic network for {\it E. coli}, generate
the bipartite matrix ${\bf A}$ following the steps outlined in section
\ref{bipartitematrix}.
Starting from the matrix ${\bf A}$ for the real {\it E. coli} metabolic network,
we generated randomized networks keeping the degree of each metabolite and reaction
node unchanged \cite{KTV1999,MS2002}.
It is important to distinguish between two kinds of links.
The entries +1 in the matrix ${\bf A}$ represent the links coming into a metabolite
node from a reaction node and the entries -1 in the matrix ${\bf A}$ represent the
links going out of a metabolite node to a reaction node.
We divided all links or edges in the bipartite graph ${\bf A}$ into these two
groups.
Two links are then randomly selected in one of these two groups and swapped.
Before swapping, we ensure that the metabolite involved in any link is not
already involved as a reactant or product in the reaction corresponding to
the other link (otherwise, we could end up with metabolites being consumed and
produced in the same reaction).
Furthermore, links corresponding to the biomass reaction are not picked for
swapping.
This process of selecting a random pair of links was repeated 18000 times.
It was verified that more than 99.9\% of the links were visited at
least once.
Starting from the real metabolic network, this procedure was repeated
1000 times (with different random number seeds), to generate 1000
randomized networks.

We determined the UP-UC clusters for each of the 1000 realizations of
the randomized network.
The cluster size distribution averaged over 1000 realizations of the
randomized network is shown by the black line in
Fig. \ref{motif}.
From the Fig. \ref{motif}, we can see that the actual metabolic network
of {\it E. coli} has its UP-UC metabolites bunched up next to each other,
forming larger clusters than may be expected in random networks with the
same local connectivity properties as the original network.
`Network motifs' have been defined as patterns of interconnections that
occur in different parts of a network at frequencies much higher than
those found in randomized networks
\cite{SMMA2002,MSIKCA2002,A2003,MIKLSASA2004}.
Thus, larger size (size $\geq 8$) UP-UC clusters are over-represented in
the real {\it E. coli} metabolic network, and may be collectively considered
as analogous to a network motif.
The smaller size ($\leq 3$) UP-UC clusters are under-represented in the
real {\it E. coli} metabolic network, and may be collectively considered
analogous to an `anti-motif'.
We obtained qualitatively similar results for the metabolic networks of
{\it S. cerevisae} and {\it S. aureus}.
Thus, real metabolic networks contain many more large UP-UC clusters than
are expected in the randomized networks with the same local connectivity.
Larger UP-UC clusters in the real network may facilitate the regulation of
certain metabolic pathways inside the organism.


\section{Essential metabolic reactions}

The metabolic network is extremely flexible to allow an organism to
survive and grow under varied environmental conditions.
Organisms in the course of evolution have developed redundancies in
their intracellular machinery in order to tolerate random failures
(e.g., random mutations, etc).
Yet certain failures in the system may turn out to be lethal for the
survival of the organism.
For example, a mutation in a gene may result in its coded protein
being nonfunctional.
Enzymes are proteins that catalyze reactions in the metabolic network.
If an enzymatic protein becomes nonfunctional due to mutation in its
coding gene, it may tantamount to a loss in the capability of the
cell to carry out certain reactions catalyzed by that enzyme.
Such a loss of ability to carry out certain biochemical reactions in
the cell may turn out to be lethal for the organism as the lost
reactions may be essential for growth under certain environmental
conditions.
A reaction is designated as `essential' for certain growth medium,
if its knockout from the metabolic network results in the organism
being unable to grow under that medium.


\subsection{Determination of essential reactions}
\label{active}

In practice, we cannot directly knockout a reaction in the metabolic
network via molecular biology techniques in order to experimentally
determine essential reactions for the survival of an organism.
However, it is possible to knockout genes via molecular biology
techniques and observe the effect of the knockout on the viability of
the organism for different growth media in the wet lab.
The knockout of a gene in effect results in the coded protein or enzyme
being eliminated from the network.
This would in effect result in the elimination of one or more reactions
from the metabolic network that are catalyzed by the enzyme.
If the knockout of a gene renders the organism unviable for certain
growth medium then the gene is deemed `essential' for that growth
medium.
The overall process of determining essential genes using in vivo
experiments is very tedious and time consuming.
Also, there is not always a one to one correspondence between genes
and reactions in the network, and it may be difficult to determine
essentiality of certain reactions using wet lab experiments.
This is because the knockout of a single gene may in turn eliminate
an enzyme that may catalyze multiple reactions in the metabolic network
which are all removed from the network as a result of the knockout
of the gene.

We have used a here computational method to determine essential
reactions in metabolic networks.
This method relies on the technique of flux balance analysis (FBA)
\cite{SP1992a,SP1992b,VP1993a,VP1993b,VBP1993a,VBP1993b,VP1994a,VP1994b,PK1997,PK1998,EP2000,EIP2001,IEP2002,SCFCEP2002,SVC2002,RVSP2003,FFFPN2003,KPE2003,DHP2004,AKVOB2004,Palssonbook}.
Flux balance analysis (FBA) is a computational modelling technique
which can be used to obtain the maximal growth rate of the organism
supported by the metabolic network for any given nutrient medium.
It also gives the steady state fluxes of all reactions in the
metabolic network for any medium (for a detailed description of FBA
see Appendix \ref{fba}).
We have used FBA to study the metabolic networks of {\it E. coli}
(version iJR904), {\it S. cerevisiae} (version iND750) and
{\it S. aureus} (version iSB619).

FBA technique was used to compute fluxes of all metabolic reactions and
optimal growth rate of {\it E. coli} for all possible aerobic minimal
media.
Any aerobic minimal media is characterized by availability of a single
carbon source and key inorganic sources (ammonium, Fe$^{2+}$, oxygen,
phosphate, potassium, proton, sodium, sulfate and water) for the uptake
of {\it E. coli}.
If for a particular medium, the optimal growth rate obtained using the
technique of FBA is zero, then under that condition the metabolic network
does not permit the organism to grow.
Using FBA, it was found that the {\it E. coli} metabolic network iJR904
supports nonzero growth under 89 aerobic minimal media
\cite{SJunpublished}.
Further, a reaction in the {\it E. coli} metabolic network was designated
as `active' if it has a nonzero flux value for at least one of the 89
minimal media, and `inactive' otherwise.
Similarly, using FBA, it was found that the metabolic networks for
{\it S. cerevisiae} and {\it S. aureus} supported nonzero growth
under 43 and 27 aerobic minimal media respectively.
These 89, 43 and 27 aerobic minimal media that supported growth in
{\it E. coli}, {\it S. cerevisiae} and {\it S. aureus}, respectively,
were designated as `feasible minimal media'.

We used FBA to computationally determine essential reactions in the
metabolic networks of {\it E. coli}, {\it S. cerevisiae} and {\it S. aureus}.
We checked the effect of `switching off' or removal of a reaction one by one
from the metabolic network on the optimal growth rate of the organism obtained
using FBA for different feasible minimal media.
In FBA, a reaction can be switched off or removed from the network by setting
the maximum flux (a parameter input in FBA) through the reaction equal to zero.
We designated a reaction as `essential' for a particular minimal medium,
if switching the reaction off resulted in a zero optimal growth rate for
that medium.
We designated a reaction as `globally essential' for an organism, if it
was essential for all its feasible minimal media under aerobic conditions.
A program to determine essential reactions in the {\it E. coli} metabolic
network iJR904 is contained in Appendix \ref{program}.

The number of essential reactions for each of the 89 minimal media
varied between 200 and 240 and the number of globally essential
reactions was 164 for the {\it E. coli} metabolic network.
The number of essential reactions for each of the 43 minimal media
varied between 165 and 187 reactions and the number of globally
essential reactions was 127 for the {\it S. cerevisiae} metabolic
network.
The number of essential reactions for each of the 27 minimal media
varied between 222 and 256 reactions and the number of globally
essential reactions was 196 for the {\it S. aureus} metabolic network.


\section{Essential metabolic reactions are largely explained by
UP/UC structure}


\subsection{Most globally essential reactions can be tagged by
a UP or UC metabolite}

We found a set of 164 metabolic reactions to be globally essential
in the {\it E. coli} metabolic network.
Similarly, the number of globally essential reactions in
{\it S. cerevisiae} ({\it S. aureus}) metabolic network was found to
be 127 (196).
We then tried to understand why certain reactions happen to be
globally essential in terms of the underlying structure of the
metabolic network.
Notice that if a UP or UC metabolite is an essential intermediate in
the production of a metabolite that are part of the biomass reaction
then that reaction responsible for the production or consumption of
that UP or UC metabolite becomes essential for the growth of the
organism.
Of the 164 globally essential reactions in the {\it E. coli}
metabolic network, 133 were found to be either UP or UC.
Similarly, a high fraction of globally essential reactions in
the metabolic networks of {\it S. cerevisiae} and {\it S. aureus}
were found to be UP or UC (see Table \ref{essentialreactions}).
This explains why the subset of 133, 86 and 157 reactions are
globally essential in {\it E. coli}, {\it S. cerevisiae} and
{\it S. aureus}, respectively, namely, there is simply no other
path around these reactions in the entire network to produce or
consume some metabolite that is presumably required for the eventual
production of biomass.

The probability of such a high overlap between the set of globally
essential reactions and set of UP/UC reactions occurring by pure
chance is very small. To quantify this, the result was compared to
a null model in which the two sets corresponding to globally
essential reactions and UP/UC reactions were considered to be
independent of each other.
The total number of reactions in the {\it E. coli} metabolic network
is 1176.
The number of globally essential reactions is 164 and the number of
UP/UC reactions is 417 for the {\it E. coli} network.
The probability that out of a set of 1176 reactions in {\it E. coli},
two independently chosen subsets of size 417 and 164 will have an
intersection of 133 or greater is $p < 10^{-37}$ (any one or both of
the subsets is chosen randomly).
Similarly, we obtained very small $p$ values for the metabolic networks
of {\it S. cerevisiae} and {\it S. aureus} (see Table \ref{essentialreactions}).

In an earlier paper \cite{MP2005}\, Mahadevan and Palsson had determined
the `lethality fraction' for each metabolite in the network.
The lethality fraction of a metabolite was defined as the fraction of
reactions in which the metabolite is involved that are essential.
Mahadevan and Palsson had observed that this lethality fraction of the
low degree metabolites is on average comparable to high degree metabolites.
In particular, they found that some metabolites with in-degree and out-degree
unity (that we have designated as UP-UC metabolites) have lethality fraction
unity.
We have presented here a stronger result regarding the role of low degree
metabolites: most globally essential reactions involve at least one UP or
UC metabolite.
The essential reactions may involve other metabolites of higher degree, but
their essentiality is due to their uniqueness in producing or consuming a
UP or UC metabolite.

\begin{table}
\centering
\begin{tabular}{|l|l|l|l|}
\hline
{\bf \small Organism} & {\bf \small \it E. coli}&{\bf \small \it S. cerevisiae}&{\bf \small \it S. aureus}\\
\hline
{\small Total number of reactions}&{\small 1176}&{\small 1579}&{\small 865}\\
\hline
{\small Number of globally essential reactions}&{\small 164 }&{\small 127}&{\small 196}\\
\hline
{\small Number of globally essential reactions}&{\small 133}&{\small 86}&{\small 157}\\
{\small that are UP or UC in the entire network}&{\tiny {($p<10^{-37}$)}}&{\tiny {($p<10^{-12}$)}}&{\tiny {($p<10^{-32}$)}}\\
\hline
{\small Number of globally essential reactions}&{\small 156}&{\small 117}&{\small 182 }\\
{\small that are UP or UC in the reduced network} & {\tiny {($p<10^{-62}$)}}&{\tiny {($p<10^{-41}$)}}&{\tiny {($p<10^{-58}$)}}\\
\hline
\end{tabular}
\caption{Almost all globally essential reactions in {\it E. coli},
{\it S. cerevisiae} and {\it S. aureus} are UP or UC. The $p$ value
represents the probability that the indicated overlap would arise
in a null model.}
\label{essentialreactions}
\end{table}


\begin{table}
\centering
\begin{tabular}{|l|c|c|c|}
\hline
{\bf \small Organism} & {\bf \small \it E. coli}&{\bf \small \it S. cerevisiae}&{\bf \small \it S. aureus}\\
\hline

{\small Number of reactions in the } & {\small 1176 } & {\small 1579 } & {\small 865 } \\
{\small original network } &  &  &  \\
\hline

{\small Number of UP reactions in the } & {\small 289 } & {\small 391 } & {\small 277 } \\
{\small original network } &  &  &  \\
\hline

{\small Number of UC reactions in the } & {\small 272 } & {\small 370 } & {\small 218 } \\
{\small original network } &  &  &  \\
\hline

{\small Number of UP/UC reactions in the } & {\small 417 } & {\small 583 } & {\small 376 } \\
{\small original network } &  &  &  \\
\hline

{\small Number of blocked reactions } & {\small 290 } & {\small 800	} & {\small 294 } \\
\hline

{\small Number of UP/UC reactions in the } & {\small 136 } & {\small 386 } & {\small 174 } \\
{\small original network that are blocked } &  &  &  \\
\hline

{\small Number of reactions in the } & {\small 886 } & {\small 779 } & {\small 571 } \\
{\small reduced network } &  &  &  \\
\hline

{\small Number of UP reactions in the } & {\small 245 } & {\small 218 } & {\small 224 } \\
{\small reduced network } &  &  &  \\
\hline

{\small Number of UC reactions in the } & {\small 245 }	& {\small 218 } & {\small 181 } \\
{\small reduced network } &  &  &  \\
\hline

{\small Number of UP/UC reactions in the }& {\small 352 } & {\small 306 } & {\small 276 } \\
{\small reduced network } &  &  &  \\
\hline

{\small Number of UP/UC reactions in the  } & {\small 71 }& {\small 109 } & {\small 74 } \\
{\small reduced network that are not } &  &  & \\
{\small UP/UC in the original network } &  &  & \\
\hline

\end{tabular}
\caption{UP(UC) reaction statistics in the original and
reduced metabolic networks of {\it E. coli},
{\it S. cerevisiae} and {\it S. aureus}.}
\label{upucstatistics}
\end{table}


\subsection{Almost all globally essential reactions are UP/UC
in the `reduced network'}
\label{reducednetwork}

It was shown above that 133 out of 164 globally essential reactions
were associated with a UP or UC metabolite in the {\it E. coli}
metabolic network.
To understand the remaining globally essential reactions, a reduced
or pruned version of the {\it E. coli} network was considered.

The databases of reconstructed metabolic networks that have been used
here contain certain reactions that can only have a zero flux value
under any steady state due to stoichiometric reasons.
Such reactions have been referred to as `strictly detailed balanced'
reactions \cite{SS1991,Heinrichbook} or `blocked' reactions \cite{BNSM2004}
(see section \ref{blocked} in Appendix \ref{fba} for more details).
The blocked reactions can be removed from the reconstructed metabolic
networks for any steady state analysis.
We used a previously described algorithm by Burgard {\it et al}
\cite{BNSM2004}\ to determine the blocked reactions in the metabolic
networks of {\it E. coli}, {\it S. cerevisiae} and {\it S. aureus}.
This algorithm to determine blocked reactions has been described in
detail in section \ref{Burgard} in Appendix \ref{fba}.
A program to determine blocked reactions in the {\it E. coli} metabolic
network is contained in Appendix \ref{program}.
The number of blocked reactions in the metabolic networks of
{\it E. coli}, {\it S. cerevisiae} and {\it S. aureus} were found to
be 290, 800 and 294, respectively (for details see Table
\ref{upucstatistics}).
We removed the 290 blocked reactions in the {\it E. coli} metabolic
network from the list of 1176 reactions to obtain the `reduced network'
of 886 reactions in {\it E. coli}.
We emphasize that the removal of blocked reactions from the original
network does not affect our results obtained using FBA for any of the 89
minimal media considered for {\it E. coli}.
The 290 blocked reactions removed from the original {\it E. coli} metabolic
network of 1176 reactions had a zero flux value for all 89 minimal media
studied here.
Further, since the blocked reactions are guaranteed to have a zero flux
value for any minimal media, they can be never essential.
So, the set of essential reactions obtained by implementing FBA on the
reduced network is exactly the same as that obtained from the original network
for any minimal media.
Thus, the set of 164 globally essential reactions is the same for the
original network of 1176 reactions and the reduced network of 886 reactions
in {\it E. coli}.
Similarly, the reduced networks for {\it S. cerevisiae} and {\it S. aureus} were
obtained by removing the corresponding blocked reactions from the original network
(see Table \ref{upucstatistics}).

The set of UP(UC) metabolites was then determined for the reduced metabolic
network obtained after removing blocked reactions from the original network.
A metabolite that was not UP(UC) in the original metabolic network can become
UP(UC) in the reduced network.
This may happen due to the removal of blocked reactions which may be contributing
to the degree of a metabolite in the original network.
The set of UP(UC) metabolites and reactions were obtained for the reduced networks
of {\it E. coli}, {\it S. cerevisiae} and {\it S. aureus}.
We found 352, 306 and 276 reactions to be UP/UC in the reduced networks of
{\it E. coli}, {\it S. cerevisiae} and {\it S. aureus}, respectively (see Table
\ref{upucstatistics}).
The set of UP/UC reactions in the reduced network turns out to smaller than that
for original network.
This is so because several reactions that were UP/UC in the original network happen
to be blocked and have been removed from the reduced network.
Also, some metabolite that was earlier not UP(UC) can now become UP(UC) in the reduced
network due to removal of blocked reactions which adds new reactions to the UP/UC set.
However, the number of reactions added turns out to be smaller than the number of
reactions removed (see Table \ref{upucstatistics}).
The new UP(UC) metabolites have, by definition, their in-degree (out-degree) unity in
the reduced network.
Even in the original network the set of UP(UC) metabolites have a low degree.
In {\it E. coli}, the average in-degree (out-degree) of the set of UP(UC) metabolites
in the reduced network was found to be 1.31(1.33) in the original network.
We remark that UP(UC) reactions in the reduced network are uniquely determined starting
from the original network.

We found 156 out of the 164 globally essential reactions (95$\%$) to be UP or UC in the
{\it E. coli} reduced network ($p<10^{-62}$).
Similarly, it was found that almost all globally essential reactions in {\it S. cerevisiae}
and {\it S. aureus} were either UP or UC in the reduced network (92$\%$ and 93$\%$,
respectively).
Thus, we have shown here that nodes with a low degree of connectivity (i.e., UP(UC)
metabolites) play an `essential' role in metabolism (see Table \ref{essentialreactions}).

The results obtained here provide some insight into the structural or topological
origin of essential reactions in metabolic networks.
It is, of course, obvious that if a metabolite is an essential intermediate for the
production of some biomass metabolite, and if this metabolite is uniquely produced
or uniquely consumed, then the corresponding production or consumption reaction will
be essential for the growth of the cell.
However, the converse of this statement that all globally essential reactions in the
network should be UP/UC is far from obvious.
The finding that about 5-8 $\%$ of globally essential reactions do not satisfy the
UP/UC property proves that the converse statement is indeed false.
Thus, the fact that the overwhelming majority (92-95 $\%$) of globally essential
reactions satisfy the UP/UC topological property in the network is a characterization
of the nature of metabolic networks found in organisms.
We remark that we do not as yet understand why the remaining globally essential reactions
happen to be essential.


\subsection{Most UP/UC reactions in the reduced network are
`conditionally essential'}

We found that there were 352 UP/UC metabolic reactions in
the {\it E. coli} reduced network.
156 of these 352 reactions were essential for all 89 minimal
media or globally essential in {\it E. coli}.
It was found that there were 400 reactions in the
{\it E. coli} network which were essential
for at least one of the 89 minimal media.
These 400 reactions that were found to be essential for some
of the 89 minimal media may be designated as
`conditionally essential' for the {\it E. coli} metabolic
network.
We found 288 of the 352 UP/UC reactions (82 \%) in the
{\it E. coli} reduced network to be
conditionally essential.
Such a large overlap is very unlikely ($p<10^{-74}$) between
the two sets of UP/UC reactions and conditionally essential
reactions.
Some of these conditionally essential UP/UC reactions were
part of the input pathways of only one carbon source and
hence, they were essential only for that minimal media.
In {\it S. cerevisiae} and {\it S. aureus}, we found the number
of conditionally essential reactions to be 269 and 331,
respectively.
In {\it S. cerevisiae}, we found 170 out of 306 UP/UC reactions
(56 \%) in the reduced network to be conditionally essential,
while in {\it S. aureus}, 257 out of 276 (93 \%) were found to
be conditionally essential.
We found the $p$ values for such large overlaps between the set
of UP/UC reactions and conditionally essential reactions in
{\it S. cerevisiae} and {\it S. aureus} to be $p<10^{-22}$ and
$p<10^{-67}$, respectively.
In {\it E. coli} and {\it S. aureus} more than 80\% of the
UP/UC reactions were conditionally essential while in
{\it S. cerevisiae} only about 56\% of the UP/UC reactions
were conditionally essential.
The substantial difference in the fraction of UP/UC reactions
which are conditionally essential in yeast as opposed to that
for the two bacteria may reflect a more evolved metabolic
structure in the eukaryote.
Yeast may have many more alternative pathways and may be more
robust to random failures as compared to the two bacteria.
This observation needs to be further investigated.


\subsection{Comparison between computationally determined
essential reactions and experimentally determined
essential genes in {\it E. coli}}
\label{essentialgenes}

We had mentioned earlier that it is not possible to always
experimentally determine essential reactions for an organism
due to lack of one to one correspondence between genes and
reactions in the metabolic network.
Experimentally one can only determine essential genes for a
given medium.
Gerdes {\it et al} \cite{GSCBRDSKAGBKDBGMFOBOO2003} have
determined experimentally the list of essential genes
in {\it E. coli} for a rich medium.
To compare the computationally predicted results with
experimental data of Gerdes {\it et al}, we implemented
FBA under rich medium for the {\it E. coli} metabolic
network iJR904.

A set of 95 reactions were found to be essential under rich
medium for the {\it E. coli} metabolic network.
89 of these 95 reactions were found to be either UP or UC
in the {\it E. coli} reduced network.
Of the 95 essential reactions in rich medium, information
about the corresponding genes (coding for the enzymes) was
available for only 85 reactions in the database iJR904
for the {\it E. coli} metabolic network.
Of these 85 reactions, 14 reactions had known isozymes,
i.e, multiple enzymes catalyzing a single reaction in
the network.
Hence, the genes corresponding to isozymes are not expected
to be essential for these 14 reactions as the effect of the
knockout of a single gene can be compensated by another gene.
Of the remaining 71 reactions, 5 had associated genes whose
essentiality was undetermined in the database by Gerdes
{\it et al} \cite{GSCBRDSKAGBKDBGMFOBOO2003}.
Of the remaining 66 reactions, a fairly high fraction of 38
reactions had associated genes that were found to be
essential in the database.

Conversely, of the 618 genes determined to be essential for
{\it E. coli} by Gerdes {\it et al} in rich medium, 158
genes were also part of the {\it E. coli} metabolic network
iJR904.
Note that there are 904 genes coding for various enzymes
catalyzing reactions in the {\it E. coli} metabolic
network.
We found that 103 of the 158 essential genes had their enzymes
catalyzing only a single reaction in the {\it E. coli}
metabolic network.
Of these 103 essential genes, 62 were associated with a UP or
UC reaction in the original {\it E. coli} metabolic network.
Further, it was found that 73 of the 103 essential genes were
associated with a UP or UC reaction in the {\it E. coli}
reduced network.
A possible reason for the difference between theoretical
prediction and experimental data could be reconciled on the
basis of incompleteness of the reconstructed metabolic network
iJR904 used for our study.
There may be alternative pathways in the real organism that are
missing in the reconstructed network iJR904.
Further, for certain reactions in the metabolic network
there may be isozymes that are presently not included in the
network.



\chapter{The regulatory network of {\it E. coli} metabolism exhibits both homeostasis
and flexibility of response}
\label{dynamics}

In the previous chapter, we studied the structure and dynamics of the metabolic networks
of {\it E. coli}, {\it S. cerevisiae} and {\it S. aureus}.
In this chapter, we have used the Boolean approach to study the system level dynamics of
the large scale transcriptional regulatory network (TRN) controlling metabolism in
{\it E. coli}.
For our study, we have used a previously published database iMC1010$^{v1}$ \cite{CKRHP2004}
representing the regulatory network controlling {\it E. coli} metabolism.
In the database iMC1010$^{v1}$, both the network connections and the Boolean functions have
been reconstructed from empirical data.
We study the trajectories and attractors of the Boolean dynamical system for multiple initial
conditions and environments, and also study the functioning of the metabolic network under the
regulatory constraints.
Based on this we find that the dynamics of the regulatory network of {\it E. coli} metabolism
leads to biologically important system level properties that endow it with robustness and
efficient functionality.

\section{The integrated regulatory and metabolic network iMC1010$^{v1}$ for {\it E. coli}}

A database iMC1010$^{v1}$ \cite{CKRHP2004} representing the integrated
regulatory and metabolic network for {\it E. coli} has been reconstructed
based on various literature sources.
This database contains the known transcriptional regulatory network (TRN) in
{\it E. coli} that determines the activity of various enzymes in its
metabolic network.
We downloaded the database iMC1010$^{v1}$ from the website \cite{GCRG}.
The regulatory network accounts for 583 genes of which 104 genes code for
103 transcription factors and 479 genes code for metabolic enzymes.
Of the former, 102 genes have a one to one correspondence between genes
and transcription factors, while two other genes code for different
subunits which together form a single transcription factor.
The 479 genes that code for metabolic enzymes are a subset of 904 genes that
code for various enzymes catalyzing reactions in the metabolic network iJR904
\cite{RVSP2003}.
Thus, the regulatory network contained in the database iMC1010$^{v1}$ controls
the activity of a subset of enzymes catalyzing various reactions in the metabolic
network database iJR904.

\begin{figure*}
\centering
\includegraphics[trim=70 0 0 0,width=19cm]{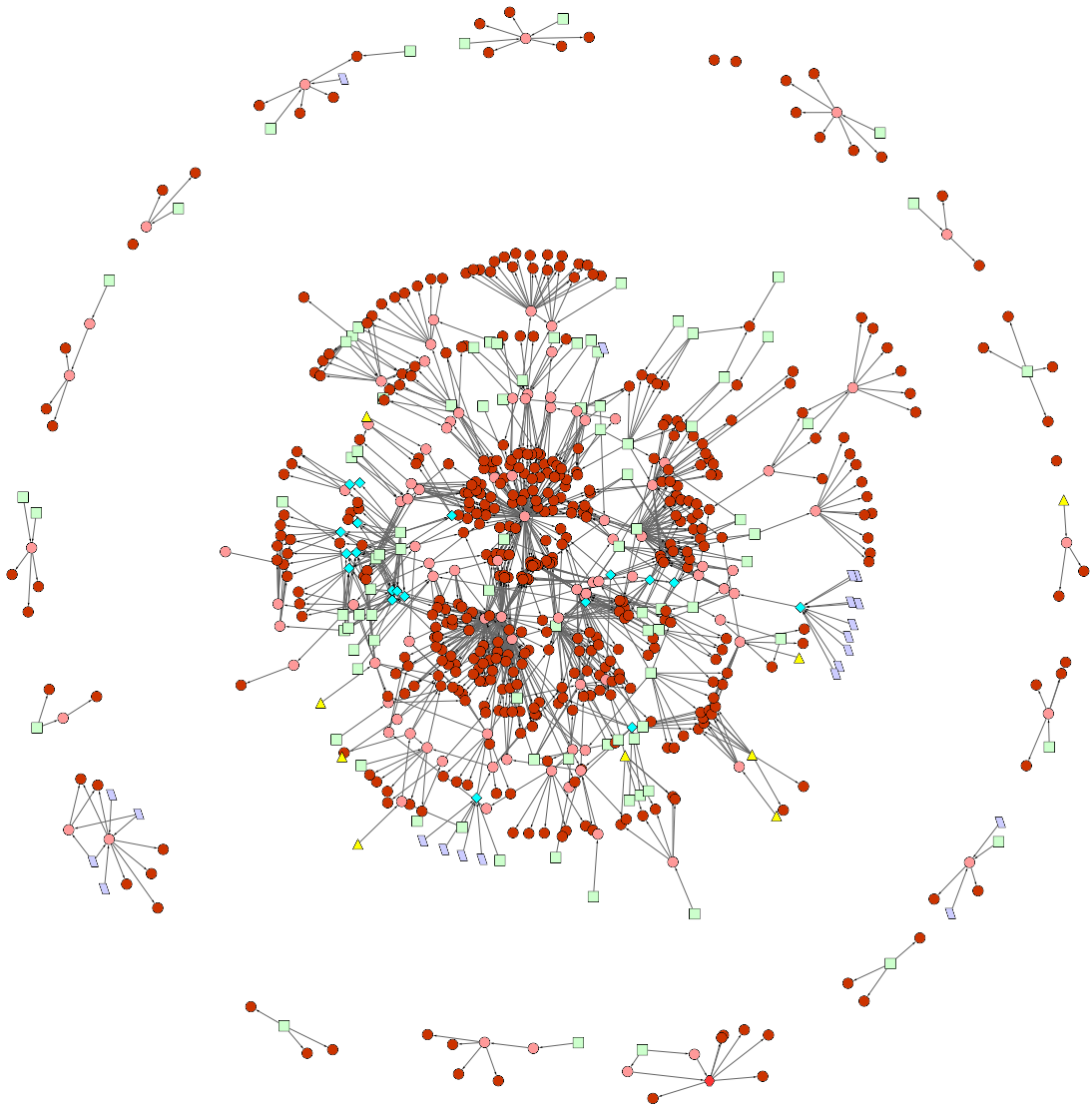}
\caption{
Map of the genetic network controlling metabolism in {\it E. coli}.
In this figure, there are genes coding for transcription factors (pink circles), genes
coding for enzymes (brown circles), external metabolites (green squares), internal
fluxes (purple parallelograms), stimuli (yellow triangles) and other conditions
(blue diamonds).
The red hexagon denotes the lone transcription factor in the network that is coded
for by two genes.
The electronic version of this figure \cite{SJ2008} (available from {\scriptsize
http://www.biomedcentral.com/1752-0509/2/21}) can be zoomed in to see arrowheads.
This picture has been drawn using the graph visualization software Cytoscape
\cite{Cytoscape}.
}
\label{trn}
\end{figure*}

The state of the 583 genes in the regulatory network is determined by the state of
103 transcription factors coded by 104 of the 583 genes.
The state of the regulatory network is also influenced by the metabolite concentrations
inside the cell.
In the database iMC1010$^{v1}$, the state of the genes is also determined by concentration
of 96 external metabolites which the cell can uptake from the environment, and the state of
21 internal fluxes of metabolic reactions which are surrogate for internal metabolite
concentrations.
The database iMC1010$^{v1}$ has been designed to study the regulatory network in conjunction
with the metabolic network iJR904 using the FBA modelling approach.
However, in FBA model, we can determine only the steady state reaction fluxes and the
method is unable to compute internal metabolite concentrations (see Appendix \ref{fba}
for details).
Therefore, in the database iMC1010$^{v1}$, the internal metabolite concentrations that
determine the state of genes have been approximated by appropriate internal fluxes of
reactions.
Further, the state of genes in the network is also determined by 9 stimuli such as
heat shock, stress, etc., and 19 other conditions such as Surplus PYR, pH, etc.
Thus, the state of the 583 genes in the regulatory network iMC1010$^{v1}$ is determined
by the state of 103 transcription factors which are products of 104 of the genes in the
set, 96 external metabolites, 21 internal fluxes of metabolic reactions, 9 stimuli and
19 other conditions.
The directed graph of regulatory network is shown in Fig. \ref{trn}, where a directed link
from one node to another denotes a regulatory interaction.

The database iMC1010$^{v1}$ not only contains the information about the connections between
genes and their inputs (transcription factors, external metabolites, internal
fluxes, stimuli and conditions) forming the network but also the Boolean rule
for each gene node based on the state of the input
nodes of that gene.
For example, the rule
\begin{center}
b2720 = IF (FhlA AND RpoN AND (NOT (o2[e]$>$0)))
\end{center}
indicates that the gene b2720 is active if transcription factors FhlA and RpoN are active
and the external metabolite oxygen is absent from the external environment;
otherwise gene b2720 is inactive.
Such information is provided in the database for each of the 583 genes.


\section{The regulatory network of {\it E. coli} metabolism
as a Boolean dynamical system}
\label{imc1010}

We wanted to construct the simplest possible dynamical system
utilizing the information regarding
the network connections and regulatory logic in the database
iMC1010$^{v1}$ \cite{CKRHP2004} to
study the dynamics of the genetic network controlling
{\it E. coli} metabolism.
In the absence of kinetic data such as rate constants, etc.,
a differential equation based simulation
of large scale genetic networks is not feasible at present and
the large number of unknown parameters
would also render the results of such a simulation difficult to interpret.
Due to absence of kinetic information, the discrete time Boolean approach
\cite{K1969a,K1969b,T1973,DW1986,Kauffmanbook,BS1998,HSWK2002,SDKZ2002,ACK2003,KPST2003,KPST2004,SVA2004,KB2005a,GD2005}
is a natural choice to study the dynamics of large scale
genetic networks.
Boolean simulations of smaller biological networks with a few nodes
have also provided useful insights
\cite{ST2001,AO2003,LLLOT2004,EPA2004,LAA2006}.
The discrete-time Boolean approach does not take into account the multiple
time scales in the real system.
It also does not describe large variations in concentrations as well as
stochastic effects due to small numbers of molecules.
Thus, the Boolean approach cannot be used for a detailed
prediction of time courses and concentrations.
We do not have the detailed empirical knowledge at the present time
to treat these aspects of the system satisfactorily on a large scale.
The Boolean approach can provide useful information about some
qualitative features of the dynamics,
e.g., the nature of the attractors of the system, and through
that, insights regarding overall organization of the system.

\begin{figure*}
\centering
\includegraphics[width=14cm]{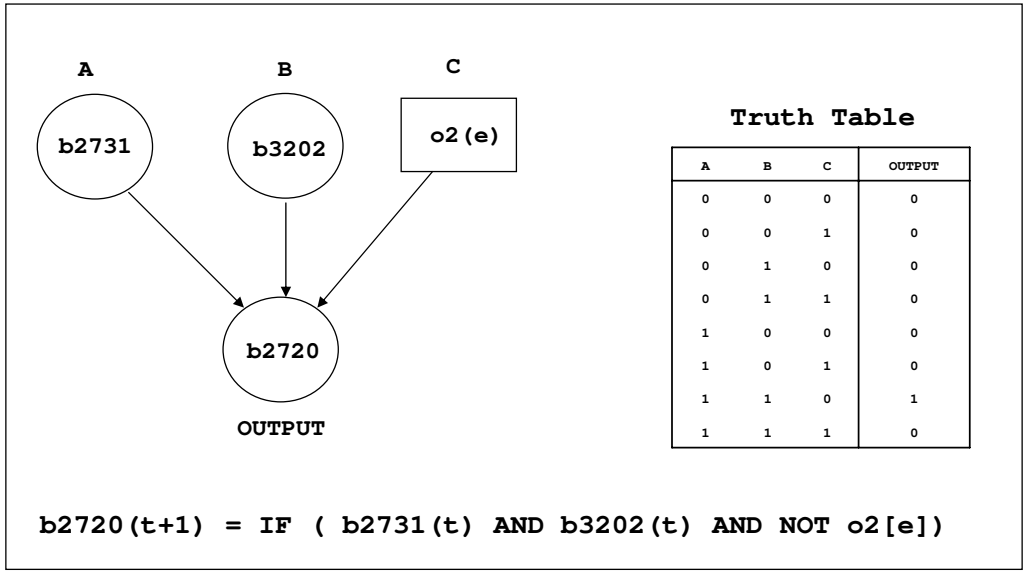}
\caption{
Example of a boolean function $G_i$ representing the regulatory logic at the
promoter region of gene b2720 that determines its expression.
The gene b2720 is on at a particular time instant if and only if both the
genes b2731 and b3202 are on at the previous time instant and oxygen is
absent in the buffered external environment.}
\label{booleanrule}
\end{figure*}

We have used the information contained in the database iMC1010$^{v1}$ to
construct the following effective
Boolean dynamical system of genes and external metabolites representing
the genetic network controlling
{\it E. coli} metabolism \cite{SJ2008} (see section \ref{construct} for details):
\begin{equation}
\label{dynamicalsystem}
g_i(t+1) = G_i({\bf g}(t), {\bf m}); \quad \quad i = 1, 2, \ldots, 583.
\end{equation}
Here $g_i(t+1)$ is the configuration of gene $i$ at time $t+1$.
In this approach, time is taken as discrete, i.e.,
$t = 0,1,2,\ldots$.
Further, at any given time $t$, a gene can be either on or off.
So, $g_i(t)$ = 1 (0) represents that at time $t$ gene $i$ is in
on(off) state.
The vector ${\bf g}(t)$, whose $i^{th}$ component is $g_i(t)$,
collectively denotes the configurations of all the 583 genes
at time $t$.
The vector ${\bf m}$ denotes the configuration of the 96 external
metabolites in the environment.
The $i^{th}$ component of vector ${\bf m}$ is $m_i$
($i=1,2,\ldots,96$).
$m_i$ = 1 if metabolite $i$ is present in the external environment
for uptake into the cell, and $m_i = 0$ if metabolite $i$ is absent.
The vector ${\bf g}(t)$ gives us the state of the genetic network
at time $t$ and the vector ${\bf m}$ represents the state of the
external environment.
The functions $G_i$ contain all the information about the
internal wiring of the network (i.e., the input nodes that
determine the state of gene $i$) as well as the Boolean logic
of each gene's regulation (i.e., given the configuration of
all of gene $i$'s inputs at time $t$, whether gene $i$ will be
on or off at time $t+1$).
Each function $G_i$ typically depends only upon those components
of ${\bf g}$ and ${\bf m}$ that directly affect the expression
of gene $i$ (see Fig. \ref{booleanrule} for an example).
Equation \ref{dynamicalsystem} expresses the fact that the on-off
state of a gene at any time instant is controlled by the state of
the genes at the previous time instant as well as the state of
the external environment.
Since the interaction of genes is mediated by transcription factors,
a single time unit corresponds to the average time between the
initiation of transcription of a gene coding for a transcription
factor and the initiation of transcription of a gene regulated by
that transcription factor.
In our study, we update the state of all genes in the network
synchronously.
In general, note that the state of the external environment given
by vector ${\bf m}$ is a function of time $t$.
The concentration of metabolites in the external environment can
change as the metabolic network uptakes or excretes some the
metabolites across the cell boundary.
However, for our study, we have considered external environment
corresponding to buffered minimal media which are characterized
by vectors ${\bf m}$ that are constant in time.
The treatment of external metabolites deciding the state of the
environment is discussed in detail in section \ref{media}.

Stuart Kauffman introduced the framework of Boolean networks to
study the dynamics of genetic networks about four decades ago
\cite{K1969a,K1969b}.
Since then Kauffman and others \cite{K1969a,K1969b,T1973,DW1986,Kauffmanbook,BS1998,HSWK2002,SDKZ2002,ACK2003,KPST2003,KPST2004,SVA2004,KB2005a,GD2005}
have extensively studied Boolean dynamical systems of the form:
\begin{equation}
\label{kauffmansystem}
g_i(t+1) = G_i({\bf g}(t)),
\end{equation}
where $g_i(t)$ denotes the state of the gene $i$ at time $t$ and
the vector ${\bf g}(t)$ denotes the state
of all genes at time $t$.
In the above system, the state of genes is determined by only the
state of other genes in the network.
In the absence of detailed molecular data on real genetic regulatory
networks, Kauffman and others used the
Boolean approach to study biologically motivated random Boolean
networks in order to gain insights about the
system level dynamics of genetic networks.
A random Boolean network is a system of $N$ binary nodes with $K$
inputs per node representing the regulatory
mechanism.
The state of any gene node at time $t$ in a random Boolean network
is determined by the the state of its $K$
input nodes at previous time $t-1$ based on any one of the 2$^{2^K}$
possible Boolean functions with $K$
inputs.
The Boolean function at each gene node is randomly chosen from the
set of 2$^{2^K}$ possible Boolean functions.

Random Boolean networks have been studied extensively over the years
by Kauffman and others.
The study of random Boolean networks has led to several important
insights regarding the dynamics of these
systems.
In particular, Kauffman found that random Boolean networks with
large number of nodes possess an ordered regime
where attractors have short periods and large basins \cite{Kauffmanbook}.
In the ordered regime, the system was found to possess the property
of homeostasis or robustness to perturbations
of gene configurations.
Recently, Kauffman and colleagues have also studied the Yeast regulatory
network as a Boolean dynamical system
where the connections between genes was based on real data while the
Boolean function at each node determining
the state of the gene at a given time instant based on the state of its
input genes at previous time instant is
randomly chosen out of a set of biologically plausible Boolean
functions \cite{KPST2003,KPST2004}.
In references \cite{ST2001,AO2003,LLLOT2004,EPA2004,LAA2006,BB2007},
the Boolean approach has been applied to
study specific biological networks where detailed genetic data is
available.
These networks are smaller than the ones studied by Kauffman and
colleagues, and have up to 40 distinct genes,
proteins and other molecules
\cite{ST2001,AO2003,LLLOT2004,EPA2004,LAA2006,BB2007}.
In reference \cite{AO2003}, where a Boolean network of 180 nodes is
considered, the network contains 15 distinct
genes and proteins (with 12 nodes for each of them corresponding to
12 distinct cells).
These models, apart from reproducing several observed phenomena of
these biological systems, have also found that the
networks possess the property of homeostasis, as well as robustness
to genetic mutations.

This study is inspired by the work of Kauffman and extends his work
in two important ways.
First, it studies the genetic network controlling {\it E. coli}
metabolism which has been reconstructed based on real data.
The network studied here has 583 genes and 96 external metabolites
which is much larger than the real biological systems studied earlier
using the Boolean approach in references
\cite{ST2001,AO2003,LLLOT2004,EPA2004,LAA2006,BB2007}.
The regulatory network contained in the database iMC1010$^{v1}$ accounts
for about half the genes presently believed to be involved in
{\it E. coli} metabolism.
The network considered here is more than an order of magnitude larger
(in terms of the number of genes involved) than other
real genetic networks considered as Boolean system which allows us a
qualitatively different systemic view of the organization
of the genetic network of an organism.
Second, the present network is able to account for the effect of the
external environment on the TRN dynamics through the
vector ${\bf m}$ in Eq. \ref{dynamicalsystem}.
Note that the system studied by Kauffman is described by the
Eq. \ref{kauffmansystem} instead of Eq. \ref{dynamicalsystem},
which takes into account the effect of genes on other genes but not the
effect of the external environment.
Other works that investigated real biological systems as Boolean
networks had only a few environmental signals
\cite{ST2001,AO2003,LLLOT2004,EPA2004,LAA2006,BB2007}.
As a consequence of the database iMC1010$^{v1}$, one is able to take
into account the effect of external
environment in a much more systematic and extensive way than before.
We have studied here how the attractors of the Boolean dynamical system
given by Eq. \ref{dynamicalsystem} depend upon the
initial condition of the genes and the state of the external environment.


\section{Construction of the Boolean dynamical system}
\label{construct}

We will now describe in detail the construction of the Boolean dynamical system given
by Eq. \ref{dynamicalsystem} starting from the database iMC1010$^{v1}$.
As mentioned earlier, the database iMC1010$^{v1}$ accounts for 583 genes whose state
is determined by the state of 103 transcription factors, 96 external metabolites,
21 internal fluxes, 9 stimuli and 19 other conditions in the network.
Further, the database iMC1010$^{v1}$ gives us the Boolean input-output map for each
gene node in the network in terms of the state of its input nodes.
In this regulatory network, the state of various genes, transcription factors,
external metabolites, internal fluxes, stimuli and conditions is represented as
Boolean variables.
Thus, the overall system contains 823 Boolean variables (583 genes, 103 transcription
factors, 96 external metabolites, 21 internal fluxes, 9 stimuli and 19 conditions).
In the present study, the 9 stimuli are assumed to be always absent.
The state of the 583 genes, 103 transcription factors, 96 external metabolites, 19
conditions and 21 internal fluxes are respectively denoted by the vectors ${\bf g,t,m,c,v}$.
The vectors ${\bf g,t,m,c,v}$ can all in principle depend upon time $t$.
Here, $g_i(t)$ ($i=1,2,\ldots,583$), the $i^{th}$ component of ${\bf g}(t)$, equals
unity if the gene $i$ is active or on in the cell at time $t$ and zero if the gene
$i$ is inactive or off.
$t_i(t)$ ($i=1,2,\ldots,103$), the $i^{th}$ component of ${\bf t}(t)$, equals unity
if the transcription factor $i$ is present in the cell at time $t$ and zero if the
transcription factor $i$ is absent.
$c_i(t)$ ($i=1,2,\ldots,19$), the $i^{th}$ component of ${\bf c}(t)$, equals unity
if the $i^{th}$ condition holds at time $t$ and zero if not.
$v_i(t)$ ($i=1,2,\ldots,21$), the $i^{th}$ component of ${\bf v}(t)$, equals unity
if the $i^{th}$ metabolic reaction in the above mentioned set of internal metabolic
reactions is happening inside the cell at time $t$ (with a flux greater
than a specified value) and zero if not.
The state of a gene $i$ in the regulatory network at time $t$ is given by the equation:
\begin{equation}
\label{Fi}
g_i(t) = F_i({\bf t}(t),{\bf m}(t),{\bf c}(t),{\bf v}(t)), \quad
\quad i = 1,2,\ldots,583.
\end{equation}
The database iMC1010$^{v1}$ gives us the form of the functions $F_i$ in terms of AND,
OR and NOT operations on the Boolean arguments as mentioned at the end of section
\ref{imc1010}.
The 103 transcription factors are coded for by a subset of 104 genes in the regulatory
network.
Of the 104 genes coding for transcription factors, 102 genes code for a single
transcription factor each while two other genes together code for one transcription
factor.
The on-off state of the genes coding for transcription factors at the previous time
instant decides the state of transcription factors at the given time instant.
A single time step therefore corresponds to the average time for transcription and
translation.
The state of transcription factors at time $t$ is given by:
\begin{equation}
\label{Ti}
t_i(t) = T_i({\bf g}(t-1)), \quad \quad i = 1,2,\ldots,103,
\end{equation}
where the function
\begin{equation}
T_i({\bf g}) = g_i
\end{equation}
for 102 transcription factors that are coded for by single genes while for the transcription
factor coded for by two genes
\begin{equation}
T_i({\bf g}) = g_{i_1}\ AND\ g_{i_2}.
\end{equation}
Substituting Eq. \ref{Ti} in Eq. \ref{Fi} gives us the equation
\begin{equation}
g_i(t) = F_i({\bf T}({\bf g}(t-1)),{\bf m}(t),{\bf c}(t),{\bf v}(t)).
\end{equation}
The above equation provides the dynamical rule for updating the gene configurations from one
instant to the next, provided the status of the variables ${\bf m,c,v}$ corresponding to
external metabolites, conditions, internal fluxes, respectively, are known.

\subsection{Treatment of external metabolites ${\bf m}$}
\label{media}

In this study, we have considered external environment corresponding to buffered minimal media characterized
by $m_i$ that are constant in time.
For buffered media, we have
\begin{equation}
{\bf m}(t) = {\bf m}
\end{equation}
independent of time $t$ which represents a constant external environment for the cell.
For each buffered medium considered, the components of {\bf m} corresponding to the metabolites present in the external
environment were set to unity and the remaining components were set to zero.
The {\it E. coli} metabolic network iJR904 is capable of transporting 143 metabolites into and out of the cell.
The 143 external metabolites in the metabolic network iJR904 include 131 organic molecules and 12 inorganic molecules.
Further, 96 of the 143 external metabolites in the metabolic network iJR904 are also included in the regulatory network
part of the database iMC1010$^{v1}$.
The 96 external metabolites accounted in the regulatory network can be divided into 86 organic molecules and 10 inorganic
molecules.
In this study, we have considered the following classes of minimal media:

\begin{itemize}

\item[(a)] {\bf List of 93 minimal media}: Using the 96 external
metabolites (86 organic and 10 inorganic external metabolites)
contained in the regulatory network, we can construct in principle
86 aerobic and 86 anaerobic minimal media (i.e, in total 172
minimal media).
A minimal medium is characterized by the availability of single
organic source of carbon and the ions of ammonium, sulphate,
phosphate, hydrogen, iron, potassium and sodium.
The components of {\bf m} corresponding to these metabolites were
set to unity and others were set to zero in a given minimal
medium.
Oxygen was set to unity in the aerobic media and to zero in anaerobic
media.
FBA was performed for each of the 172 possible minimal media for the
{\it E. coli} metabolic network iJR904 to obtain the optimal
growth rate for each of the 172 minimal media
(see Appendix \ref{fba} for details on FBA) .
Using FBA, it was found that only 62 aerobic and 31 anaerobic minimal
media supported nonzero growth.
The list of 93 minimal media (62 aerobic and 31 anaerobic) is provided
in Table \ref{minimalmedialist}.
Note that in obtaining this list of 93 minimal media, we have used the
FBA model without incorporating any regulatory
constraints.
We have used this list of 93 minimal media for obtaining most our results.

\item[(b)] {\bf Library of 109732 media}: Following Barrett {\it et al}
\cite{BHRP2005}, we obtained a list of 109732 media.
The 143 external metabolites in the {\it E. coli} metabolic network were
divided into the four different sets:
carbon sources, nitrogen sources, sulphur sources and phosphorus sources.
Note that the four sets are not mutually exclusive.
For example, an amino acid is a carbon source as well as a nitrogen source.
A large library of 109732 different media was obtained by picking a single
metabolite from each of the four sets corresponding to
carbon sources, nitrogen sources, sulphur sources and phosphorus sources.
In addition, each media may or may not contain oxygen.
Some of the work reported here uses this larger library of 109732 media.

\end{itemize}


{\footnotesize
\begin{center}
\begin{longtable}{|c|l|l|}
\caption{List of minimal media considered as environmental conditions
to study the network regulating {\it E. coli} metabolism.
The 62 minimal media listed here are considered in aerobic conditions.
The first 31 media are considered also in anaerobic conditions.
Each carbon source is provided along with ammonium, sulphate, phosphate,
hydrogen, iron, potassium and sodium ions for uptake.
Oxygen is provided in aerobic conditions.}
\label{minimalmedialist} \\

\hline
\multicolumn{1}{|c}{\textbf{\tiny Serial number}} &
\multicolumn{1}{|l}{\textbf{\tiny Carbon Source}} &
\multicolumn{1}{|l|}{\textbf{\tiny Abbreviation of the carbon source}} \\
\hline
\endfirsthead

\multicolumn{3}{c}{\tiny {\tablename} \thetable{} -- Continued} \\
\hline
\multicolumn{1}{|c}{\textbf{\tiny Serial number}} &
\multicolumn{1}{|l}{\textbf{\tiny Carbon Source}} &
\multicolumn{1}{|l|}{\textbf{\tiny Abbreviation of the carbon source}} \\
\hline
\endhead

\hline
\multicolumn{3}{l}{{\tiny Continued on Next Page\ldots}} \\
\endfoot

\hline
\endlastfoot

{\tiny 1} &	{\tiny 2-Dehydro-3-deoxy-D-gluconate} & {\tiny 2ddglcn} \\
{\tiny 2} &	{\tiny	N-acetyl-D-glucosamine} &	{\tiny 	acgam} \\
{\tiny 3} &	{\tiny 	L-Arabinose} &	{\tiny 	arab-L} \\
{\tiny 4} &	{\tiny	Cytidine} &	{\tiny 	cytd} \\
{\tiny 5} &	{\tiny	D-Fructose} &	{\tiny 	fru} \\
{\tiny 6} &	{\tiny	L-Fucose} &	{\tiny 	fuc-L} \\
{\tiny 7} &	{\tiny	D-Glucose 6-phosphate} &	{\tiny 	g6p} \\
{\tiny 8} &	{\tiny		D-Galactose} &	{\tiny	 	gal} \\
{\tiny 9} &	{\tiny		D-Galactarate} &	{\tiny	 	galct-D} \\
{\tiny 10} &	{\tiny		D-Galactonate} &	{\tiny	 	galctn-D} \\
{\tiny 11} &	{\tiny		Galactitol} &	{\tiny	 	galt} \\
{\tiny 12} &	{\tiny		D-Glucosamine} &	{\tiny	 	gam} \\
{\tiny 13} &	{\tiny		D-Glucose} &	{\tiny	 	glc-D} \\
{\tiny 14} &	{\tiny		D-Gluconate} &	{\tiny	 	glcn} \\
{\tiny 15} &	{\tiny		D-Glucarate} &	{\tiny	 	glcr} \\
{\tiny 16} &	{\tiny		L-idonate} &	{\tiny	 	idon-L} \\
{\tiny 17} &	{\tiny		Inosine} &	{\tiny	 	ins} \\
{\tiny 18} &	{\tiny		Lactose} &	{\tiny	 	lcts} \\
{\tiny 19} &	{\tiny		Maltose} &	{\tiny	 	malt} \\
{\tiny 20} &	{\tiny		Maltohexaose} &	{\tiny	 	malthx} \\
{\tiny 21} &	{\tiny		Maltopentaose} &	{\tiny	 	maltpt} \\
{\tiny 22} &	{\tiny		Maltotriose} &	{\tiny	 	malttr} \\
{\tiny 23} &	{\tiny		Maltotetraose} &	{\tiny	 	maltttr} \\
{\tiny 24} &	{\tiny		D-Mannose} &	{\tiny	 	man} \\
{\tiny 25} &	{\tiny		Melibiose} &	{\tiny	 	melib} \\
{\tiny 26} &	{\tiny		D-Ribose} &	{\tiny	 	rib-D} \\
{\tiny 27} &	{\tiny		L-Rhamnose} &	{\tiny	 	rmn} \\
{\tiny 28} &	{\tiny		D-Sorbitol} &	{\tiny	 	sbt-D} \\
{\tiny 29} &	{\tiny		 Trehalose} &	{\tiny		 tre} \\
{\tiny 30} &	{\tiny		Xanthosine} &	{\tiny		xtsn} \\
{\tiny 31} &	{\tiny		D-Xylose} &	{\tiny		xyl-D} \\
{\tiny 32} &	{\tiny		3-(3-hydroxy-phenyl)propionate} &	{\tiny		3hpppn} \\
{\tiny 33} &	{\tiny		Acetate} &	{\tiny		ac} \\
{\tiny 34} &	{\tiny		Acetoacetate} &	{\tiny		acac} \\
{\tiny 35} &	{\tiny		D-Alanine} &	{\tiny		ala-D} \\
{\tiny 36} &	{\tiny		L-Alanine} &	{\tiny		ala-L} \\
{\tiny 37} &	{\tiny		L-Arginine} &	{\tiny		arg-L} \\
{\tiny 38} &	{\tiny		L-Asparagine} &	{\tiny		asn-L} \\
{\tiny 39} &	{\tiny		L-Asparate} &	{\tiny		asp-L} \\
{\tiny 40} &	{\tiny		Citrate} &	{\tiny		cit} \\
{\tiny 41} &	{\tiny		Fumarate} &	{\tiny		fum} \\
{\tiny 42} &	{\tiny		L-Glutamine} &	{\tiny		gln-L} \\
{\tiny 43} &	{\tiny		L-Glutamate} &	{\tiny		glu-L} \\
{\tiny 44} &	{\tiny		Glycine} &	{\tiny		gly} \\
{\tiny 45} &	{\tiny		Glycerol} &	{\tiny		glyc} \\
{\tiny 46} &	{\tiny		Glycolate} &	{\tiny		glyclt} \\
{\tiny 47} &	{\tiny		Hexadecanoate (n-C16:0)} &	{\tiny	hdca} \\
{\tiny 48} &	{\tiny		D-Lactate} &	{\tiny		lac-D} \\
{\tiny 49} &	{\tiny		L-Lactate} &	{\tiny		lac-L} \\
{\tiny 50} &	{\tiny		L-Malate} &	{\tiny		mal-L} \\
{\tiny 51} &	{\tiny		D-Mannitol} &	{\tiny		mnl} \\
{\tiny 52} &	{\tiny		Octadecanoate (n-C18:0)} &	{\tiny		ocdca} \\
{\tiny 53} &	{\tiny		Phenylpropanoate} &	{\tiny		pppn} \\
{\tiny 54} &	{\tiny		L-Proline} &	{\tiny		pro-L} \\
{\tiny 55} &	{\tiny		Pyruvate} &	{\tiny		pyr} \\
{\tiny 56} &	{\tiny		D-Serine} &	{\tiny		ser-D} \\
{\tiny 57} &	{\tiny		L-Serine} &	{\tiny		ser-L} \\
{\tiny 58} &	{\tiny		Succinate} &	{\tiny		succ} \\
{\tiny 59} &	{\tiny		L-tartrate} &	{\tiny		tartr-L} \\
{\tiny 60} &	{\tiny		L-Threonine} &	{\tiny		thr-L} \\
{\tiny 61} &	{\tiny		L-Tryptophan} &	{\tiny		trp-L} \\
{\tiny 62} &	{\tiny		Tetradecanoate (n-C14:0)} &	{\tiny	  	ttdca} \\

\end{longtable}
\end{center}
}

\subsection{Treatment of conditions ${\bf c}$}

The 19 Boolean variables $c_i(t)$ corresponding to conditions can be divided as follows:

\begin{itemize}

\item 15 of the 19 Boolean variables $c_i(t)$ depend upon the
configuration of a subset of transcription
factors and external metabolites at time $t$, i.e.,
\begin{equation}
\label{Ci}
c_i(t) = C_i({\bf t}(t),{\bf m(t)}), \quad \quad i = 1, 2, \ldots, 15,
\end{equation}
where the Boolean functions $C_i$ are given by the database iMC1010$^{v1}$.
The functions given by Eq. \ref{Ci} can be substituted in Eq. \ref{Fi}
which eliminates these 15 $c_i$ variables
from the dynamical system at the expense of more complicated effective
dependence of $g_i(t)$ on ${\bf t}(t)$ and ${\bf m}$.

\item Another condition variable $c_i(t)$ corresponds to the growth
of the cell.
This variable representing the growth of the cell is fixed to unity
as most media considered in this study give
a nonzero growth rate for the cell.

\item Another condition variable $c_i(t)$ corresponds to the pH of the
external environment.
In this study, pH is taken to be between 5.5 and 7, i.e., weakly acidic.
For example, one of the habitats of {\it E. coli} is the human gut where
the pH is weakly acidic.
The pH condition affects only 3 genes in the network.
For two of the genes that are affected by pH condition, the operative
regulatory clause is `pH $<$ 4'.
The Boolean variable $c_i$ corresponding to pH is fixed to zero (false)
for these two genes.
For the third gene the clause is `pH $<$ 7'.
For this gene, the Boolean variable $c_i$ corresponding to pH is fixed
to unity (true).

\item The remaining two condition variables $c_i(t)$ correspond to
`surplus FDP' and `surplus PYR' in the database
iMC1010$^{v1}$.
They represent whether surplus amounts of fructose 1,6-bisphosphate
and pyruvate are being produced in the cell.
These two conditions depend upon the values of some of the internal
fluxes $v_i$ and the presence of an external metabolite,
fructose, through specified Boolean functions in the database.
The variable corresponding to external fructose is treated as unity,
if the minimal medium includes fructose and zero otherwise.
The treatment of the internal fluxes $v_i$ is discussed next.

\end{itemize}


\subsection{Treatment of internal fluxes ${\bf v}$}

The state of some of the genes in the regulatory network is influenced
by the state of 21 components of vector ${\bf v}$ representing
fluxes of internal metabolic reactions.
As described in Covert {\it et al} \cite{CSP2001}, the fluxes of
internal reactions in the database are surrogate for
internal metabolite concentrations inside the cell that determine
the state of genes in the regulatory network.
The internal metabolite concentrations have been approximated by
internal fluxes in the database iMC1010$^{v1}$ as the database
is intended to model the integrated metabolic and regulatory system
f {\it E. coli} using the framework of FBA.
Since FBA cannot determine the internal metabolite concentrations,
the internal metabolite concentrations have been approximated by
internal fluxes of reactions in the network.
In this study, we have treated the variables corresponding to the
internal fluxes in two distinct ways leading to two slightly
different dynamical systems as discussed below.

\begin{itemize}

\item[(a)] {\bf System A}: In the first approach, for a given external environment
${\bf m}$, we determined whether each of the
21 internal reactions in the {\it E. coli} metabolic network whose fluxes correspond
to the 21 components of vector ${\bf v}$
was `blocked' or not \cite{SS1991,BNSM2004,SSGKRJ2006}.
A reaction is said to be blocked in a particular environmental condition, if under
that medium no steady state flux is possible
through it \cite{SS1991,Heinrichbook,BNSM2004}.
The algorithm to determine blocked reactions for a given environment ${\bf m}$ is
described in detail in section \ref{blocked} in
Appendix \ref{fba}.
An internal flux variable $v_i(t)$ was set equal to zero for a medium
(specified by vector ${\bf m}$), if the reaction was found to
be blocked for that medium, and unity otherwise.
Thus, in this approach, the  internal fluxes $v_i$ were not dynamical variables,
but rather fixed parameters (albeit fixed with an
eye on self-consistency).

\item[(b)] {\bf System B}: In the second approach, the Boolean variables corresponding
to internal fluxes $v_i$ were allowed to be
dynamical by making a simplifying assumption about their dynamics.
In the metabolic network, the flux of a reaction is determined by the concentrations of
the participating metabolites and the
catalyzing enzymes.
The enzyme concentrations are determined by the activity or state of their respective genes.
In a discrete-time approximation, an enzyme is present at time $t$ if the genes coding for
it were active at the previous time $t-1$.
Thus, the internal flux $v_i(t)$ is set equal to unity if the genes coding for the enzyme
catalyzing that metabolic reaction were
active at the previous time $t-1$, and zero otherwise.
This could be done for a subset of 10 out of 21 internal reaction fluxes, since the genes of
their catalyzing enzymes were part of the
583 genes in the database.
Genes coding for the enzymes of the remaining 11 internal reaction fluxes were not part of the
regulatory network database and hence
the corresponding $v_i$ could not be made dynamical variables in this fashion.
These latter $v_i$ were fixed as in part (a) for each medium by checking their blocked status
for the medium under consideration.
The approach (b) introduces feedbacks in the genetic regulatory network.

\end{itemize}

\noindent The above treatment defines the substitutions to be made in
Eq. \ref{Fi} for the variables ${\bf c}(t)$ and ${\bf v}(t)$.
Each component of {\bf c} in Eq. \ref{Fi} is either a specified Boolean function
of ${\bf t}(t)$, {\bf m}, and ${\bf v}(t)$, or is
a suitably chosen Boolean constant.
Each component of ${\bf v}(t)$ is, in turn, either a suitably chosen
Boolean constant or a specified Boolean function of ${\bf g}(t-1)$.
The above mentioned substitutions together with Eq. \ref{Ti} make the right hand
side of Eq. \ref{Fi} a function of only ${\bf g}(t-1)$
and ${\bf m}$, i.e., Eq. \ref{Fi} reduces to
\begin{equation}
g_i(t) = G_i({\bf g}(t-1), {\bf m}),
\end{equation}
which is the same as Eq. \ref{dynamicalsystem}.
The functions $G_i$ define the final dynamical system, and include information
coming from the functions $F_i$, as well as the dependence
of ${\bf t}$, ${\bf c}$ and ${\bf v}$ on ${\bf g}$ and ${\bf m}$.

Note that the choices (a) and (b) for the ${\bf v}$ variables yield different
dynamical systems for Eq. \ref{dynamicalsystem} which we
denote as system A and system B, respectively.
In system B, 6 out of 583 genes have additional links from other genes in the
set compared to system A.
Programs implementing the two dynamical systems A and B are discussed in
Appendix \ref{program}.


\section{Homeostasis: The attractor is insensitive to any
perturbation of the genes}

\begin{figure*}
\centering
\includegraphics[width=14cm]{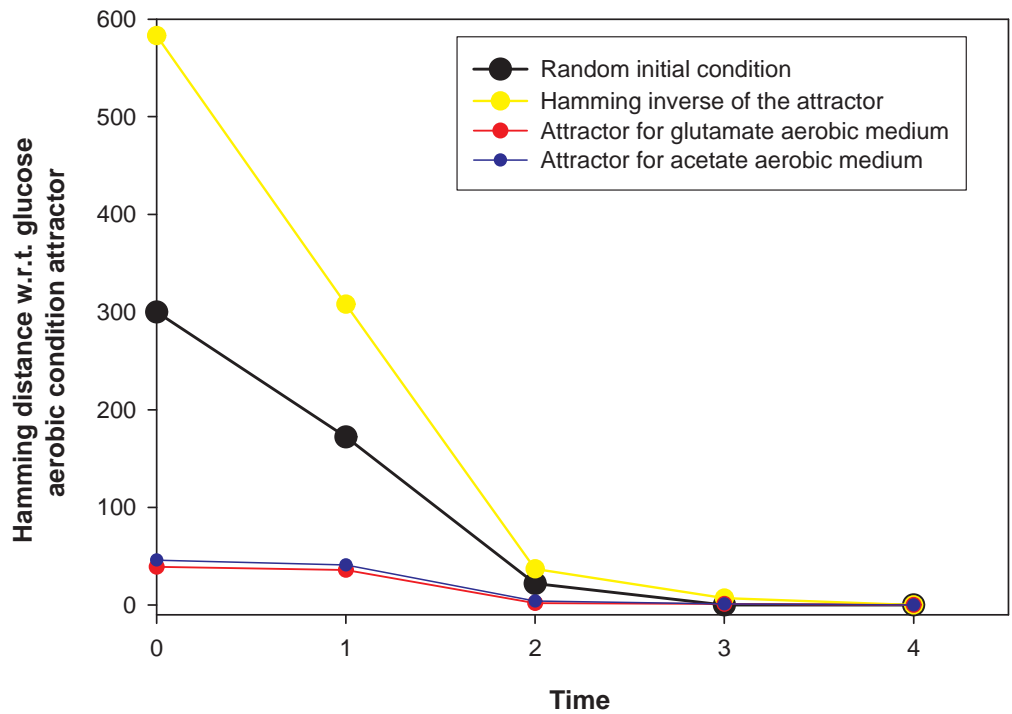}
\caption{
Dynamical behaviour of the genetic network controlling {\it E. coli}
metabolism for a fixed environment
corresponding to glucose aerobic minimal medium.
For all initial conditions of the genes, the system is attracted to a
fixed point whose configuration
depends upon the medium.
The plots depict, as a function of time, the hamming distance of the
configuration from the fixed point
attractor corresponding to the medium.
We have shown here simulations for 4 different initial conditions of
the genes.
One is a randomly chosen initial condition.
Another is the `hamming inverse' of the attractor (in which the
configuration of every gene is reversed
with respect to the attractor).
Two other initial conditions are the attractor configurations of
other minimal media.}
\label{homeostasis}
\end{figure*}

The trajectories of the Boolean dynamical system given by
Eq. \ref{dynamicalsystem}
depend on the initial state of the genes and the state of
the environment.
In this section, we investigate the dependence of the attractors on the initial
condition of the genes for a given fixed environmental condition.
Starting from 10000 randomly chosen initial conditions of the genes, i.e., ${\bf g}$
vectors, we determined the attractors of the dynamical system A for a fixed environment
corresponding to glucose aerobic minimal medium.
For glucose aerobic minimal medium, the system reached a fixed point attractor for
each of the 10000 initial conditions of the genes.
For each of the 10000 initial conditions, the fixed point attractor was reached in a
maximum of 4 time steps.
Further, the same fixed point attractor was obtained for each of the 10000 initial
conditions of the genes, under glucose aerobic minimal medium.
This seems to suggest that a unique fixed point is the global attractor for the
genetic network under glucose aerobic minimal medium regardless of the initial
conditions of the genes.
Examples are shown in Fig. \ref{homeostasis} for glucose aerobic minimal medium for four
different initial conditions of the genes.
Similarly, we determined the attractors of the dynamical system A for each of the 93
buffered minimal media listed in Table \ref{minimalmedialist} starting from 10000
randomly chosen initial conditions of the genes in the network.
For each of the 93 different buffered media, we found that the system reached a fixed
point attractor in maximum of 4 time steps independent of the initial conditions of
the genes in the network.
The attractor was also determined for each of the 109732 media in the larger library
mentioned in section \ref{media} starting from a single randomly chosen initial condition
of the genes.
A fixed point attractor was found for each of the 109732 buffered media.
In principle, there are 2$^{583}$ initial conditions of the genes and one may argue that
only a small amount of the state space has been sampled in these simulations.
In the next chapter, an analytical argument will be presented based on the architecture
of the regulatory network as to why a unique final configuration or attractor independent
of the initial condition of the genes is inevitable for each fixed environment ${\bf m}$.
Thus, we found that as long as the external environment remains fixed, the genetic network
controlling {\it E. coli} metabolism will revert to a unique configuration of its genes
after any perturbation of gene states.

The dynamical system B includes some additional links between genes compared to system A
as was mentioned earlier.
The attractors of the system B were obtained for each of the 93 buffered minimal media
listed in Table \ref{minimalmedialist} starting from 1000 randomly chosen initial conditions
of the genes for each minimal media.
For system B, we obtained 36 distinct attractors (8 fixed point attractors and 28 two-cycles)
for 89 of the 93 buffered minimal media.
For the remaining $4$ minimal media, we obtained $10$ distinct attractors ($4$ fixed point
attractors and $6$ two-cycles).
The attractor was again reached in a maximum of 4 time steps.
For each of the two-cycles, it was found that most of the genes (562 to 567 out of 583)
were in fact frozen in a fixed configuration, and only 16 to 21 genes oscillated back and
forth between zero and one with period two.
These 21 genes are listed in Table \ref{oscillate}.
Furthermore, for any given medium, it was found that each of the 562 frozen genes had
the same configuration across all the attractors (36 or 10).
This means that for any given medium, most genes (562 or more out of 583) end up in the
same fixed configuration independent of the initial conditions of the genes.
One can show that there are no other attractors of the system B, using the structural
properties of the regulatory network.
We have also checked that the 562 frozen genes end up in the same configuration in both
system A and system B for any given medium.

Together, the results obtained here for dynamical systems A and B imply that the regulatory
network controlling {\it E. coli} metabolism exhibits a high degree of homeostasis, in that
it is highly insensitive to initial conditions of the genes.
Further, for any given medium, all genetic perturbations die out very quickly, restoring an
overwhelming majority of the genes to a configuration that is independent of the perturbation.


\begin{table}
\centering
\begin{tabular}{c|c}

{\small \bf Gene} &	{\small \bf bnumber} \\
\hline

{\small nagC} &	    {\small b0676} \\
{\small nagA} &		{\small b0677} \\
{\small nagB} &		{\small b0678} \\
{\small nagE} &		{\small b0679} \\
{\small deoR} &		{\small b0840} \\
{\small uxaB} &		{\small b1521} \\
{\small kdgR} &		{\small b1827} \\
{\small uxaA} &		{\small b3091} \\
{\small uxaC} &		{\small b3092} \\
{\small exuT} &		{\small b3093} \\
{\small exuR} &		{\small b3094} \\
{\small kdgK} &		{\small b3526} \\
{\small glmU} &		{\small b3730} \\
{\small kdgT} &		{\small b3909} \\
{\small uxuA} &		{\small b4322} \\
{\small uxuB} &		{\small b4323} \\
{\small uxuR} &		{\small b4324} \\
{\small deoC} &		{\small b4381} \\
{\small deoA} &		{\small b4382} \\
{\small deoB} &		{\small b4383} \\
{\small deoD} &		{\small b4384} \\
\hline

\end{tabular}
\caption{List of 21 genes whose configuration can
oscillate in dynamical system B.}
\label{oscillate}
\end{table}

\section{Flexibility: The system mounts a highly flexible response
to changed environments}

In the last section, we have seen that the genetic network
controlling {\it E. coli} metabolism exhibits the property
of homeostasis or robust adaptation to internal perturbations
of gene configurations for a fixed external environment.
While this property of homeostasis is very useful for any
given environmental condition, the organism also needs to
exhibit a robust and flexible response to external
perturbations corresponding to changes in the environment.
In this section, we investigate the sensitivity of the
attractors of the genetic network to changes in the
environmental conditions.
We studied the flexibility of the genetic network to
environmental changes in two ways.

First, we determined the hamming distance between attractor
states of the system A corresponding to pairs of minimal media.
For the set of 93 buffered minimal media listed in
Table \ref{minimalmedialist}, we had obtained the attractors
of the system A starting from different initial conditions of
the genes.
The attractor was found to be a fixed point for any given
minimal medium independent of initial conditions of the genes.
Thus, a set of 93 fixed point attractors was obtained for
the set of 93 minimal media listed in
Table \ref{minimalmedialist}.
We found the largest hamming distance between any two
attractors out of this set for 93 minimal media to be 114.

\begin{figure*}
\centering
\includegraphics[width=14cm]{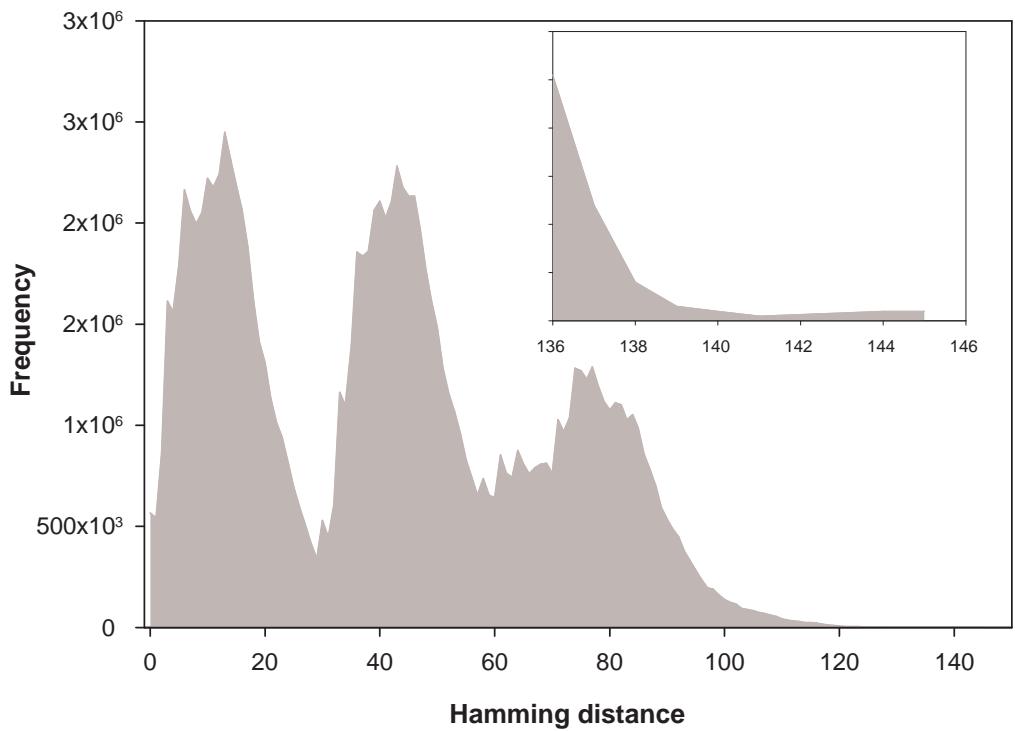}
\caption{
The genetic network controlling {\it E. coli} metabolism is flexible in its response
to changed environmental conditions.
Changing the environmental condition can lead to a wide range of hamming distances
among the attractors.
In the figure, the distribution of pair-wise hamming distances between attractors for
15427 different environmental conditions is shown.
{\bf Inset:} Enlargement of the graph for large hamming distances.
The largest hamming distance obtained between attractors for two different environmental
conditions is 145.}
\label{flexibility}
\end{figure*}

We had also obtained the attractors of the dynamical
system A for each of the 109732 media contained in the
larger library mentioned in section \ref{media}.
For each of the 109732 media, a fixed point attractor
was obtained.
We then ran `constrained FBA' for each of the 109732
attractors for the larger library of 109732 buffered
media to determine which of them support a nonzero
growth rate for their corresponding buffered media
(for details of constrained FBA, see section
\ref{constrainedfba} in Appendix \ref{fba}).
We obtained a nonzero growth rate for the attractors
corresponding to 15427 buffered media.
We then computed the pairwise hamming distances among
this set of 15427 attractors also.
The largest hamming distance between any pair of
attractors was found to be 145.
The distribution of these hamming distances is trimodal
as shown in Fig. \ref{flexibility}.
The trimodal distribution obtained here is similar to
that found and discussed by Barrett {\it et al}
\cite{BHRP2005}.
Thus, although the attractor for a fixed environmental
condition is unique, the attractors for two different
environmental conditions can be quite far apart.
Therefore, while the system is insensitive to
fluctuations of gene configurations for a fixed external
environment, it can move to quite a different attractor
when it encounters a changed environment.
Thus, the system shows flexibility of response to
changed environmental conditions.

\begin{figure*}
\centering
\includegraphics[width=14cm]{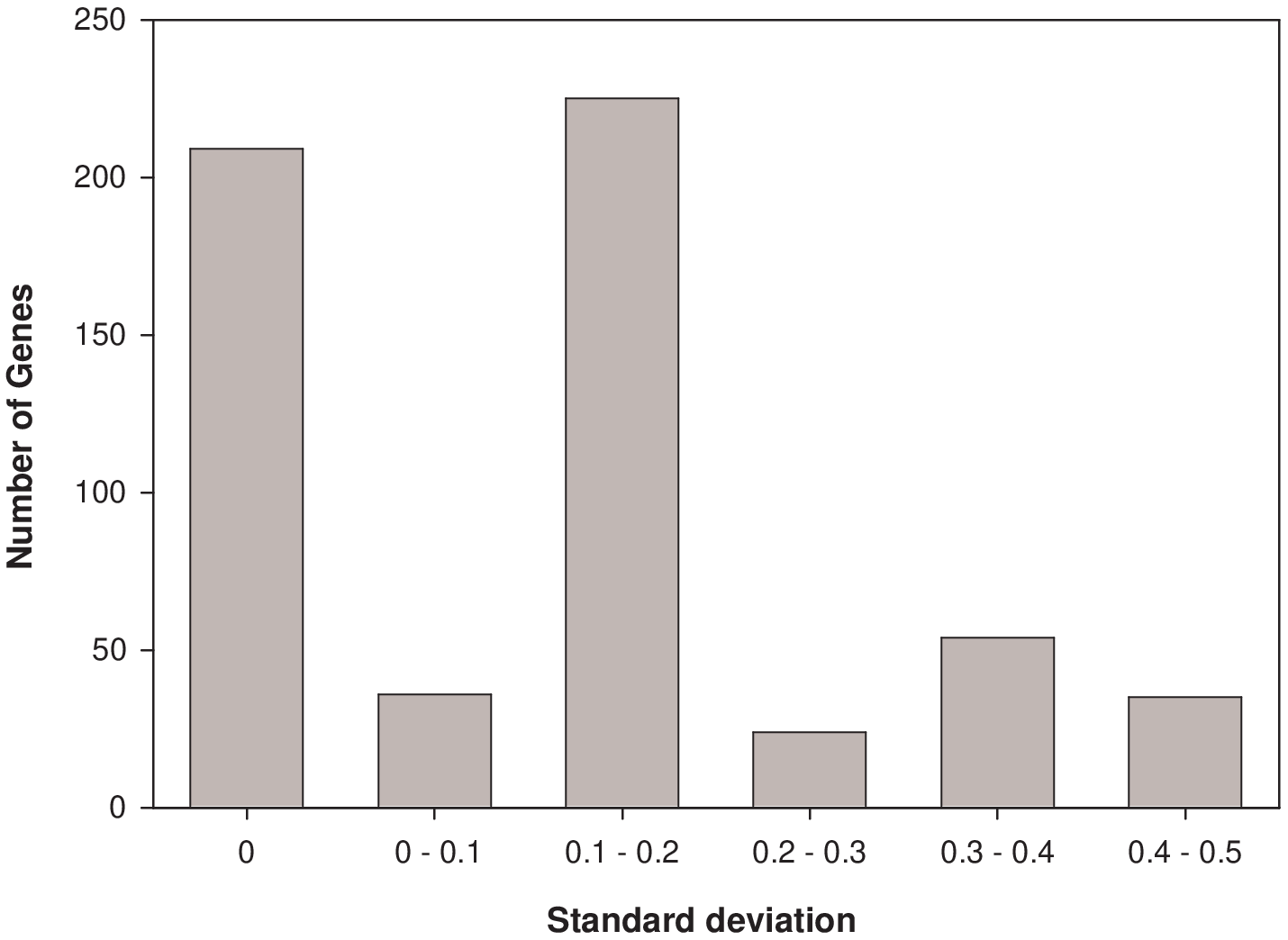}
\caption{
The histogram of standard deviation of a gene's configurations across 15427 attractors for different environmental conditions.
The left-most bar corresponds to 209 genes whose configuration remains unchanged.}
\label{variability}
\end{figure*}

Second, we determined the number of media in which any
gene is on across the set of attractors for 15427
buffered media.
This number for any gene can range from 0 to 15427.
We found that across these 15427 conditions the genes
that had a configuration which differed between any pair
of attractors were drawn from a set of 374 out of the
583 genes in the network.
Of these 374 genes, 66 genes code for transcription
factors and 308 genes code for metabolic enzymes.
The remaining 209 genes had the same configuration
(75 off and 134 on) in all the 15427 attractors.
The variability of a gene's configuration across
different environmental conditions can be
characterized by the standard deviation of its
value (0 or 1) across this set.
We found this standard deviation to range from zero
to close to its maximum possible value 0.5.
The mean of the standard deviations for the 374 genes
whose configuration can differ across the set of
15427 attractors is 0.20.
The histogram of standard deviation values is shown
in Fig. \ref{variability}.
These observations quantify the considerable variety in a
gene's variability across environmental conditions.

\section{Adaptability: The regulatory network's response to changed media
increases metabolic efficiency}
\label{adaptability}

\begin{figure*}
\centering
\includegraphics[width=14cm]{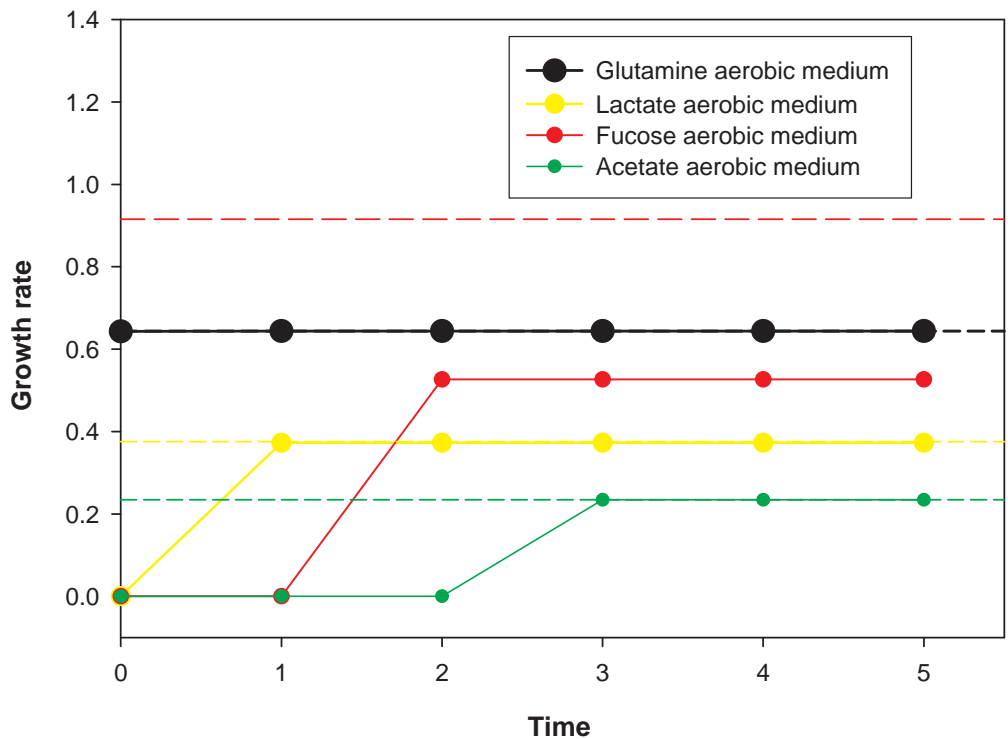}
\caption{
Metabolic efficiency due to regulation.
The figure shows the adaptation of the {\it E. coli} regulatory network towards higher growth rate in response to change of medium.
Growth rate obtained using constrained FBA is plotted for 4 trajectories of the regulatory network corresponding to aerobic minimal
media with glutamine, lactate, fucose or acetate as the carbon source.
The initial condition of the genetic network in each case is the attractor for the glutamate aerobic minimal medium.
Dotted lines show the pure FBA growth rate in the 4 minimal media.
The growth rate increases in three and remains constant in one of these trajectories.}
\label{optimality}
\end{figure*}

In this section, we tracked how the metabolic response of the
cell, as measured by its growth rate computed using FBA, changes
when its environment changes.
A reaction in the metabolic network can be assumed to be off
if none of the enzymes catalyzing it are being produced, or,
equivalently, in our dynamical system, if the genes coding for
those enzymes are in the off state.
For any configuration of the metabolic genes, FBA can thus be
used to compute the growth rate of the cell by turning off
all reactions whose corresponding genes are in the off state
in that configuration, thereby capturing the effect of gene
regulation on metabolic function (see section \ref{constrainedfba}
in Appendix \ref{fba}).
We computed this `constrained FBA' growth rate for each of the
attractors of the dynamical system A for the 93 minimal media
listed in Table \ref{minimalmedialist}.
81 of them, listed in Table \ref{optimalitytable}, gave a nonzero
growth rate.
Starting from an initial condition of the genes in dynamical
system A that corresponds to the attractor for any one of the
81 buffered minimal media listed in Table \ref{optimalitytable},
say X, we computed the time course of the genetic network
configuration in another buffered medium, say Y, until the
system reached the attractor corresponding to minimal medium Y.
For each of the genetic network configurations in the trajectory,
i.e., from the attractor for medium X to the attractor for
medium Y, we computed the growth rate using constrained FBA.
This effectively tracks how the constrained growth rate of the
cell changes with time after its environment changes suddenly
from X to Y.
The result is shown in Fig. \ref{optimality} for the cases
where the carbon source in X is glutamate and in Y is glutamine,
lactate, fucose or acetate.
In the attractor of X the growth rate is low for the medium Y.
We find that genetic network configuration changes with time
so as to typically increase the growth rate.

We found that for the 81 minimal media listed in Table
\ref{optimalitytable}, the growth rate in the attractor
configuration of the medium was greater than the average growth
rate in the other 80 attractors by a factor of 3.5 (averaged
over the 81 media).
Moreover, the average time to move to the attractor configuration
from initial conditions corresponding to attractors for 80
remaining media was only 2.6 time steps.
In other words, regulatory dynamics enables the cell to adapt
to its environment to increase its metabolic efficiency very
substantially, fairly quickly.

\begin{figure*}
\centering
\includegraphics[width=14cm]{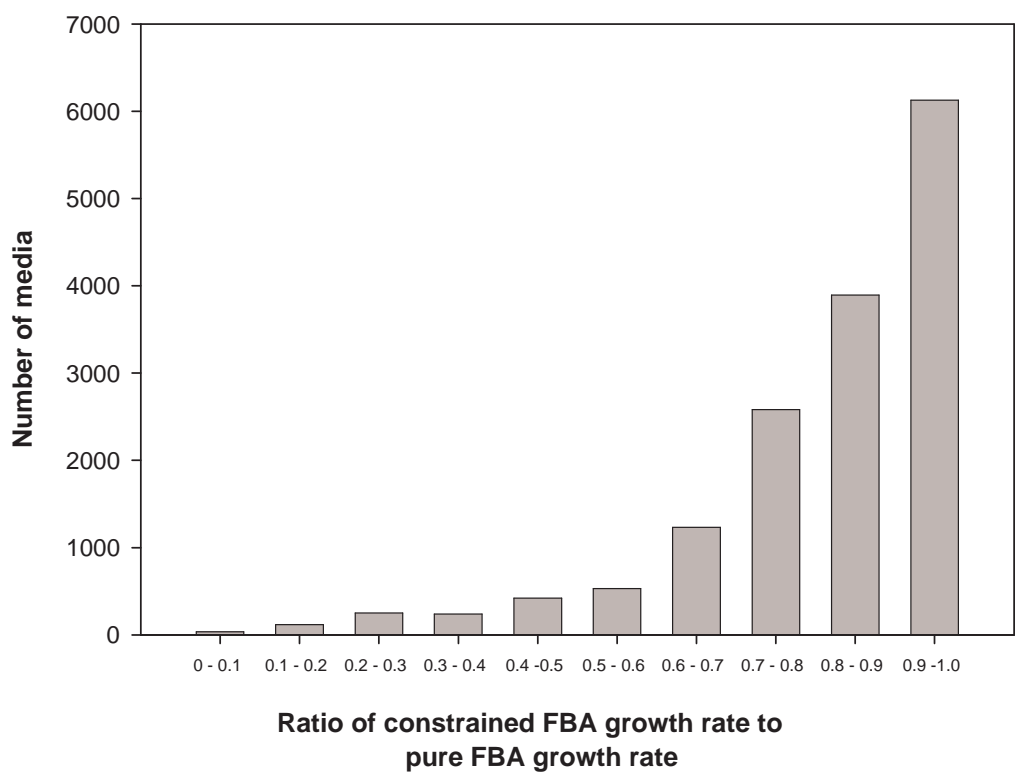}
\caption{
Histogram of the ratio of constrained FBA growth rate in the attractor of each of 15427 minimal media discussed in text to the pure
FBA growth rate in that medium.
This is peaked in the bin with the largest ratio ($\geq 0.9$).}
\label{optimalityratio}
\end{figure*}

We then considered the set of 15427 media whose respective
attractor configurations gave a nonzero growth rate following
constrained FBA.
We also computed the pure FBA growth rate for each of these
15427 media without imposing any regulatory constraints from
the respective attractors for these media.
We determined the ratio of the constrained FBA growth rate to
the pure FBA growth rate for each of the 15427 media.
The average value of this ratio was found to be as high as
0.815 and was less than 0.5 for only 7\% of the media.
The histogram of these ratios is shown in
Fig. \ref{optimalityratio}.
This shows that the regulatory dynamics results in a
close-to-optimal metabolic functioning under a large set
of environments.
This observation also lends support to the usefulness of FBA
in probing metabolic organization.

We remark that in a dynamical system of the type given
by Eq. \ref{dynamicalsystem}, it is of course not surprising
that the attractor of the genes' configuration ${\bf g}$
depends upon the external metabolite configuration ${\bf m}$.
Our results related to flexibility and adaptability are an
attempt to quantify the change in the attractors as the
external environment is varied and to show that the
change is functionally useful in the survival of the organism.

{\footnotesize
\begin{center}
\begin{longtable}{|c|c|c|c|c|c|}
\caption{Comparison of growth rate obtained using pure (unconstrained)
FBA with that obtained using constrained FBA for various minimal media.
For each media, the maximum uptake rate of the carbon source was set to
10 in appropriate units and the uptake rates of all other metabolites
(inorganics) in the media were left unconstrained.
The data points in columns 4, 5 and 6 have been rounded off to three
decimal places.
The appropriate units are mM g-DCW$^{-1}$ hr$^{-1}$ (milli moles per gram
dry carbon weight per hour), see \cite{CSP2001}.}
\label{optimalitytable} \\

\hline
\multicolumn{1}{|c}{\textbf{\tiny Serial}} &
\multicolumn{1}{|c}{\textbf{\tiny Minimal}} &
\multicolumn{1}{|c|}{\textbf{\tiny Oxygen}} &
\multicolumn{1}{|c|}{\textbf{\tiny Growth Rate with }} &
\multicolumn{1}{|c|}{\textbf{\tiny Growth Rate with no}} &
\multicolumn{1}{|c|}{\textbf{\tiny Ratio}} \\
\multicolumn{1}{|c}{\textbf{\tiny number}} &
\multicolumn{1}{|c}{\textbf{\tiny media}} &
\multicolumn{1}{|c|}{\textbf{\tiny availability}} &
\multicolumn{1}{|c|}{\textbf{\tiny regulatory constraints}} &
\multicolumn{1}{|c|}{\textbf{\tiny regulatory constraints}} &
\multicolumn{1}{|c|}{\textbf{\tiny (GRreg/GRpure)}} \\
\multicolumn{1}{|c}{} &
\multicolumn{1}{|c}{} &
\multicolumn{1}{|c|}{} &
\multicolumn{1}{|c|}{\textbf{\tiny (GRreg)}} &
\multicolumn{1}{|c|}{\textbf{\tiny (GRpure)}} &
\multicolumn{1}{|c|}{} \\
\hline
\endfirsthead

\multicolumn{6}{c}{\tiny {\tablename} \thetable{} -- Continued} \\
\hline
\multicolumn{1}{|c}{\textbf{\tiny Serial}} &
\multicolumn{1}{|c}{\textbf{\tiny Minimal}} &
\multicolumn{1}{|c|}{\textbf{\tiny Oxygen}} &
\multicolumn{1}{|c|}{\textbf{\tiny Growth Rate with }} &
\multicolumn{1}{|c|}{\textbf{\tiny Growth Rate with no}} &
\multicolumn{1}{|c|}{\textbf{\tiny Ratio}} \\
\multicolumn{1}{|c}{\textbf{\tiny number}} &
\multicolumn{1}{|c}{\textbf{\tiny media}} &
\multicolumn{1}{|c|}{\textbf{\tiny availability}} &
\multicolumn{1}{|c|}{\textbf{\tiny regulatory constraints}} &
\multicolumn{1}{|c|}{\textbf{\tiny regulatory constraints}} &
\multicolumn{1}{|c|}{\textbf{\tiny (GRreg/GRpure)}} \\
\multicolumn{1}{|c}{} &
\multicolumn{1}{|c}{} &
\multicolumn{1}{|c|}{} &
\multicolumn{1}{|c|}{\textbf{\tiny (GRreg)}} &
\multicolumn{1}{|c|}{\textbf{\tiny (GRpure)}} &
\multicolumn{1}{|c|}{} \\
\hline
\endhead

\hline
\multicolumn{6}{l}{{\tiny Continued on Next Page\ldots}} \\
\endfoot

\hline
\endlastfoot

{\tiny 1} &	{\tiny	ac} &	{\tiny	aerobic} &	{\tiny	0.234} &	{\tiny	0.234} &	{\tiny	0.998} \\
{\tiny 2} &	{\tiny	ala-D} &	{\tiny	aerobic} &	{\tiny	0.416} &	{\tiny	0.423} &	{\tiny	0.985} \\
{\tiny 3} &	{\tiny	ala-L} &	{\tiny	aerobic} &	{\tiny	0.416} &	{\tiny	0.423} &	{\tiny	0.985} \\
{\tiny 4} &	{\tiny	arab-L} &	{\tiny	aerobic} &	{\tiny	0.785} &	{\tiny	0.786} &	{\tiny	0.998} \\
{\tiny 5} &	{\tiny	arab-L} &	{\tiny	anaerobic} &	{\tiny	0.220} &	{\tiny	0.222} &	{\tiny	 0.992} \\
{\tiny 6} &	{\tiny	arg-L} &	{\tiny	aerobic} &	{\tiny	0.743} &	{\tiny	0.784} &	{\tiny	0.948} \\
{\tiny 7} &	{\tiny	asn-L} &	{\tiny	aerobic} &	{\tiny	0.452} &	{\tiny	0.452} &	{\tiny	0.999} \\
{\tiny 8} &	{\tiny	asp-L} &	{\tiny	aerobic} &	{\tiny	0.451} &	{\tiny	0.451} &	{\tiny	0.998} \\
{\tiny 9} &	{\tiny	cytd} &	{\tiny	aerobic} &	{\tiny	0.826} &	{\tiny	0.872} &	{\tiny	0.948} \\
{\tiny 10} &	{\tiny	cytd} &	{\tiny	anaerobic} &	{\tiny	0.282} &	{\tiny	0.394} &	{\tiny	 0.716} \\
{\tiny 11} &	{\tiny	ddglcn} &	{\tiny	aerobic} &	{\tiny	0.830} &	{\tiny	0.831} &	{\tiny	 0.998} \\
{\tiny 12} &	{\tiny		ddglcn} &	{\tiny		anaerobic} &	{\tiny		0.226} &	{\tiny		 0.228} &	{\tiny		0.993} \\
{\tiny 13} &	{\tiny		fru} &	{\tiny		aerobic} &	{\tiny		0.955} &	{\tiny		0.957} &	 {\tiny		0.998} \\
{\tiny 14} &	{\tiny		fru} &	{\tiny		anaerobic} &	{\tiny		0.297} &	{\tiny		0.299} &	 {\tiny		0.992} \\
{\tiny 15} &	{\tiny		fuc-L} &	{\tiny		aerobic} &	{\tiny		0.526} &	{\tiny		0.915} &	 {\tiny		0.575} \\
{\tiny 16} &	{\tiny		fuc-L} &	{\tiny		anaerobic} &	{\tiny		0.156} &	{\tiny		 0.158} &	{\tiny		0.993} \\
{\tiny 17} &	{\tiny		fum} &	{\tiny		aerobic} &	{\tiny		0.439} &	{\tiny		0.439} &	 {\tiny		0.998} \\
{\tiny 18} &	{\tiny		g6p} &	{\tiny		aerobic} &	{\tiny		0.990} &	{\tiny		0.992} &	 {\tiny		0.998} \\
{\tiny 19} &	{\tiny		g6p} &	{\tiny		anaerobic} &	{\tiny		0.377} &	{\tiny		0.380} &	 {\tiny		0.992} \\
{\tiny 20} &	{\tiny		gal} &	{\tiny		aerobic} &	{\tiny		0.944} &	{\tiny		0.946} &	 {\tiny		0.998} \\
{\tiny 21} &	{\tiny		gal} &	{\tiny		anaerobic} &	{\tiny		0.270} &	{\tiny		0.272} &	 {\tiny		0.992} \\
{\tiny 22} &	{\tiny		galct-D} &	{\tiny		aerobic} &	{\tiny		0.663} &	{\tiny		0.664} &	 {\tiny		0.998} \\
{\tiny 23} &	{\tiny		galct-D} &	{\tiny		anaerobic} &	{\tiny		0.209} &	{\tiny		 0.210} &	{\tiny		0.993} \\
{\tiny 24} &	{\tiny		galctn-D} &	{\tiny		aerobic} &	{\tiny		0.830} &	{\tiny		0.831} &	 {\tiny		0.998} \\
{\tiny 25} &	{\tiny		galctn-D} &	{\tiny		anaerobic} &	{\tiny		0.226} &	{\tiny		 0.228} &	{\tiny		0.993} \\
{\tiny 26} &	{\tiny		galt} &	{\tiny		aerobic} &	{\tiny		1.007} &	{\tiny		1.009} &	 {\tiny		0.998} \\
{\tiny 27} &	{\tiny		galt} &	{\tiny		anaerobic} &	{\tiny		0.253} &	{\tiny		0.255} &	 {\tiny		0.993} \\
{\tiny 28} &	{\tiny		gam} &	{\tiny		aerobic} &	{\tiny		0.955} &	{\tiny		0.957} &	 {\tiny		0.998} \\
{\tiny 29} &	{\tiny		gam} &	{\tiny		anaerobic} &	{\tiny		0.297} &	{\tiny		0.299} &	 {\tiny		0.992} \\
{\tiny 30} &	{\tiny		glc-D} &	{\tiny		aerobic} &	{\tiny		0.955} &	{\tiny		0.957} &	 {\tiny		0.998} \\
{\tiny 31} &	{\tiny		glc-D} &	{\tiny		anaerobic} &	{\tiny		0.297} &	{\tiny		 0.299} &	{\tiny		0.992} \\
{\tiny 32} &	{\tiny		glcn} &	{\tiny		aerobic} &	{\tiny		0.876} &	{\tiny		0.877} &	 {\tiny		0.998} \\
{\tiny 33} &	{\tiny		glcn} &	{\tiny		anaerobic} &	{\tiny		0.241} &	{\tiny		0.243} &	 {\tiny		0.990} \\
{\tiny 34} &	{\tiny		glcr} &	{\tiny		aerobic} &	{\tiny		0.663} &	{\tiny		0.664} &	 {\tiny		0.998} \\
{\tiny 35} &	{\tiny		glcr} &	{\tiny		anaerobic} &	{\tiny		0.209} &	{\tiny		0.210} &	 {\tiny		0.993} \\
{\tiny 36} &	{\tiny		gln-L} &	{\tiny		aerobic} &	{\tiny		0.644} &	{\tiny		0.644} &	 {\tiny		0.999} \\
{\tiny 37} &	{\tiny		glu-L} &	{\tiny		aerobic} &	{\tiny		0.670} &	{\tiny		0.674} &	 {\tiny		0.994} \\
{\tiny 38} &	{\tiny		glyc} &	{\tiny		aerobic} &	{\tiny		0.555} &	{\tiny		0.555} &	 {\tiny		0.998} \\
{\tiny 39} &	{\tiny		glyclt} &	{\tiny		aerobic} &	{\tiny		0.177} &	{\tiny		0.177} &	 {\tiny		0.998} \\
{\tiny 40} &	{\tiny		hpppn} &	{\tiny		aerobic} &	{\tiny		1.124} &	{\tiny		1.125} &	 {\tiny		0.999} \\
{\tiny 41} &	{\tiny		idon-L} &	{\tiny		aerobic} &	{\tiny		0.866} &	{\tiny		0.867} &	 {\tiny		0.998} \\
{\tiny 42} &	{\tiny		idon-L} &	{\tiny		anaerobic} &	{\tiny		0.207} &	{\tiny		 0.208} &	{\tiny		0.992} \\
{\tiny 43} &	{\tiny		ins} &	{\tiny		aerobic} &	{\tiny		0.888} &	{\tiny		0.889} &	 {\tiny		0.998} \\
{\tiny 44} &	{\tiny		ins} &	{\tiny		anaerobic} &	{\tiny		0.350} &	{\tiny		0.352} &	 {\tiny		0.995} \\
{\tiny 45} &	{\tiny		lac-D} &	{\tiny		aerobic} &	{\tiny		0.410} &	{\tiny		0.413} &	 {\tiny		0.992} \\
{\tiny 46} &	{\tiny		lac-L} &	{\tiny		aerobic} &	{\tiny		0.372} &	{\tiny		0.375} &	 {\tiny		0.992} \\
{\tiny 47} &	{\tiny		lcts} &	{\tiny		aerobic} &	{\tiny		1.900} &	{\tiny		1.903} &	 {\tiny		0.998} \\
{\tiny 48} &	{\tiny		lcts} &	{\tiny		anaerobic} &	{\tiny		0.566} &	{\tiny		0.571} &	 {\tiny		0.992} \\
{\tiny 49} &	{\tiny		mal-L} &	{\tiny		aerobic} &	{\tiny		0.427} &	{\tiny		0.439} &	 {\tiny		0.971} \\
{\tiny 50} &	{\tiny		malt} &	{\tiny		aerobic} &	{\tiny		1.911} &	{\tiny		1.914} &	 {\tiny		0.998} \\
{\tiny 51} &	{\tiny		malt} &	{\tiny		anaerobic} &	{\tiny		0.593} &	{\tiny		0.598} &	 {\tiny		0.992} \\
{\tiny 52} &	{\tiny		malthx} &	{\tiny		aerobic} &	{\tiny		5.826} &	{\tiny		5.835} &	 {\tiny		0.998} \\
{\tiny 53} &	{\tiny		malthx} &	{\tiny		anaerobic} &	{\tiny		1.995} &	{\tiny		 2.010} &	{\tiny		0.992} \\
{\tiny 54} &	{\tiny		maltpt} &	{\tiny		aerobic} &	{\tiny		4.824} &	{\tiny		4.832} &	 {\tiny		0.998} \\
{\tiny 55} &	{\tiny		maltpt} &	{\tiny		anaerobic} &	{\tiny		1.591} &	{\tiny		 1.603} &	{\tiny		0.992} \\
{\tiny 56} &	{\tiny		malttr} &	{\tiny		aerobic} &	{\tiny		2.867} &	{\tiny		2.871} &	 {\tiny		0.998} \\
{\tiny 57} &	{\tiny		malttr} &	{\tiny		anaerobic} &	{\tiny		0.890} &	{\tiny		 0.897} &	{\tiny		0.992} \\
{\tiny 58} &	{\tiny		maltttr} &	{\tiny		aerobic} &	{\tiny		3.822} &	{\tiny		3.828} &	 {\tiny		0.998} \\
{\tiny 59} &	{\tiny		maltttr} &	{\tiny		anaerobic} &	{\tiny		1.186} &	{\tiny		 1.195} &	{\tiny		0.992} \\
{\tiny 60} &	{\tiny		man} &	{\tiny		aerobic} &	{\tiny		0.955} &	{\tiny		0.957} &	 {\tiny		0.998} \\
{\tiny 61} &	{\tiny		man} &	{\tiny		anaerobic} &	{\tiny		0.297} &	{\tiny		0.299} &	 {\tiny		0.992} \\
{\tiny 62} &	{\tiny		melib} &	{\tiny		aerobic} &	{\tiny		1.900} &	{\tiny		1.903} &	 {\tiny		0.998} \\
{\tiny 63} &	{\tiny		melib} &	{\tiny		anaerobic} &	{\tiny		0.566} &	{\tiny		 0.571} &	{\tiny		0.992} \\
{\tiny 64} &	{\tiny		mnl} &	{\tiny		aerobic} &	{\tiny		1.020} &	{\tiny		1.025} &	 {\tiny		0.995} \\
{\tiny 65} &	{\tiny		pro-L} &	{\tiny		aerobic} &	{\tiny		0.754} &	{\tiny		0.762} &	 {\tiny		0.990} \\
{\tiny 66} &	{\tiny		pyr} &	{\tiny		aerobic} &	{\tiny		0.346} &	{\tiny		0.348} &	 {\tiny		0.992} \\
{\tiny 67} &	{\tiny		rib-D} &	{\tiny		aerobic} &	{\tiny		0.750} &	{\tiny		0.751} &	 {\tiny		0.998} \\
{\tiny 68} &	{\tiny		rib-D} &	{\tiny		anaerobic} &	{\tiny		0.139} &	{\tiny		 0.140} &	{\tiny		0.992} \\
{\tiny 69} &	{\tiny		rmn} &	{\tiny		aerobic} &	{\tiny		0.526} &	{\tiny		0.915} &	 {\tiny		0.575} \\
{\tiny 70} &	{\tiny		rmn} &	{\tiny		anaerobic} &	{\tiny		0.156} &	{\tiny		0.158} &	 {\tiny		0.993} \\
{\tiny 71} &	{\tiny		sbt-D} &	{\tiny		aerobic} &	{\tiny		1.020} &	{\tiny		1.025} &	 {\tiny		0.995} \\
{\tiny 72} &	{\tiny		sbt-D} &	{\tiny		anaerobic} &	{\tiny		0.256} &	{\tiny		 0.258} &	{\tiny		0.992} \\
{\tiny 73} &	{\tiny		ser-D} &	{\tiny		aerobic} &	{\tiny		0.346} &	{\tiny		0.348} &	 {\tiny		0.992} \\
{\tiny 74} &	{\tiny		ser-L} &	{\tiny		aerobic} &	{\tiny		0.354} &	{\tiny		0.356} &	 {\tiny		0.994} \\
{\tiny 75} &	{\tiny		succ} &	{\tiny		aerobic} &	{\tiny		0.469} &	{\tiny		0.469} &	 {\tiny		0.998} \\
{\tiny 76} &	{\tiny		tre} &	{\tiny		aerobic} &	{\tiny		1.911} &	{\tiny		1.914} &	 {\tiny		0.998} \\
{\tiny 77} &	{\tiny		tre} &	{\tiny		anaerobic} &	{\tiny		0.593} &	{\tiny		0.598} &	 {\tiny		0.992} \\
{\tiny 78} &	{\tiny		xtsn} &	{\tiny		aerobic} &	{\tiny		0.857} &	{\tiny		0.859} &	 {\tiny		0.998} \\
{\tiny 79} &	{\tiny		xtsn} &	{\tiny		anaerobic} &	{\tiny		0.346} &	{\tiny		0.348} &	 {\tiny		0.995} \\
{\tiny 80} &	{\tiny		xyl-D} &	{\tiny		aerobic} &	{\tiny		0.785} &	{\tiny		0.786} &	 {\tiny		0.998} \\
{\tiny 81} &	{\tiny		xyl-D} &	{\tiny		anaerobic} &	{\tiny		0.220} &	{\tiny		 0.222} &	{\tiny		0.992} \\

\end{longtable}
\end{center}
}

\section{Robustness of the network to gene knockouts}

\begin{figure*}
\centering
\includegraphics[width=14cm]{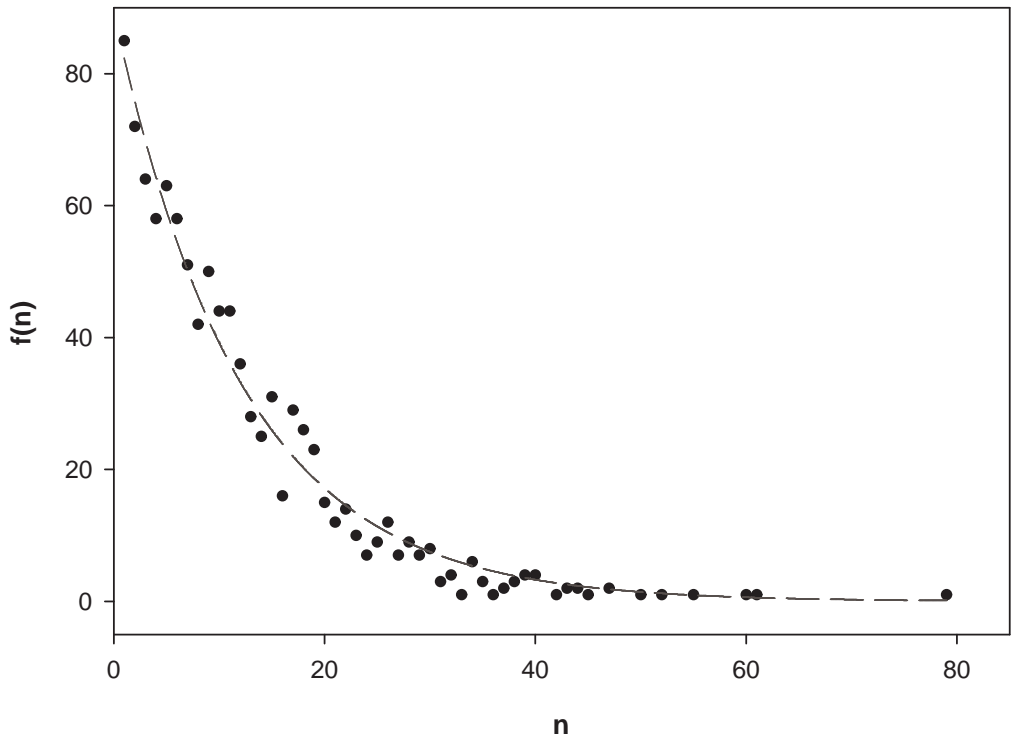}
\caption{
Frequency distribution of the number of random knockouts needed to make a cell unviable for growth for all 81 minimal media.
The dashed curve is the best fit to an exponential distribution.}
\label{robustness}
\end{figure*}

In this section, we investigate the robustness of the network functionality
to successive gene knockouts.
We considered the progressive decline of metabolic performance for an ensemble
of 1000 `random knockout trajectories'.
Each trajectory was constructed as follows: One out of 583 genes was chosen
at random and knocked out, i.e., its $g_i$ was set to be identically zero.
The constrained FBA growth rate was then determined for the attractors of the
resultant dynamical system of 582 genes for each of the 81 minimal media
listed in Table \ref{optimalitytable}.
If a nonzero growth rate was obtained for at least one of the 81 minimal media,
we continued and knocked out another gene at random from the remaining 582 genes.
This process of knocking out a randomly chosen gene from the network is repeated
until the attractors for all the 81 media in the truncated network become
dysfunctional (i.e., gave a zero growth rate).
The number of knockout steps, $n$, needed for the network to become metabolically
dysfunctional for all the 81 media was determined for
each of the 1000 random knockout trajectories considered.
Figure \ref{robustness} shows the number or frequency $f(n)$ of trajectories
with a given value of $n$.
The curve fits the exponential distribution $f(n) \sim \exp(-n/n_0)$
with $n_0 = 12.1$.
Thus, we find that the chance of survival of the organism decreases
exponentially with the number of gene knockouts.

\chapter{Design features of the genetic network controlling {\it E. coli} metabolism}
\label{design}

In chapter \ref{dynamics}, we have studied the regulatory network of
{\it E. coli} metabolism contained in the database iMC1010$^{v1}$
\cite{CKRHP2004} as a Boolean dynamical system.
We found that the regulatory network of {\it E. coli} metabolism
exhibits two biologically important system level properties:
homeostasis and flexibility of response to changed environments.
In this chapter, we explore the relationship between the structure
of the genetic network controlling {\it E. coli} metabolism
and the observed system level dynamical properties discussed
earlier.
Our study reveals that some very simple architectural features of
the genetic network controlling {\it E. coli} metabolism are
responsible for several of the observed system level dynamical
properties.


\section{The regulatory network of {\it E. coli} metabolism is
essentially an acyclic graph}

The directed graph of the regulatory network of {\it E. coli} metabolism as
contained in the database iMC1010$^{v1}$ was shown in Fig \ref{trn}.
In Fig. \ref{trn}, we can see that there is a large connected component and
few disconnected components.
The large connected component accounts for most (86\%) nodes in the regulatory
network.

\begin{sidewaysfigure}
\centering
\includegraphics[trim=20 80 20 20,width=20cm]{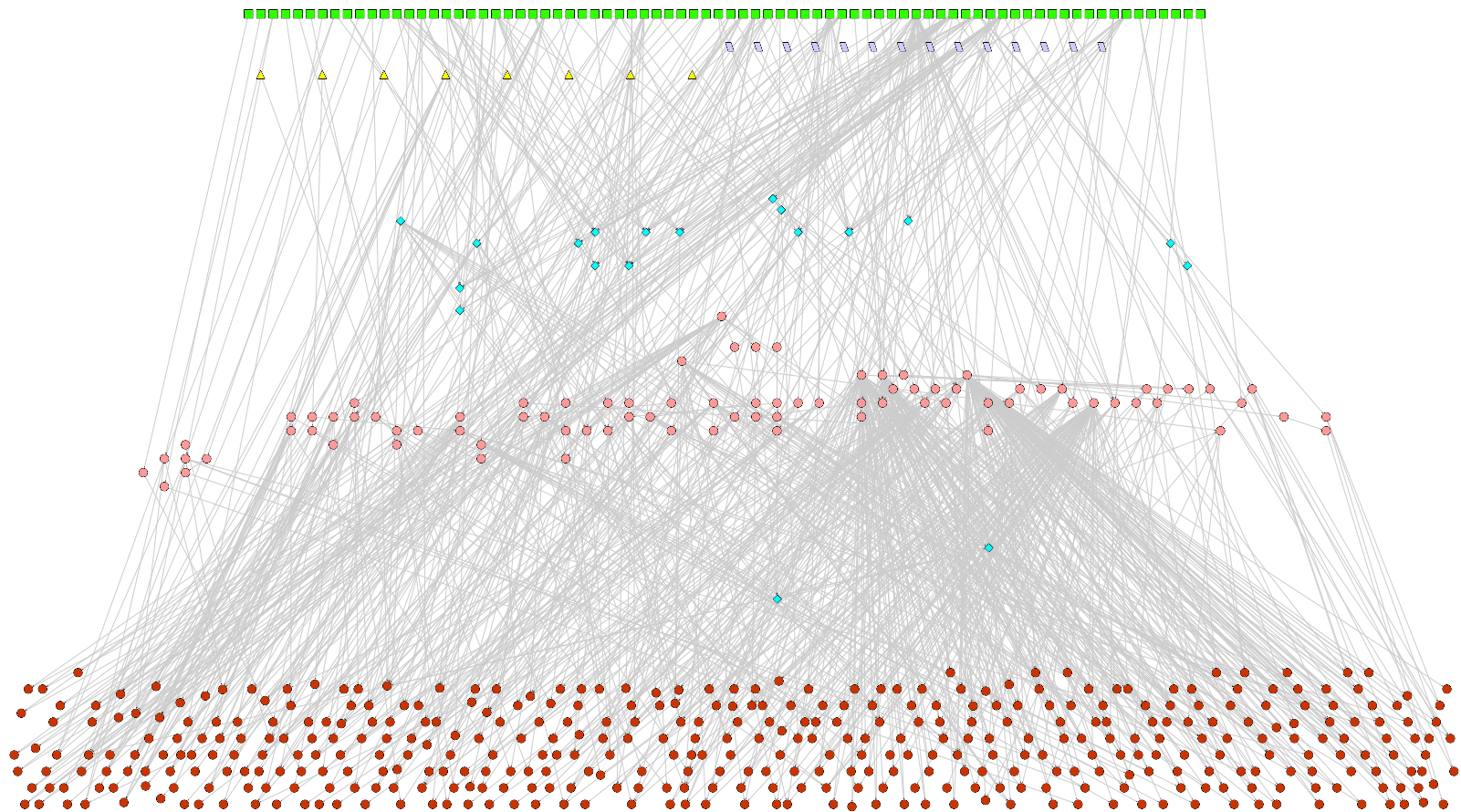}
\caption{
Largest connected component of the genetic network controlling {\it E. coli}
metabolism shown in Fig. \ref{trn}.
In this figure, there are genes coding for transcription factors (pink
circles), genes coding for enzymes (brown circles), external metabolites
(green squares), internal fluxes (purple parallelograms), stimuli (yellow
triangles) and other conditions (blue diamonds).
All links in this graph are pointing downwards.
The electronic version of this figure \cite{SJ2008} (available from {\scriptsize
http://www.biomedcentral.com/1752-0509/2/21}) can be zoomed in to see arrowheads.
This picture has been drawn using the graph visualization software Cytoscape
\cite{Cytoscape}.}
\label{tree}
\end{sidewaysfigure}

We found the regulatory network for dynamical system A to be an acyclic directed
graph with maximal depth 4.
The largest connected component of Fig. \ref{trn} is shown as a hierarchical
inverted tree in Fig. \ref{tree}, where all links are pointing downwards.
At the bottom of the hierarchy are the nodes that have no outgoing links in
the regulatory network.
We refer to the nodes at the bottom of the hierarchy as `leaves' of the acyclic
graph.
The leaves of the regulatory network are the 479 genes coding for enzymes.
The largest connected component of the regulatory network shown in Fig. \ref{tree}
has 409 genes coding for enzymes at the bottom of the hierarchy.
At the top of the hierarchy are the nodes that have no incoming links in the
regulatory network.
We refer to the nodes at the top of the hierarchy as `root nodes' of the acyclic
graph.
The depth of a node in the acyclic graph is the length of the longest path to it
from a root node.
Root nodes correspond to external metabolites and other variables that have fixed
values in the dynamical system A such as internal reaction fluxes, certain
conditions, etc.
In our simulations of the dynamical system given by Eq. \ref{dynamicalsystem} in
chapter \ref{dynamics}, we consider only buffered media that are characterized
by vector ${\bf m}$ constant in time.
Thus, the variables corresponding to external metabolites are held fixed for a
particular simulation, and by virtue of their root location, these variables act
as control variables of the dynamical system.
The genes coding for transcription factors are at intermediate levels in the
acyclic graph.

The above mentioned structural characteristics of the regulatory network for
dynamical system A explains why
\begin{itemize}
\item[(a)] there are only fixed point attractors of this dynamical system,
\item[(b)] their basin of attraction is the entire configuration space,
\item[(c)] it takes at most 4 time steps to reach the attractors from any
initial configuration, and
\item[(d)] the attractor configuration depends upon the medium.
\end{itemize}
The configuration of the external environment represented by vector ${\bf m}$
determines the configuration of the root level in the acyclic graph.
For any fixed environment, the configuration of the root nodes are fixed.
This then fixes the configurations of all nodes at the next level or depth 1
at the next time instant ($t=1$) and subsequent times irrespective of their
values at time $t=0$, because the input variables to the Boolean functions
controlling them are fixed.
This then fixes the configurations of all nodes at depth 2 at time $t=2$
irrespective of their configurations at time $t=1$, and so on, until
at time $t=4$, the configuration of the leaf nodes at maximum depth (depth 4)
are fixed irrespective of the configuration these nodes had held earlier.
Thus, in 4 time steps, all nodes in the graph have reached a fixed configuration
and the attractor is a fixed point.
The same kind of considerations apply when we start from any arbitrary initial
condition of genes as long as the configuration of the root nodes is held fixed.
This implies that for a fixed external environment, starting from any initial
configuration of the genes the system will reach a unique fixed point which is
a global attractor.
Thus, the basin of attraction is the entire configuration space of the genes.

A change in the medium or external environment is a change in the configuration
of root nodes at the top of the hierarchy.
The change of the configuration of root nodes percolates down the hierarchy in
a maximum of 4 time steps resulting in a new fixed point attractor.
Thus, the hierarchical and acyclic structure of the graph, along with external
metabolites as root nodes or control variables explains the insensitivity of the
attractors to initial condition or perturbation of the genes' configuration as
well as responsiveness of the attractors to changed environments.

The acyclicity of the transcriptional regulatory network of {\it E. coli} was noted
earlier by Shen-Orr {\it et al} \cite{SMMA2002}.
Shen-Orr {\it et al} had found that the known regulatory network of {\it E. coli}
was devoid of cycles of length $\ge$ 2.
There were only autoregulatory loops in the known regulatory network of {\it E. coli}
compiled largely from RegulonDB database \cite{RegulonDB}.
The autoregulatory loops are not included in the database iMC1010$^{v1}$ studied here.
Ma {\it et al} \cite{MBZ2004} also observed the hierarchical acyclic structure of
the {\it E. coli} regulatory network.
They found the maximum depth of the acyclic regulatory network to be 5.
The regulatory network studied by Ma {\it et al} \cite{MBZ2004} included parts of
the network that regulate systems other than metabolism.
Here, we have studied the database iMC1010$^{v1}$ that represents only the part of
the known {\it E. coli} regulatory network that controls metabolism.
Balaszi {\it et al} \cite{BBO2005} had earlier observed that the control of the graph
representing the {\it E. coli} regulatory network is in the hands of environmental
signals.
We remark that the acyclic structure of the {\it E. coli} regulatory network along
with root control by environmental variables was noted by previous studies mentioned
above.
However, the present work is the first one that brings these facts about the structure
of the {\it E. coli} regulatory network together to study its dynamics and elaborate upon
the consequences of the observed architecture for homeostasis and response flexibility
of the system.

Note that the graph of the regulatory network representing the dynamical system B is
not completely acyclic.
We allow the internal fluxes that affect the state of genes in the regulatory network
to be dynamical in system B.
Our treatment of internal fluxes in the dynamical system B effectively leads to some of
the genes coding for enzymes which are leaves in the dynamical system A to have outgoing
links that feedback to genes coding for transcription factors.
Thus, the feedback from internal fluxes results in a few cycles in the regulatory network
representing the dynamical system B.
However, the feedback due to internal fluxes affect only 5 genes coding for transcription
factors and 16 genes downstream of these coding for enzymes (21 genes in all).
Due to the small depth and the highly disconnected structure of the regulatory network at
the level of the genes coding for transcription factors in system A, the few cycles in the
network of system B turn out to be short and localized.
This results in the low period attractors and small `twinkling islands' in the dynamics
described in chapter 3 for system B.
The oscillating genes are drawn from the set of 21 genes mentioned above.
The rest of the 562 genes get fixed to the same configuration as in system A for any
medium.
Thus, in this study, we find that the genetic network regulating {\it E. coli} metabolism
is largely an acyclic graph where the root control lies with environmental variables.


\section{Disconnected structure of the `reduced dynamical system':
modularity, flexibility and evolvability}

\begin{figure*}
\centering
\includegraphics[trim=80 0 50 10,width=16cm]{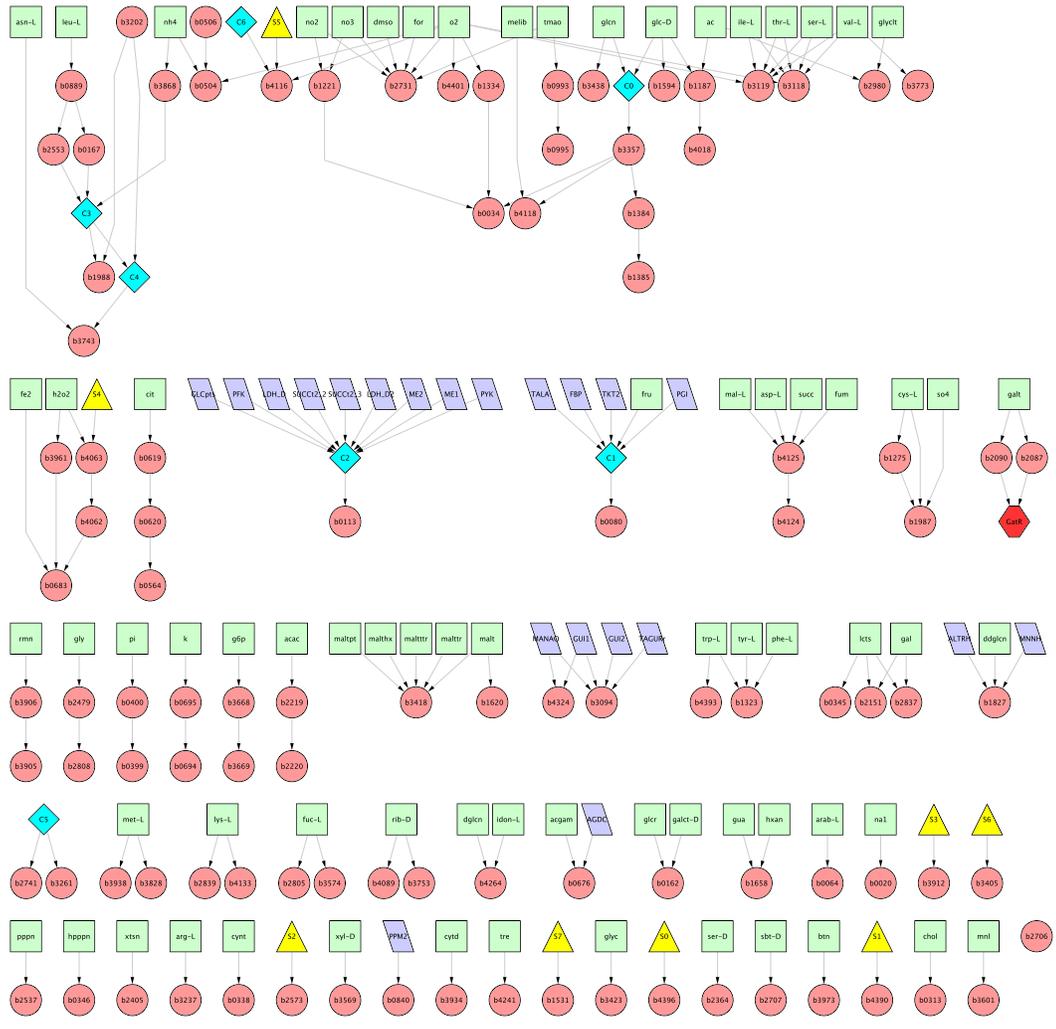}
\caption{
Picture of the regulatory network obtained when all leaf nodes in the
network of Fig. \ref{trn} are removed along with all their links.
In this figure, there are genes coding for transcription factors
(pink circles), genes coding for enzymes (brown circles), external
metabolites (green squares), internal fluxes (purple parallelograms),
stimuli (yellow triangles) and other conditions (blue diamonds).
The red hexagon denotes the lone transcription factor in the network that
is coded for by two genes.
The nomenclature for conditions C0 to C6 and stimuli S0 to S7 is given in
Table \ref{nodelabel}.
The electronic version of this figure \cite{SJ2008} (available from {\scriptsize
http://www.biomedcentral.com/1752-0509/2/21}) can be zoomed in to read node names.
This picture has been drawn using the graph visualization software
Cytoscape \cite{Cytoscape}.}
\label{tf}
\end{figure*}

The leaf nodes corresponding to the genes coding for enzymes in the acyclic graph
representing the regulatory network of {\it E. coli} metabolism have no outgoing
links and their states do not determine the state of other nodes in the network.
So, it is worthwhile to investigate the dynamics of the `reduced dynamical system'
obtained from the full system of Fig. \ref{trn} by removing the leaves corresponding
to the enzyme coding genes at the bottom of the hierarchy.
When leaf nodes in the system are removed along with all their links, we are left
with Fig. \ref{tf} representing the reduced dynamical system.
The reduced dynamical system shown in Fig. \ref{tf} is a disconnected graph with
unexpectedly large number of disconnected components.
The largest connected component of the full network shown in Fig. \ref{tree} has
broken up into 38 disconnected components in Fig. \ref{tf}.
There are several small components containing upto only 4 nodes at depth $\geq 1$
and one component with 27 genes coding for transcription factors at depth $\geq 1$
in the reduced dynamical system.
The latter component is regulated by oxygen, some inorganic sources of nitrogen,
certain amino acids and sugars.
The smaller components in the reduced dynamical system are typically regulated by
single metabolites or groups of biochemically related metabolites.
This procedure reduces the number of outgoing links for global regulators drastically.
For example, the gene b3357 coding for Crp is left with only 3 outgoing links in the
reduced system of Fig. \ref{tf} instead of 105 in the full system of Fig. \ref{trn}.

Two components of a dynamical system that are disconnected from each other are
dynamically independent as two nodes belonging to different disconnected components
do not affect the states of each other.
Thus, the dynamics of each disconnected component can be analyzed independent of the
other disconnected components.
The dynamics of the `reduced dynamical system' shown in Fig. \ref{tf}, in particular
its attractors and basins of attraction, can be reconstructed from those of its
disconnected components.
Such a disconnected or `product' structure of a dynamical system greatly simplifies
its mathematical analysis.
Modularity of biological systems refers to the existence of subsystems that are
relatively independent of each other.
We may regard each disconnected component in Fig. \ref{tf} as defining a core of
a module.
The modularity of the present genetic regulatory network is then nothing but the
property that it is composed of disconnected components at this level of
description.

Starting from a disconnected component of Fig. \ref{tf} (or a module core) the
genes downstream from it which code for enzymes (and do not feedback into the
upper layers) may be regarded as constituting the rest of the module.
Thus, a single genetic module is a disconnected component of Fig. \ref{tf}
together with all its downstream leaves.
Restoring the leaves and their links in Fig. \ref{tf} will take us back to Fig.
\ref{trn} which contains the large connected component shown in Fig. \ref{tree}.
As the full system shown in  Fig. \ref{trn} has a large connected component and
a few disconnected components as opposed to many disconnected components in the
reduced system shown in Fig. \ref{tf}, this means that leaf nodes typically receive
links from more than one module core of the reduced dynamical system.
The structure of the acyclic graph of the regulatory network resembles that of a
banyan tree which has multiple trunks emanating from independent roots and in which
leaves receive sustenance from more than one root.
In the present architecture, there is no direct crosstalk between the module cores
of the reduced dynamical system at the level of transcription factor coding genes
but the module cores can affect common leaf nodes corresponding to enzyme coding
genes.
This enables many leaf nodes to be influenced by several root nodes corresponding
to environmental variables.
This `multitasking' adds to the complexity of cellular response to different
environments and possibly contributes to greater metabolic efficiency.
When the environment is changed from one minimal medium to another, it corresponds
to replacement of one carbon source by another which may belong to a different module
core of the reduced dynamical system.
The genetic network needs to respond to the change of medium by activating genes
coding for enzymes that catalyze metabolic reactions needed to break down the new
carbon source and process its moieties.
The connections of the leaf nodes to the modules above them must be such that the
genes coding for enzymes catalyzing the input metabolic pathway of the new carbon
source get activated, given our finding that the constrained FBA growth rate increases
as the new attractor is reached (see section \ref{adaptability} in chapter
\ref{dynamics}).

The location and dynamical autonomy of the modules in the reduced dynamical system
could also contribute to evolvability.
If a new module is added to the reduced dynamical system of Fig. \ref{tf}, it would
not affect the dynamics of the existing modules.
Thus, the organism can explore new environmental niches characterized by new food
sources without jeopardizing existing capabilities.
This may be a particular case of the more general observation that the architectural
features of organisms responsible for their flexibility to environmental conditions
also contribute to their evolvability \cite{AF2000,Kirschnerbook}.


\begin{table}
\centering
\begin{tabular}{c|l}

{\small \bf Abbreviation} &	{\small \bf Name} \\
\hline

{\small C0} & {\small CRP noGLC}\\

{\small C1} & {\small Surplus FDP}\\

{\small C2} & {\small Surplus PYR}\\

{\small C3} & {\small NRI\_low}\\

{\small C4} & {\small NRI\_hi}\\

{\small C5} & {\small Growth}\\

{\small C6} & {\small pH}\\

{\small S0} & {\small Dipyridyl}\\

{\small S1} & {\small High NAD}\\

{\small S2} & {\small Heat shock}\\

{\small S3}	& {\small Stress} \\

{\small S4} & {\small Oxidative stress}\\

{\small S5} & {\small LBMedia}\\

{\small S6} & {\small High osmolarity}\\

{\small S7} & {\small Salicylate}\\
\hline

\end{tabular}
\caption{Abbreviations used to label nodes corresponding to conditions and
stimuli in Fig. \ref{tf} and their corresponding names.}
\label{nodelabel}
\end{table}

\section{Almost all input functions are canalyzing in the {\it E. coli}
regulatory network}

Kauffman and colleagues studied extensively the dynamics of random Boolean
networks in an attempt to gain understanding of the system level behaviour
of real genetic regulatory networks \cite{Kauffmanbook}.
They constructed random Boolean networks with $N$ binary nodes and $K$
inputs per node.
The Boolean function that decides the state of each node at time $t$ based
on the state of the input nodes at time $t-1$ was chosen at random from the
set of possible Boolean functions of $K$ inputs in a random Boolean network.
By studying random Boolean networks, Kauffman and others have found that the
system can be in two different dynamical regimes: ordered and chaotic.
Kauffman found that random Boolean networks with $K=1$ and large $N$ are deep
in the ordered regime for random choice of Boolean functions.
For $K>2$, the random Boolean networks for random choice of Boolean functions
are in the chaotic regime.
Further, it was found that random Boolean networks with $K=2$ and large $N$
are exactly in the phase transition between the ordered and chaotic regimes
for random choice of Boolean functions.
It was shown that biasing the choice of Boolean functions in the generation of
random Boolean networks can drive networks with $K>2$ into the ordered regime
\cite{Kauffmanbook}.
It has been shown by Kauffman and his colleagues that stability to perturbations
in the random Boolean networks with $K>2$ can arise due to the canalyzing
property of input functions \cite{Kauffmanbook,HSWK2002,KPST2003,KPST2004}.
A canalyzing Boolean function has at least one input such that at least one of
the two values of this input determines the output of the function
\cite{Kauffmanbook}.
For example, the AND, OR and NOT functions satisfy the canalyzing property while
the XOR function does not satisfy the canalyzing property.
For a given number of inputs, $K$, the fraction of Boolean functions that are
canalyzing decreases as $K$ increases.

On the basis of above mentioned findings, Kauffman had suggested that the
distribution of Boolean functions in real genetic regulatory networks
should be biased towards the set of canalyzing functions.
In reference \cite{HSWK2002}, Harris {\it et al} compiled Boolean rules for
eukaryotes from the available literature and found all rules in their compilation
to be canalyzing.
For the present regulatory network of {\it E. coli} metabolism, the frequency
distribution of the number of genes with $K$ regulatory inputs is given in
Table \ref{canalyzing}.
We found that Boolean functions for $579$ of the $583$ genes in the regulatory
network of {\it E. coli} metabolism to possess the canalyzing property.
Only $4$ genes had input functions that were not canalyzing.
The above results establish the preponderance of canalyzing Boolean functions in the
regulatory network of {\it E. coli} metabolism.


\begin{table}
\centering
\begin{tabular}{c|c}

{\small \bf Number of regulatory inputs} &	{\small \bf Number of Genes} \\
{\small \bf $K$} &	\\
\hline

{\small 1} & {\small 259}\\

{\small 2} & {\small 189}\\

{\small 3} & {\small 68}\\

{\small 4} & {\small 39}\\

{\small 5} & {\small 10}\\

{\small 6} & {\small 4}\\

{\small 8} & {\small 2}\\
\hline

\end{tabular}
\caption{The table shows the number of genes in the genetic network controlling
{\it E. coli} metabolism with $K$ regulatory inputs.}
\label{canalyzing}
\end{table}

\section{The dynamical system achieves flexibility even though it is far from
the `edge of chaos'}

As mentioned earlier, by studying extensively the dynamics of random Boolean networks,
Kauffman had found that these systems can exist in two broad regimes: ordered and chaotic.
In the ordered regime, the attractors of the dynamical system have a large `frozen core'
of genes locked in fixed configuration along with a few `twinkling islands' of genes
that may switch states between the two allowed values.
In the chaotic regime, the number of genes in the frozen core is much less than the number
of genes in the twinkling islands.
One might expect that a dynamical system whose attractors have large frozen cores and
very small twinkling islands is rather rigid and therefore unlikely to be adaptable to
the external environment and also unlikely to be evolvable.
This expectation led Kauffman to the conjecture that genetic regulatory systems ought to be
close to the `edge of chaos', the boundary that separates the ordered regime from the
chaotic regime in the space of dynamical systems (see \cite{Kauffmanbook}).

However, in the previous chapter, we have shown the regulatory network of {\it E. coli}
metabolism to be deep in the ordered regime,
since it always falls into the same attractor that is a fixed point or has isolated low
period cycles for all initial conditions of genes in a few time steps.
Thus, all or most genes in the the regulatory network of {\it E. coli} metabolism get
frozen for a fixed environment.
In other words, the dynamical system studied by us is far from the edge of chaos.
We have seen above that this property of homeostasis is an inevitable consequence of the
hierarchical, largely acyclic architecture of the regulatory network.
Simultaneously, we have shown in the last chapter that the system is also highly
responsive to changes in the environment.
How have these two properties managed to co-exist?
The answer lies in the observation that root nodes of the hierarchical, acyclic graph of
the regulatory network are primarily the environmental variables (i.e., the external
metabolites in the present network).
The attractor configuration is thus a function of the external environment, specified
by the vector ${\bf m}$ representing the state of the 96 external metabolites in the
medium.
For any fixed vector ${\bf m}$ corresponding to a buffered medium, there is a global attractor
in which most or all genes in the network have frozen configurations.
However, when vector ${\bf m}$ changes the genes `unfreeze' and move to a new attractor
configuration.
The modular organization of the network with a lot of crosstalk between modules at the leaf
level (enzyme coding genes) ensures that the melting and refreezing is quite substantial.
Furthermore, the same architecture that produces this flexibility of response to the
external environment can also enhance evolvability.

The present architecture as an alternative to the edge of chaos hypothesis for
simultaneously producing homeostasis and flexibility has not been noticed earlier
because the earlier literature has primarily focussed on the abstract genetic
network itself without much reference to the environmental control variables
that abound in the real systems.
We have observed this architecture in the present work because the dynamical system
studied here systematically accounts for the effect of external metabolites on the
state of the genetic network controlling {\it E. coli} metabolism.
This possibility has become evident because the database iMC1010$^{v1}$
brings together, within the same network, genes as well as nodes describing
external environmental signals, on a large scale.



\chapter{Discussion and Future Outlook}
\label{discussion}


In this thesis, we have studied the large scale structure and system level dynamics
of certain biological networks.
Further, we have tried to explain the observed system level dynamical properties of
the biological networks studied in terms of their underlying structure.
In this chapter, we now summarize some of the results reported in this thesis and
discuss possible implications as well as limitations associated with our work.
We also mention some future directions of research and speculations based on work
reported here.

\section{Metabolic networks: low degree metabolites, essential reactions, functional
modules}

\subsection{Summary of the results, discussion and speculations}

In chapter 2, we have introduced the notion of `uniquely produced' (`UP') and `uniquely
consumed' (`UC') metabolites in the metabolic network.
We have designated a metabolite that is both UP and UC as a `UP-UC' metabolite in the
network.
A UP-UC metabolite has only one reaction producing it and one reaction consuming it in
the network.
We have designated a set of reactions connected by UP-UC metabolites as a `UP-UC cluster'
in the metabolic network.
Under any steady state, the flux of the reaction producing a UP-UC metabolite is
proportional to the flux of the reaction consuming it.
Hence, in any steady state, the fluxes of all reactions that are part of a single UP-UC
cluster in the metabolic network are always proportional to each other.
We have determined the list of UP(UC) metabolites and UP-UC clusters of
reactions in the metabolic networks of {\it E. coli}, {\it S. cerevisiae} and
{\it S. aureus}.
Further, we have used the computational technique of flux balance analysis (FBA) to
determine essential reactions for growth in the metabolic networks of {\it E. coli},
{\it S. cerevisiae} and {\it S. aureus}.

In chapter 2, we have shown that UP-UC metabolites in the metabolic network lead to
correlated UP-UC clusters of reactions in the network.
We found that large UP-UC clusters are over-represented in the real metabolic network
as opposed to randomized networks with same local connectivity as the real network.
We then showed that UP-UC clusters at the metabolic level correspond, with a high
probability, to sets of genes forming expression modules at the regulatory level
in {\it E. coli}.
We have used here only the operon information in determining the correspondence between
UP-UC clusters in the metabolic network and sets of coexpressed genes in regulatory
network.
It is possible for genes that do not belong to the same operon to be coexpressed inside
the cell.
Thus, an extended analysis of {\it E. coli} gene expression data may find a better
correspondence between UP-UC clusters at the metabolic level and sets of coexpressed
genes at the regulatory level.
Further, the analysis of gene expression data to check coregulation of genes corresponding
to UP-UC clusters of reactions should be extended to other organisms where operon information
is not available.

Metabolic networks inside different organisms have been shown to follow a
power law degree distribution \cite{JTAOB2000,WF2001}.
It has been suggested that the power law degree distribution of real networks
contributes towards its robustness \cite{AJB2000}.
In particular, it has been emphasized that networks with power law degree
distribution are vulnerable to selective attack on its high degree nodes or
hubs, while removal of many low degree nodes from these networks have
negligible effect on the functionality of the system \cite{AJB2000}.
For example, in the case of the internet, the removal of high degree nodes
corresponding to routers with many connections can turn out to be fatal for
the communication system.
Further, for the {\it S. cerevisiae} protein-protein interaction network,
the essentiality of a protein was also found to be correlated with the degree
of the protein in the network \cite{JMBO2001}.
However, in case of the metabolic network, the metabolites participate in
reaction processes where they are produced or consumed.
We can only control the reaction process through its catalyzing enzyme that
is a product of some gene.
It is unclear how a biological process can lead to removal of metabolites from
the network.
A removal of high degree metabolite from the metabolic network would require
elimination of all reactions in which the metabolite participates or knockout
of all genes associated with reactions in which the metabolite is involved.
Further, genetic mutations effectively correspond to removal of biochemical
reactions from the metabolic network.
Thus, in case of the metabolic network, one is interested in determining the
effect of removing a reaction rather than the effect of removing a metabolite.
In chapter 2, we have shown that most globally essential reactions in the
metabolic networks of {\it E. coli}, {\it S. cerevisiae} and {\it S. aureus}
either produce or consume a low degree UP or UC metabolite.
The essential reactions may involve other metabolites of higher degree, but
their essentiality is due to their uniqueness in producing or consuming an
intermediate low degree UP or UC metabolite that is needed for the eventual
production of biomass.
Thus, we have shown here that, in a consideration of robustness or fragility
of metabolic networks to naturally occurring perturbations, it is the role
of low degree metabolites that needs to be considered rather than high degree
metabolites.
This is an example of how the study of dynamical properties (in this case,
flows) and functionality at the system level can alter one's
perspective on what is significant for robustness or fragility of real world
complex systems.
We remark that the importance of low degree nodes in determining the robustness
or fragility of complex networks has also been observed elsewhere
\cite{JK2002a,JK2002b}.
There the crashes of an evolving catalytic network were found to be due to
`core-shifts' caused by changes in low degree nodes of the network.

The existence of essential reactions in the metabolic network is an indicator
of the fragility in the system.
Even though the metabolic networks we have studied have many reaction nodes,
a small perturbation such as the removal of a single essential reaction node
from the network, destroys the functionality of the complete metabolic network.
One way of dealing with this fragility is by introducing redundancy at the
level of genetic network by associating genes coding for isozymes with
essential reactions.
Isozymes are multiple enzymes catalyzing a single reaction in the metabolic
network.
We have to knockout multiple genes at the same time in order to remove an
essential reaction with associated isozymes from the metabolic network.
Hence, we may expect essential reactions in the metabolic network to have a
greater probability of being associated with isozymes than non-essential
reactions.
However, in an earlier paper \cite{PPH2004}, Papp {\it et al} showed that both
essential and non-essential reactions in the metabolic network of
{\it S. cerevisiae} are equally likely to be associated with isozymes.
In that paper, isozymes were found to be associated with greater probability
with reactions carrying high flux rather than with essential reactions in the
metabolic network \cite{PPH2004}.
This finding suggests that organisms in the course of evolution have developed
redundancies in the genetic network predominantly to associate isozymes with
metabolic reactions carrying high flux rather than with essential reactions in
the network.

This raises the question: why has evolution tolerated the fragility associated
with essential reactions?
The finding that essential reactions can be tagged by UP or UC metabolites may
provide some insight into this.
UP or UC metabolites that participate in very few reactions perhaps do so in
part because some feature of their chemical structure prohibits ready association
with other molecules in nature.
If so, the low degree of UP or UC metabolites is a consequence of constraints
coming from chemistry.
Then evolution seems to tolerate essential reactions that produce or consume such
metabolites because chemistry has left it with no other choice.

Alternatively, the fragility associated with low degree metabolites may be a
byproduct of some other desirable property that contributes to evolvability or
robustness of the system, such as modularity.
In chapter 2, we have shown that UP-UC metabolites which have low degree in the
network lead to correlated clusters of reactions in the metabolic network.
This raises the question: if UP-UC metabolites contribute to modularity, could it
be that the evolutionary advantages of that have outweighed the disadvantage of
the above mentioned fragility caused by the same low degree metabolites?
Is it the case that evolution has preferred `chemically constrained' low degree
metabolites in spite of the fragility they cause because they contribute to
modularity?

A goal in biology is to understand highly evolved biological organization in terms
of simpler and more inevitable structures \cite{M1999}.
In chapter 2, we have presented evidence that certain regulatory modules, in
particular certain operons, mirror the UP-UC structure of the metabolites whose
production and consumption they regulate in {\it E. coli}.
This could be an example of how the origin of certain regulatory structure can be
traced to simple chemical constraints.
In future, we hope to shed light on this by investigating metabolic and regulatory
networks inside many other organisms.

The fragility caused by low degree metabolites in metabolic networks can have potential
applications in medicine.
We have observed that essential reactions are explained by UP/UC structure in three
organisms.
Thus, it is likely that UP/UC structure explains essential reactions in other organisms
that are pathogens for humans.
This generates candidate targets for therapeutic intervention.
It is conceivable that drugs could be found that incapacitate the enzymes of one or
the other essential reactions in pathogens.

\subsection{Caveats regarding metabolic networks}

We now discuss some limitations and caveats associated with our results reported in chapter 2.
The results have been obtained by using the metabolic network databases iJR904 \cite{RVSP2003},
iND750 \cite{DHP2004} and iSB619 \cite{BP2005} for {\it E. coli}, {\it S. cerevisiae} and
{\it S. aureus}, respectively.
These databases were reconstructed using the annotated genome sequences for the three organisms
and the available biochemical information in the literature.
Since the annotation of the fully sequenced genomes of {\it E. coli}, {\it S. cerevisiae} and
{\it S. aureus} is not yet complete, the list of reactions contained in the three metabolic
network databases iJR904, iND750 and iSB619 is still incomplete.

We may expect additions to the list of biochemical reactions in the reconstructed metabolic
network databases in future based on more complete annotation information.
The addition of reactions to the reconstructed metabolic network databases can introduce
alternate pathways for certain crucial metabolites required for the eventual production of
biomass metabolites in the databases for the three organisms studied here.
Thus, the introduction of alternate pathways may affect the set of essential metabolic reactions
obtained using FBA for the three organisms.
Similarly, the inclusion of additional reactions in the metabolic network databases would render
some of the presently UP(UC) metabolites non-UP(non-UC).

In this context, it is worth noting that for the metabolic network databases as they stand, the
present definition of UP(UC) metabolites allows us to establish a connection between distinct
properties of the metabolic network.
In chapter 2, we have shown that UP-UC metabolites lead to UP-UC clusters which are correlated
sets of reactions in the metabolic network.
The genes corresponding to UP-UC clusters were shown to predict regulatory modules in {\it E. coli}.
We have also shown that most essential metabolic reactions are explained by their association with
a UP or UC metabolite.
Also, most of our results have been shown to hold for metabolic networks inside three distinct
organisms.
This suggests that our definition of UP(UC) metabolites and reactions does capture a certain useful
pattern in the metabolic network.
We remark that a reduction in the number of UP(UC) metabolites and essential reactions would only
improve the correspondence of UP-UC clusters and regulatory modules described in section \ref{operon}
and between the theoretically predicted and experimentally observed essential genes discussed in
section \ref{essentialgenes}.

\section{The genetic regulatory network controlling metabolism in {\it E. coli}}

\subsection{Summary of results}

In chapter 3, we have studied the genetic network controlling {\it E. coli} metabolism as represented in the
database iMC1010$^{v1}$ \cite{CKRHP2004} as a Boolean dynamical system.
We have constructed an effective Boolean dynamical system of genes and external metabolites using the information
contained in the database iMC1010$^{v1}$.
We have studied the dependence of the attractors of the Boolean dynamical system representing the genetic network
controlling {\it E. coli} metabolism to changes in initial conditions of genes and configuration of the external
environment.
Robustness is an inherent property of living systems which enables them to maintain their functionalities in the
face of diverse perturbations \cite{SSSDD2004,K2004,Wagnerbook}.
Biological systems encounter externally induced perturbations in the form of changes in the environment
and internal perturbations such as fluctuations in gene configurations.
In chapter 3, we have shown that the genetic network controlling {\it E. coli} metabolism for a fixed external
environment has essentially a unique attractor for the configuration of the genes regardless of the initial
conditions or perturbations of gene configurations.
So, we found the attractors of the dynamical system to be fixed points or low period cycles for any given fixed
environmental condition and the system exhibits the property of homeostasis.
However, robustness is a more general property than homeostasis which concerns itself with maintaining system
functionality rather than system state.
In chapter 3, we have then shown that when the dynamical system encounters a changed external environment, the
system again flows and stabilizes to another unique attractor state regardless of the initial configuration of
genes; however, the attractor for the changed environment may be widely separated from the attractor for the
previous environment.
We found the genetic network controlling {\it E. coli} metabolism as a dynamical system to be flexible to different
environmental conditions as their attractors show a wide variation.
In chapter 3, using the technique of FBA, we have further shown that the attractors of the genetic network for most
environmental conditions allow optimal functioning of the underlying metabolic network.
Thus, we have shown that the robustness of the genetic network controlling {\it E. coli} metabolism manifests
itself in two different ways.
When the system encounters internal perturbations in form of changes in genes' configuration for a fixed external
environment, the system returns to the present attractor.
When the system encounters external perturbations in form of changed environment, the system moves to a new attractor
that maintains the systems functionality in the sense of maintaining growth rate of the organism.
Further, we have shown that the robust behaviour of the genetic network in moving to a different stable attractor
state in response to a changed external environment also ensures metabolic efficiency for the organism.

In chapter 4, we have studied in detail the design features of the genetic network controlling {\it E. coli}
metabolism in order to understand the origin of the observed system level dynamical properties of homeostasis
and response flexibility.
We have found the genetic network controlling {\it E. coli} metabolism as represented in the database iMC1010$^{v1}$
to be hierarchical and an largely acyclic graph.
Cycles where they did exist were found to be of small length and localized.
The nodes with no incoming links in the genetic network are referred to as the root nodes of the hierarchy which
correspond to different external metabolites.
The nodes with no outgoing links in the genetic network are referred to as the leaves of the hierarchy which
correspond to genes coding for metabolic enzymes.
The nodes at the intermediate level of the hierarchical acyclic graph have both incoming and outgoing links in the
genetic network and correspond to genes coding for transcription factors.
The observed hierarchical architecture of the genetic network controlling {\it E. coli} metabolism with root control
in the hands of external metabolites endows the system with the twin dynamical properties of homeostasis and response
flexibility.
Thus, the robust behaviour of the dynamical system has its origin in the underlying architecture or topology of the
genetic network.

As mentioned above, the genetic network controlling {\it E.coli} metabolism as represented in the database iMC1010$^{v1}$
lacks feedback from genes to other genes via transcription factors.
The models studied originally by Kauffman \cite{K1969a,Kauffmanbook} were random Boolean networks.
Those networks had substantial feedbacks between genes and hence had more complicated attractors and dynamics.
One of our main results is that the genetic network controlling {\it E.coli} metabolism as a real biological system is
structured (and hence departs from random networks) in such a way that it has simple attractors and dynamics.
Thus, while at the abstract level the Boolean dynamical system studied by us is not very different from Kauffman's
(apart from the inclusion of the external environment), our dynamical results are quite different because the underlying
network has a very different structure from the one considered by Kauffman.

In chapter 4, we have shown that removing the leaf nodes at the bottom of the hierarchy corresponding to the
genes coding for metabolic enzymes along with their links from the full graph shown in Fig. \ref{trn} leads to
a subgraph shown in Fig. \ref{tf} with many disconnected components.
The various disconnected components shown in Fig. \ref{tf} are dynamically independent of each other and may be
regarded as modules of the genetic network.
The different disconnected components shown in Fig. \ref{tf} are regulated by different external metabolites.
Thus, for any given environmental condition, only a subset of the disconnected components shown in Fig. \ref{tf}
get switched on.
The highly disconnected structure at the level of genes coding for transcription factors compared to the highly
connected structure of the full network implies that the leaf nodes corresponding to the genes coding for
metabolic enzymes get incoming links from many different disconnected components shown in Fig. \ref{tf}.
Thus, although there is a lack of crosstalk between disconnected components shown in Fig. \ref{tf}, there is a
lot of crosstalk between the modules at the leaf level, that of genes coding for metabolic enzymes in the
genetic network.

In chapter 4, we have argued that the above mentioned disconnected structure at the level of genes coding for
transcription factors can contribute towards the evolvability of the genetic network.
Since the different disconnected components are regulated by different subsets of external metabolites in
the environment, if a population of such cells encounters an environment where a particular set of external
metabolites is consistently absent, then the species can discard the disconnected components regulated by
these external metabolites from the genetic network without a loss of functionality with respect to other
external metabolites.
Conversely, if the population encounters a new set of external metabolites in a sustained manner, it can develop a
new module with pathways controlled by this new set without affecting the existing functionality, thereby adapting
itself to the new niche.

\subsection{Caveats and speculations regarding regulatory networks}

The results reported in chapters 3 and 4 have been obtained using the integrated regulatory and metabolic network
database iMC1010$^{v1}$ \cite{CKRHP2004} for {\it E. coli} and there are limitations that stem from the
reconstructed database itself.
The database iMC1010$^{v1}$ for {\it E. coli} has been reconstructed using various literature sources.
The database iMC1010$^{v1}$ describes the regulation of 583 genes in {\it E. coli}.
Of the 583 regulated genes in the database iMC1010$^{v1}$, 479 genes code for metabolic enzymes in {\it E. coli}.
These 479 enzyme coding genes are a subset of the 904 enzyme coding genes associated with various reactions in the
{\it E. coli} metabolic network database iJR904.
Thus, the database iMC1010$^{v1}$ does not describe the regulation of a large fraction of genes coding for metabolic
enzymes in {\it E. coli}.
In the database iMC1010$^{v1}$, the set of 583 genes are regulated by a set of 96 external metabolites.
In the {\it E. coli} metabolic network iJR904, there are 143 external metabolites which
can be transported across the cell boundary.
The set of 96 external metabolites accounted in the database iMC1010$^{v1}$ are a subset of the 143 external
metabolites contained in the metabolic network database iJR904.
Thus, the database iMC1010$^{v1}$ does not completely account for the effect of all possible external
metabolites that the {\it E. coli} can uptake or excrete on the state of genes in the regulatory network.
Further, the present incomplete genetic network database iMC1010$^{v1}$ of 583 genes and 96 external metabolites
could have many connections as false positives or false negatives, especially the latter.
It is also possible that some of the Boolean rules in the database iMC1010$^{v1}$ may be incorrect or incomplete
due to lack of detailed experimental data for different possible environmental conditions.
Hence, we expect modifications and expansion of the present database iMC1010$^{v1}$ describing the genetic network
controlling {\it E. coli} metabolism in the near future.
This could modify our results on the dynamics of the genetic network controlling {\it E. coli} metabolism reported
in this thesis.

We next ask the question: How would the future introduction of additional gene nodes and connections into the
existing network database iMC1010$^{v1}$ affect the results reported in this thesis?
If new nodes and connections corresponding to genes coding for metabolic enzymes are added to the present database
iMC1010$^{v1}$, it is unlikely to affect our qualitative conclusions about the nature of attractors of the genetic
network controlling {\it E. coli} metabolism significantly.
The reason is that most of the genes coding for metabolic enzymes are likely to be leaves of the genetic network like
the nodes at the bottom of Fig. \ref{tree}, in which case they would have no outgoing links and would not affect the
dynamics of other gene nodes in the network.
However, the inclusion of genes coding for metabolic enzymes as well as additional connections to existing genes in
the genetic network would add to the constraints on FBA.
It would be then interesting to see the extent to which regulatory dynamics enhances metabolic efficiency in different
environmental conditions for an enlarged genetic network accounting for the regulation of most genes coding for
enzymes in the {\it E. coli} metabolic network.

If new nodes and connections corresponding to genes coding for transcription factors are added to the present database
iMC1010$^{v1}$, it can affect our qualitative conclusions about the dynamics and the nature of the attractors
of the genetic network controlling {\it E. coli} metabolism.
In particular, the introduction of feedback loops or cycles between genes coding for transcription factors could lead to
longer cycles as attractors of the dynamical system.
This is a question to be ultimately settled by more detailed knowledge of the empirical facts.
Several regulatory systems with feedbacks are known, e.g., the Yeast cell-cycle network and the {\it Drosophila} segment
polarity network \cite{AO2003,LLLOT2004}.
Several genes in {\it E. coli} are known to have autoregulatory self-loops \cite{RegulonDB} that are not included in
the present database iMC1010$^{v1}$.
The introduction of autoregulatory loops in the present network could produce two-cycles at the individual nodes even
at constant input.
Earlier works have observed that apart from self-loops, the transcriptional regulatory network of {\it E. coli} is
largely acyclic \cite{SMMA2002,MBZ2004,BBO2005} and has a small depth of about 5.
We remark that the lack of feedback from genes to other genes via transcription factors is not an assumption on our part,
rather it reflects the way this biological system actually is as captured in the present database iMC1010$^{v1}$ and also
in other studies \cite{SMMA2002,MBZ2004,BBO2005}.

In addition to the feedback from genes to other genes via transcription factors, discussed above, there can be another
kind of feedback in the genetic network controlling {\it E.coli} metabolism.
This is the feedback of the concentrations of internal metabolites in the metabolic network into the genetic network.
The feedback of internal metabolite concentrations could affect both the genes coding for enzymes and genes coding
for transcription factors.
In the database iMC1010$^{v1}$, the effect of internal metabolite concentrations on the state of the genes has been
approximated via fluxes of internal reactions in the metabolic network.
In the present database iMC1010$^{v1}$, there are 21 internal fluxes affecting the state of genes in the network.
The feedback of these 21 internal fluxes from the metabolic network affect the state of only 5 genes coding for
transcription factors and 16 genes downstream of these coding for metabolic enzymes in the present network.
In chapter 3, we found that these 21 genes in the present network which get affected by metabolic feedback via
internal fluxes may have their states undetermined across the different attractors for a given fixed environment.
The state of these 21 genes can oscillate back and forth between zero and one in the two-cycle attractors obtained for
any given fixed environment.
This set of 21 genes in the genetic network controlling {\it E.coli} metabolism forms the twinkling island.

We may expect introduction of additional feedback from internal metabolite concentrations into the genetic network
during the future network expansion.
If the additional feedback from internal metabolite concentrations affect enzyme coding genes in the genetic
network, then it is unlikely to change our qualitative conclusions about the observed dynamics of the system.
Since the enzyme coding genes happen to be the leaf nodes of the hierarchical acyclic graph, they do not affect
other nodes in the genetic network controlling {\it E.coli} metabolism.
Thus, the effect of feedback from internal metabolite concentrations on to an enzyme coding gene is unlikely to affect
the dynamics of other nodes in the genetic network.
If the additional feedback from internal metabolite concentrations affect genes coding for transcription factors in
the genetic network, then it can change our qualitative conclusions about the observed dynamics of the system.
It is possible that there is a larger set of genes forming the twinkling island, and possibly longer cycles as
attractors of the genetic network controlling {\it E. coli} metabolism.
In chapter 4, we have shown that the genetic network controlling {\it E. coli} metabolism has a highly disconnected
structure at the level of genes coding for transcription factors.
This disconnected architecture of the network at the level of genes coding for transcription factors as well as the
overall small depth ($\sim$4) of that graph suggests that
the introduction of addition feedback loops at the level of genes coding for transcription factors may only lead to
cyclic attractors which are of low period and localized.
We further showed that most Boolean functions in the present database iMC1010$^{v1}$ satisfy the canalyzing property.
These canalyzing functions may provide additional stability to the network when additional feedback loops get
introduced.

The metabolic network needs to function at all times as it is responsible for utilizing the nutrients available in the
external environment to produce the key molecules required for growth and maintenance of the cell.
The lack of feedback between genes coding for transcription factors in the genetic network controlling {\it E. coli}
metabolism observed here may be expected.
However, it is known that there are many feedback loops at the level of the metabolic network where the concentration
of an internal metabolite that is the end product of a metabolic pathway regulates the activity of the enzyme or the
protein catalyzing the reaction at the start of the pathway.
Such processes where the activity of the enzymes or proteins that are products of genes at the leaf level in the
acyclic graph representing the genetic network controlling {\it E.coli} metabolism are controlled by internal
metabolite concentrations are not included in the database iMC1010$^{v1}$.
Further, the regulation of enzyme activity by internal metabolite concentrations occurs at a much faster time scale
than transcription processes inside the cell.
{\it E. coli} is known to double its population in about 20 minutes in rich medium while transcriptional process inside
cells occur in the order of minutes to hours.
Thus, it is possible that metabolism being a functionality that needs to be active whenever food is available is largely
regulated without cycles at the genetic level, with feedbacks typically entering at the level of metabolites regulating
enzymes to ensure efficient functioning on a faster time scale.
Feedback loops or cycles are expected in systems exhibiting explicit temporal phenomena such as cell cycle or circadian
rhythms.
Nevertheless it would be important to explore these questions with an enlarged database representing the genetic network
controlling {\it E.coli} metabolism.
We further need to check the universality of the observed architecture for the genetic network controlling {\it E. coli}
metabolism in other organisms as and when an integrated metabolic and regulatory network database like iMC1010$^{v1}$
becomes available for them.

In chapter 3, we have approximated the nonzero flux of an internal reaction by the activity of the gene coding for the
enzyme catalyzing the reaction in the metabolic network.
Hence, we have converted the metabolic feedback via internal fluxes into effective feedbacks from enzyme coding genes
to other genes.
This was done in order to have a simplified dynamics and a closed system of the genes alone (along with external
metabolites).
We remark that there exist in the literature alternative ways of treating metabolic feedback on regulation and
in particular the flux variables.
These include the regulatory FBA \cite{CSP2001,CKRHP2004,HLPP2006} and dynamic FBA \cite{MED2002,LLTLZLL2006} in which the
fluxes and the genes are dynamically coupled to each other.
However, in these methods one makes an arbitrary choice of the flux vector out of many alternative flux vectors satisfying
the constraints.
Another method, SR-FBA \cite{SESR2007}, has been proposed that systematically accounts for multiple optimal metabolic steady
states given a regulatory state.
However, SR-FBA cannot be used for dynamical simulations since it only yields the various steady states for the
metabolic-regulatory system.
In chapter 3, our treatment of the internal fluxes is simpler compared to the above mentioned methods in that it eliminates
the flux variables in favour of an effective feedback of the genes upon other genes.
In the context of the present database, we believe that our broad conclusions would not change significantly because of our
simplified approach to the treatment of the internal fluxes since the feedbacks from the fluxes affect only 5 genes coding for
transcription factors and 16 genes downstream of these coding for enzymes, thus affecting only 21 genes out of 583.
A better treatment of the feedbacks from internal metabolites than is achieved by our approach and the other approaches
mentioned above requires metabolite concentrations which are difficult to compute at the present time due to the paucity of
kinetic information for such large scale networks.

We end this chapter with a comment relating our results presented in chapters 3 and 4 to earlier works by Kauffman and
a speculation.
Kauffman \cite{K1969a,Kauffmanbook} has found biologically motivated random Boolean networks to possess multiple attractors
that he has interpreted as different cell types of a multicellular organism.
In chapters 3 and 4, we have studied the nature of the attractors of the genetic network regulating {\it E. coli} metabolism.
Since {\it E. coli} is a prokaryote, perhaps it is not surprising that we get a much simpler picture of its genetic network
exhibiting a much higher degree of order in the dynamics than the systems investigated by Kauffman.
While we have shown in chapter 3 that the configuration of the genetic network regulating {\it E. coli} metabolism can
settle down into different attractors, yet, unlike Kauffman, for whom different attractors were realized via different
initial conditions of the genes, in our case the different attractors are realized when the control variables corresponding
to external metabolites have different configurations.
In chapter 3, we have shown that when the control variables are held fixed, there is none (or very little) multiplicity of
attractors irrespective of the initial conditions of genes.
The architecture and dynamics we have found in chapters 3 and 4 is probably quite suitable for prokaryotic lifestyle and
evolution.
The question remains open whether for eukaryotes and especially multicellular ones, the hypothesis that associates different
cell types with different attractors of the regulatory network is valid.
While that hypothesis remains an enticing possibility, it is worth noting that the simple architecture of the genetic network
regulating {\it E. coli} metabolism described in detail in chapter 4 could have its uses in the eukaryotic case as well.
Environmental control through the root nodes corresponding to external metabolites of the cellular attractors can itself cause
a cell to differentiate into another type, the environment being determined in the multicellular case by the state of other
cells in the organism.
The modular structure of the the genetic network regulating {\it E. coli} metabolism at the intermediate level of the essentially
acyclic graph described in chapter 4 would even permit a cell to hop through several attractors in the course of development of
the organism as the environmental cues to this cell change.
Such an architecture could thus contribute to developmental flexibility and, potentially, evolvability of eukaryotes as well.
The multiplicity of internal attractor basins pointed out by Kauffman would be an asset in keeping the cell in its new attractor
after a transient environmental cue has caused it to shift from one basin to another.
It would be interesting to investigate these questions when a database similar to iMC1010$^{v1}$ becomes available for a
multicellular organism.



\appendix

\chapter{List of UP-UC clusters in the {\it E. coli} metabolic network}

\label{upuclist}

The table below lists the 85 UP-UC clusters of various sizes
in the {\it E. coli} metabolic network iJR904 \cite{RVSP2003}.
The abbreviations of reactions and metabolites, the description of
the metabolic pathway where the reaction occurs, the reaction equation
and Gene-Protein-Reaction (GPR) association is taken from the database
iJR904.
After identifying the gene(s) for each reaction in every cluster using
the GPR association, we determined which of the genes associated with
a cluster are in the same operon.
For a given cluster, genes in the same operon are coloured with the same
shade (red, brown or pink).
Furthermore, we have added information obtained using flux balance analysis
(FBA) as to whether each reaction is `active' or `inactive'.
Active reactions in the {\it E. coli} metabolic network are those that
were found to have a non-zero flux for at least one of the 89 flux vectors
each corresponding to a different minimal medium (see section \ref{active}
in chapter \ref{lowdeg}).
Inactive reactions are those that had a zero flux for all the 89 flux
vectors.
Reactions for which the corresponding gene name was not available
in the database have been labelled as NA in the GPR association column.
Of these 85 UP-UC clusters, 69 clusters are such that genes are
identified for at least two distinct reactions in the cluster.
The remaining 16 clusters are shaded blue.

\begin{center}
\includegraphics[trim=100 0 100 50,angle=90,width=18cm]{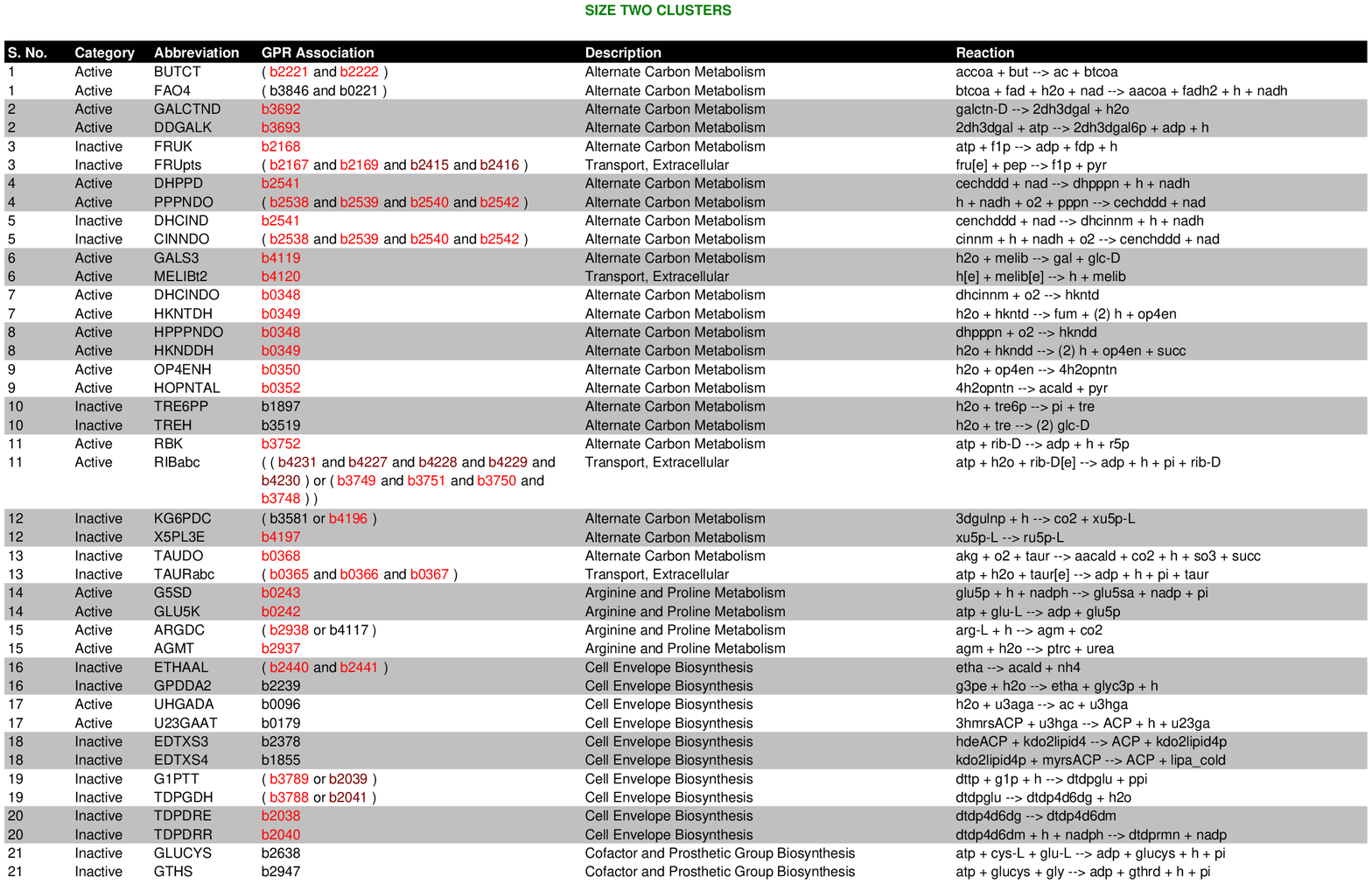}
\end{center}
\begin{center}
\includegraphics[trim=100 0 100 50,angle=90,width=18cm]{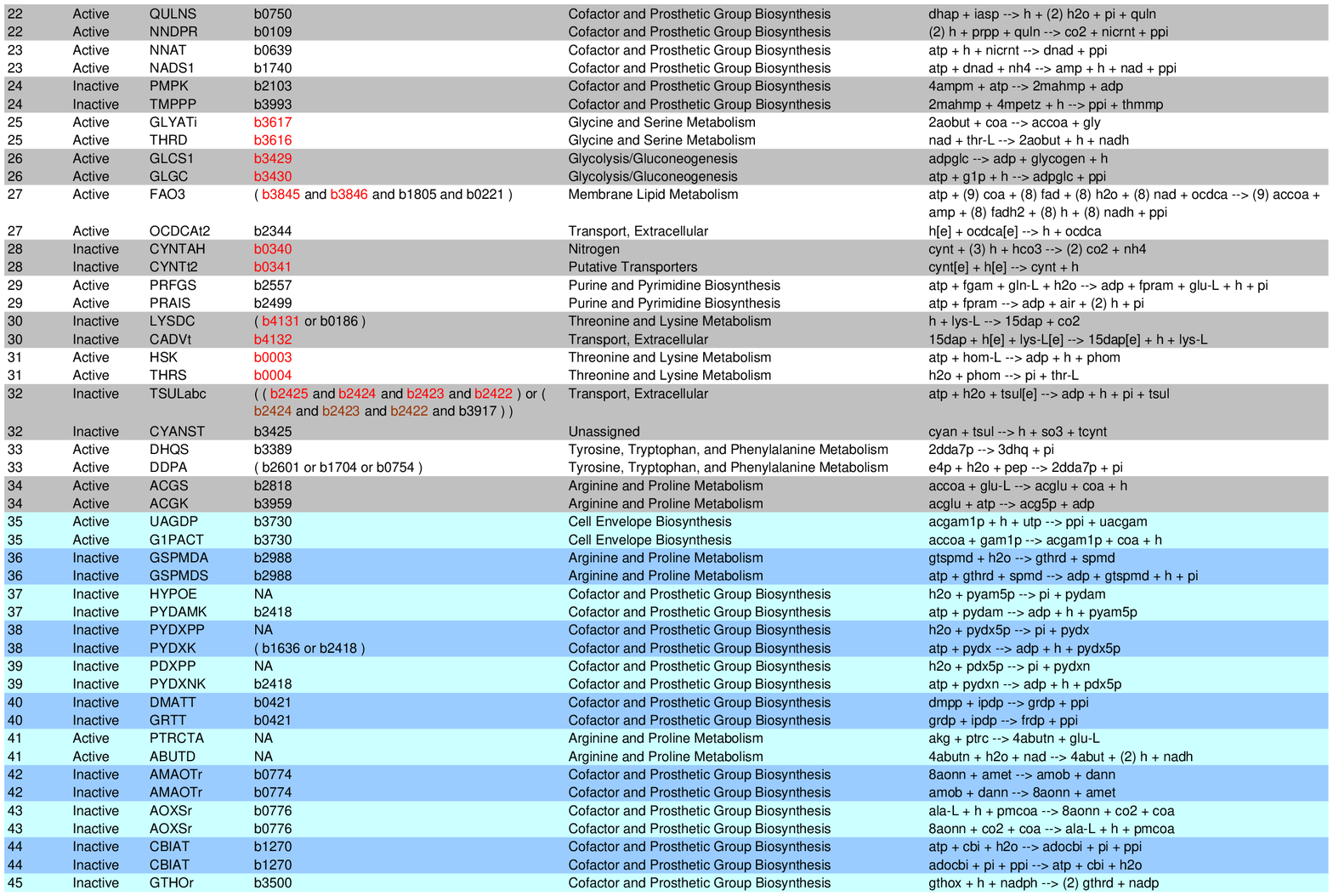}
\end{center}
\begin{center}
\includegraphics[trim=100 0 100 50,angle=90,width=18cm]{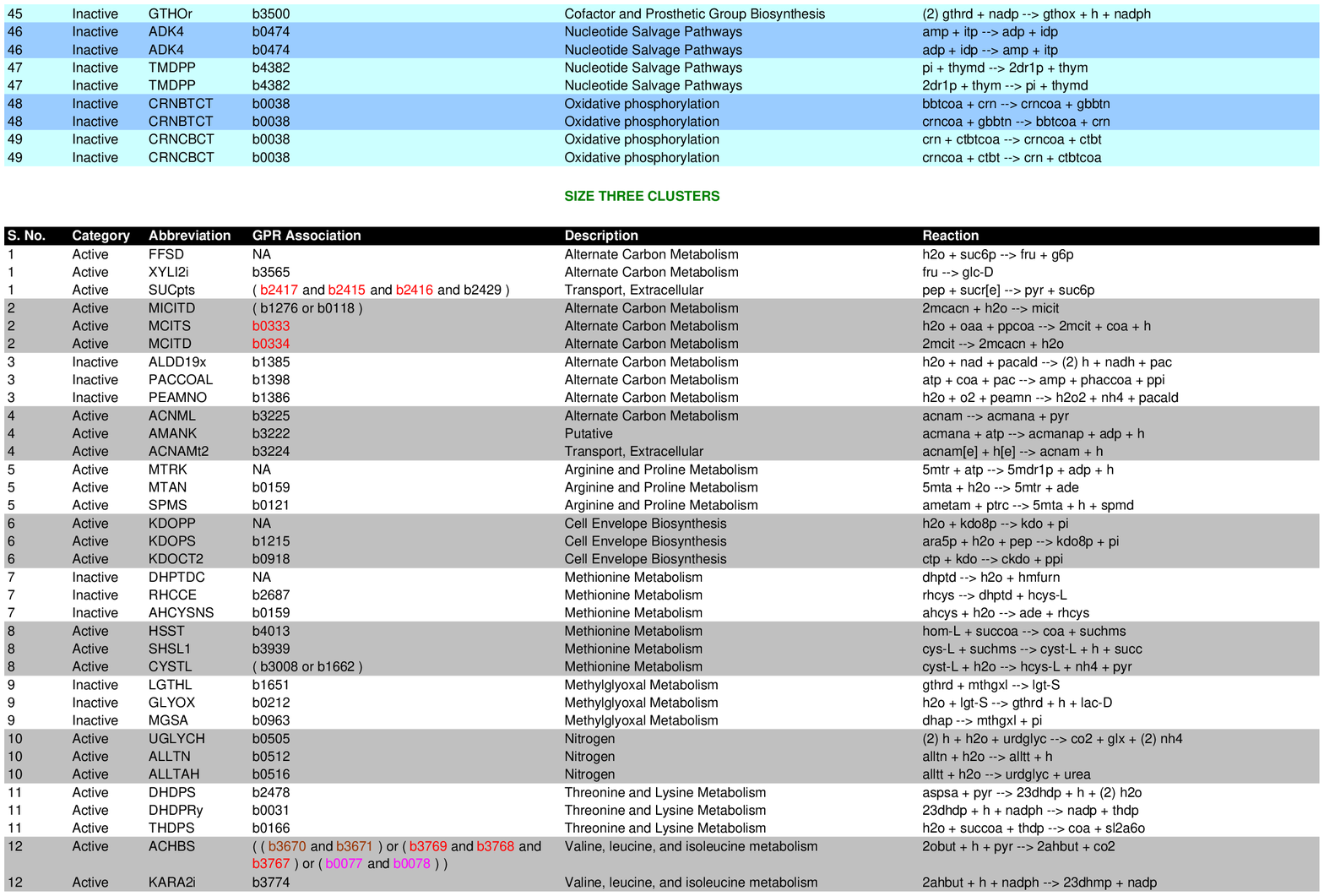}
\end{center}
\begin{center}
\includegraphics[trim=100 0 100 50,angle=90,width=18cm]{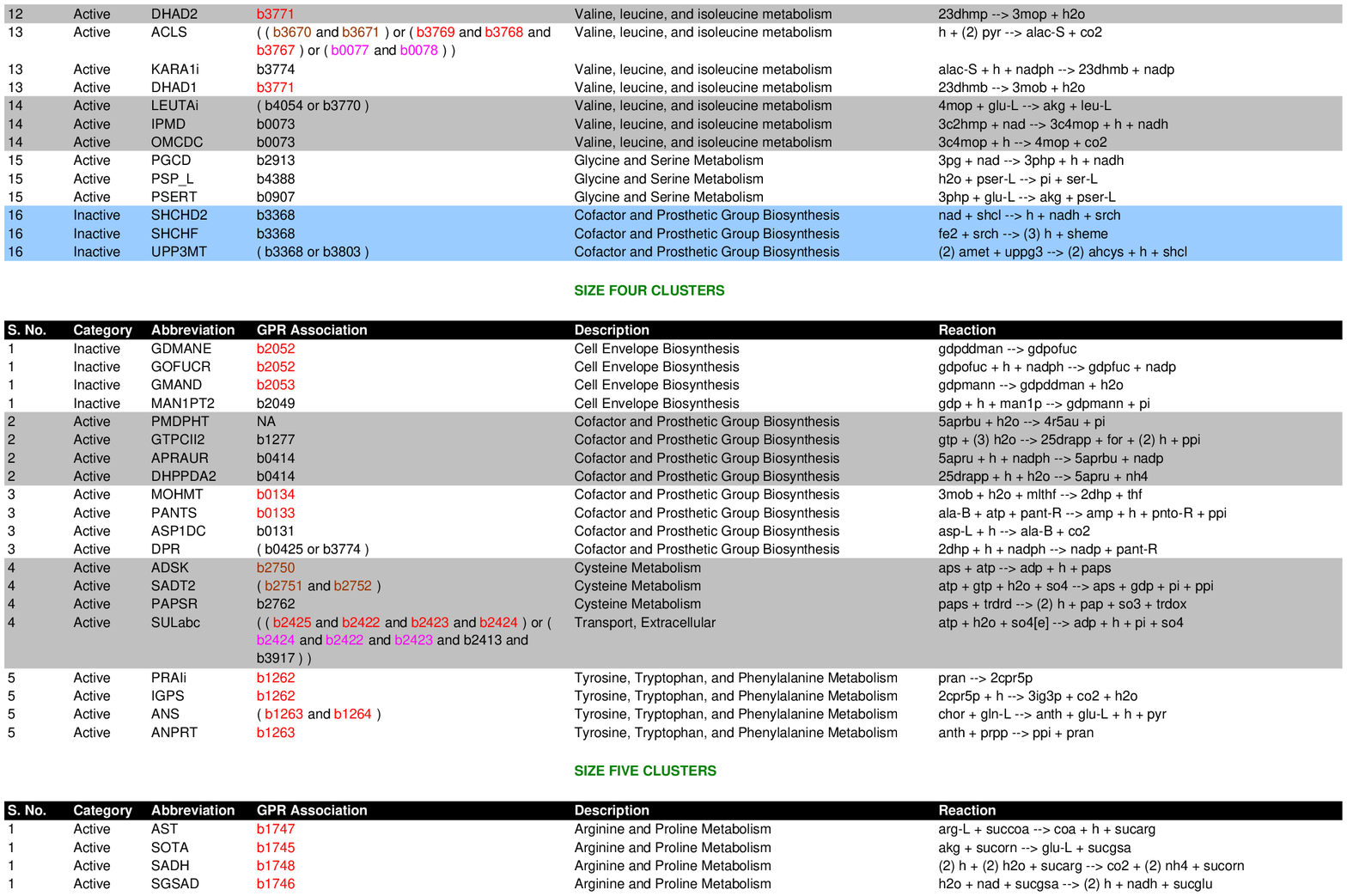}
\end{center}
\begin{center}
\includegraphics[trim=100 0 100 50,angle=90,width=18cm]{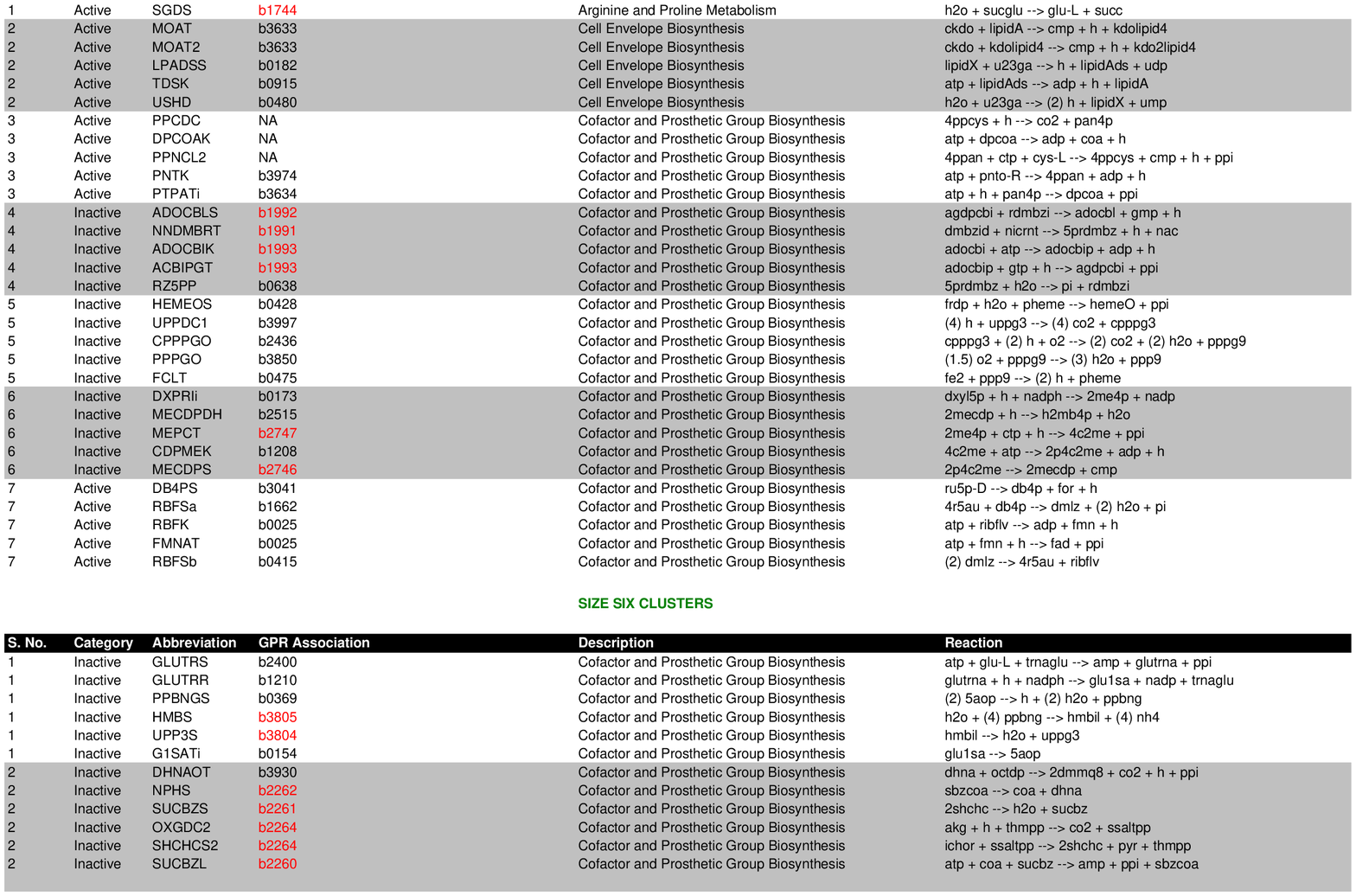}
\end{center}
\begin{center}
\includegraphics[trim=100 0 100 50,angle=90,width=18cm]{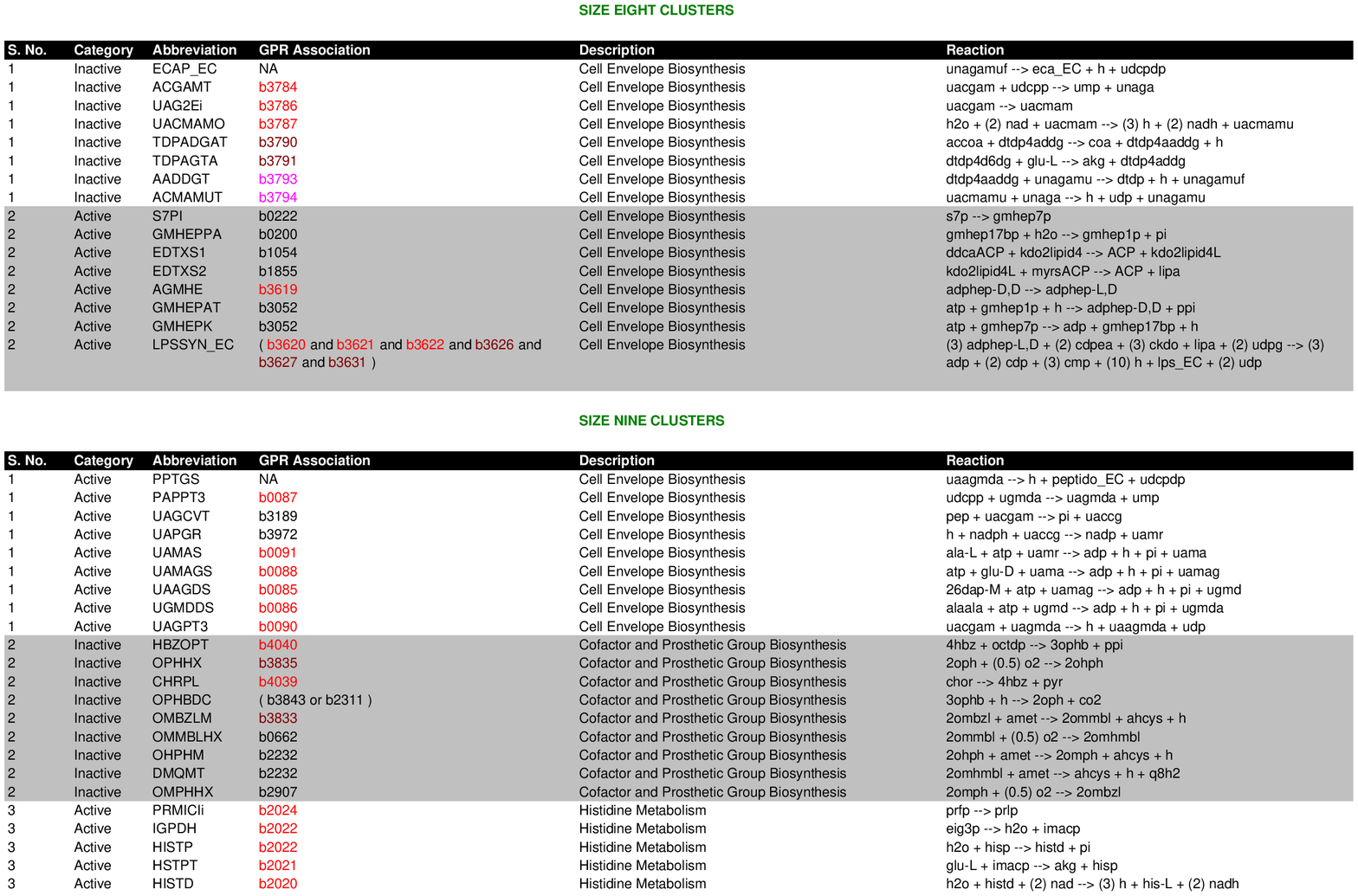}
\end{center}
\begin{center}
\includegraphics[trim=100 0 100 50,angle=90,width=18cm]{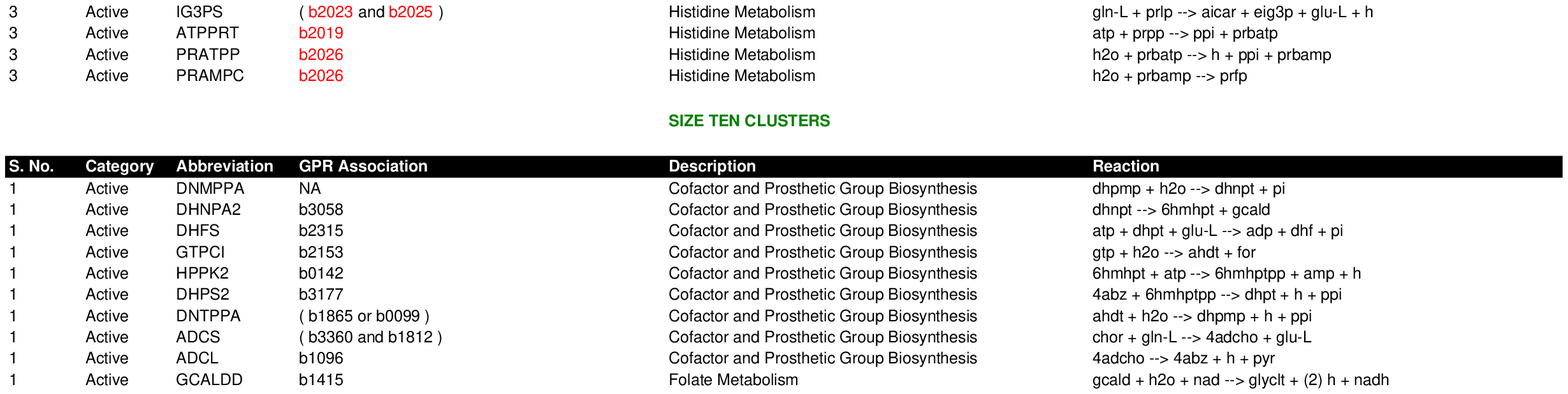}
\end{center}


\chapter{Flux Balance Analysis (FBA)}
\label{fba}

Flux balance analysis (FBA) is a computational modelling technique that can be used to obtain a prediction for the fluxes
of all reactions in the metabolic network and growth rate of the organism under the assumption of steady state without the
detailed knowledge of rate constants
\cite{SP1992a,SP1992b,VP1993a,VP1993b,VBP1993a,VBP1993b,VP1994a,VP1994b,PK1997,PK1998,EP2000,EIP2001,IEP2002,SCFCEP2002,SVC2002,RVSP2003,FFFPN2003,KPE2003,DHP2004,AKVOB2004,Palssonbook}.
The FBA approach has been primarily developed by Palsson group \cite{GCRG} and will be described here in some detail.


\section{FBA modelling approach}


\subsection{Inputs for FBA model}

\begin{enumerate}

\item {\bf List of metabolic reactions along with stoichiometric coefficients of metabolites:}
The key requirement for modelling an organism's metabolism using FBA is the list of all biochemical reactions along with
stoichiometric coefficients of the involved metabolites that can happen inside the cell.
The list of reactions must include all enzyme catalyzed internal reactions that can occur within the cell boundary and all
transport processes across the cell boundary.
The transport reactions must include both the diffusion and active transport mechanisms across the membrane.
The list of reactions that can happen within a cell is largely known for many organisms due to availability of their annotated
genomes and biochemical literature \cite{GCRG,KEGG,Ecocyc}.

\item {\bf Reversibility of reactions:}
Due to thermodynamic constraints certain reactions inside the cell are practically irreversible while other reactions are
reversible.
The net flux of a reversible reaction can be either in forward direction or backward direction for different environmental
conditions.
However, a reaction that is practically irreversible can have net flux in only one direction.
Hence, the knowledge of reversibility of a reaction limits the allowable range of flux through a reaction.

\item {\bf Flux capacity constraints:}
Based on limitations on the association rates of certain enzymes that catalyze reactions in  the metabolic network, we can impose
limitations on the maximum flux that can flow through a given reaction.
Such constraints for certain reactions will limit the solution space of attainable flux distributions.

\item {\bf Growth medium:}
In principle, an organism has the capability to uptake certain metabolites from its environment.
These are the external metabolites that can be transported across the cell boundary.
However, all external metabolites are not available for uptake in any given growth medium or environment.
Hence, we need to define the growth medium for the organism.
The growth medium essentially constrains the maximum uptake rates of various transport reactions across the cell boundary.
If an external metabolite is not available for uptake in the growth medium then the maximum flux of the reaction that transports the
external metabolite into the cell is set equal to 0.
If an external metabolite is available for uptake in the growth medium then the flux of the reaction that transports the external
metabolite into the cell is allowed to have a nonzero value.

\item {\bf Biomass composition:}
We require the biomass composition of the organism in terms of ratios of key metabolic precursors that contribute to the unit production
of biomass.
The biomass composition is added to the metabolic network as a fictitious reaction.
The stoichiometric coefficients of the biomass metabolites are based on experimentally derived proportions of the metabolite precursors
that contribute to unit biomass production.

\end{enumerate}


\subsection{Assumptions in FBA model}

\begin{enumerate}

\item {\bf Steady state condition:}
The most important assumption made in the FBA modelling technique is that under any given environmental condition the organism reaches
a steady state.
A steady state condition is defined as one wherein the concentration of all metabolites and velocities of all reactions in the network
are constant.

\item {\bf Optimal metabolic functioning:}
The second assumption made in FBA is that the cell tries to adjust its intracellular machinery or reaction fluxes so as to maximize
its growth rate or biomass production.

\end{enumerate}


\begin{figure*}
\centering
\includegraphics[width=14cm]{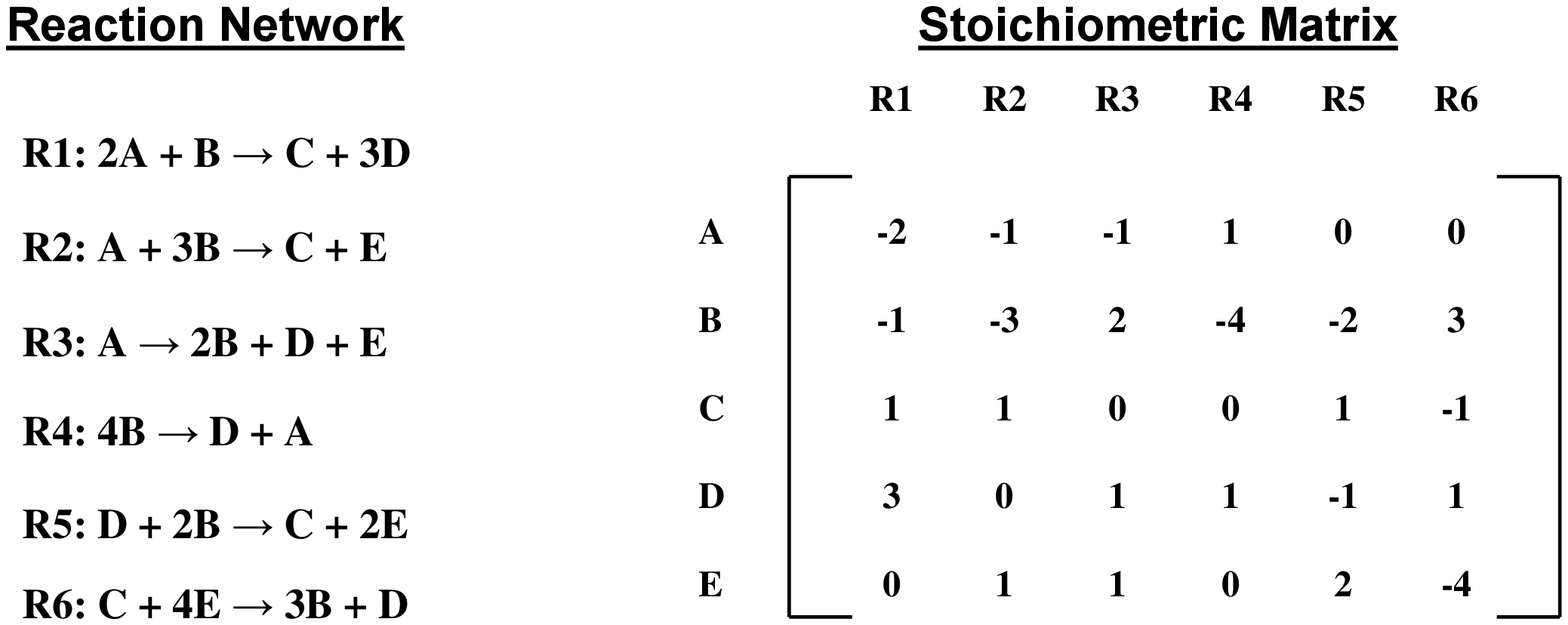}
\caption{Example of stoichiometric matrix for a hypothetical reaction network}
\label{smatrix}
\end{figure*}


\subsection{Computation of reaction fluxes}

\begin{enumerate}

\item {\bf Transformed network:}
Starting from the original reconstructed network (containing both reversible and irreversible reactions), we can obtain a transformed network
by replacing each reversible reaction with two irreversible reactions (one each for the forward and backward direction).
Such a transformed network has the advantage that the fluxes of all reactions in the transformed network can take only positive values, i.e.,
\begin{equation}
v_j \ge 0 \quad \quad \forall j=1,2,\ldots,n,
\end{equation}
where $v_j$ is the flux of reaction $j$ and $n$ is the number of reactions in the network.
All the results presented in this thesis have been obtained using the transformed metabolic network.

\item {\bf Stoichiometric matrix:}
The list of reactions along with the stoichiometric coefficients of the involved metabolites can be compactly represented in the form of the stoichiometric matrix.
The stoichiometric matrix ${\bf S}$ is a matrix of dimensions $m$ $\times$ $n$ where $m$ denotes the number of metabolites and $n$ denotes the number of reactions in the metabolic network.
Thus, the rows of the stoichiometric matrix correspond to various metabolites and the columns correspond to various reactions in the metabolic
network.
The stoichiometric matrix for a hypothetical reaction network is shown in Fig. \ref{smatrix}.

\item {\bf Mass balance:}
The rate of change of concentration of a metabolite $i$ is given by the equation
\begin{equation}
\frac{dX_i}{dt} = \sum_{j=1}^{n} S_{ij}v_j,
\end{equation}
where $X_i$ is the concentration of metabolite $i$, $S_{ij}$ is the stoichiometric coefficient of metabolite $i$ in reaction $j$, $v_j$ is the
flux of reaction $j$ and $n$ is the number of reactions in the network.
The above equation states that the rate of change of concentration of a metabolite $i$ is the difference between the sum of the rates at which
the metabolite is produced and the sum of the rates at which the metabolite is consumed.
Thus, each metabolite achieves a dynamic mass balance in the network and no mass is lost during the process.
In any steady state, the above equation can be written as
\begin{equation}
\frac{dX_i}{dt} = \sum_{j=1}^{n} S_{ij}v_j = 0
\end{equation}
and the stoichiometric constraints can be represented by the equation
\begin{equation}
\label{steadystate}
{\bf S}.{\bf v} = 0,
\end{equation}
where ${\bf S}$ is the stoichiometric matrix and ${\bf v}$ is the column vector representing the reaction fluxes, i.e., ${\bf v}$ = ($v_1$,$v_2$,\ldots,$v_n$)$^T$.
In FBA, we are interested in determining the steady state flux through reactions in the metabolic network.
For the metabolite A in the hypothetical reaction network shown in Fig. \ref{smatrix}, the steady state condition gives the equation
\begin{equation}
- 2v_{R1} - v_{R2} - v_{R3} + v_{R4} = 0,
\end{equation}
where $v_{Rj}$ is the flux of reaction $Rj$.
Hence, under steady state condition, we will get a linear equation relating various fluxes for each metabolite in the network.
We can determine the fluxes of reactions in the network by solving the system of linear equations relating various fluxes.
However, the number of metabolites $m$ in an organism's metabolic network is typically much less than the number of reaction fluxes $n$
to be determined.
Hence, the steady state solution of reaction fluxes ${\bf v}$ satisfying Eq. \ref{steadystate} is underdetermined.
Hence, we get a space of possible solutions satisfying the stoichiometric constraints given by Eq. \ref{steadystate}.
The solution space can be further limited by imposing flux capacity constraints for some reactions, i.e.,
\begin{equation}
\label{fluxcapacity}
v_j \le \alpha_j,
\end{equation}
where $\alpha_j$ is the upper limit of reaction flux $v_j$.

\item {\bf Linear optimization:}
To obtain a particular solution or flux distribution from the space of possible solutions given by Eq. \ref{steadystate} and flux capacity constraints for a given growth medium, linear optimization is used to find a flux distribution is the solution space that gives the optimal
value for an objective function \cite{P1984,FS1986,MD1990,SP1992a,SP1992b,VP1993a,VP1993b}.
The most commonly used objective function is maximization of biomass production rate.
Other objective functions that have been tried include maximization of ATP production and minimization of ATP utilization.
Hence, one obtains a particular solution or flux distribution ${\bf v}$ for a given growth medium satisfying Eq. \ref{steadystate} and the flux
capacity constraints such that the growth rate reaction flux has the maximum value.
Some of the internal reaction fluxes and growth rate predicted using FBA for few growth media in {\it E. coli} have been shown to be consistent
with experimental data \cite{EIP2001,IEP2002,SVC2002}.

\end{enumerate}


\subsection{Limitations of the FBA model}

\begin{enumerate}

\item Using the FBA model, we can determine only the steady state reaction fluxes in the network for a given growth medium.
We cannot use the technique to predict various internal metabolite concentrations in the network.

\item The regulation of metabolic reactions or pathways is completely neglected in FBA and linear optimization is used to determine the
reaction or pathway flux.

\item Using linear optimization, we obtain only a particular flux vector or distribution as the solution that maximizes the growth rate or
biomass reaction.
There are usually multiple optimal solutions that satisfy the governing constraints due to redundancies in the metabolic network
\cite{LPDG2000,PLZKAGD2001}.
However, the FBA modelling approach is able to fix the value of most reaction fluxes in the network \cite{KS2002} and only a few reaction
fluxes remain underdetermined for any given growth medium.

\end{enumerate}


\section{Blocked reactions}
\label{blocked}

`Strictly detailed balanced' reactions \cite{SS1991,Heinrichbook} or `blocked' reactions
\cite{BNSM2004} are reactions that can have only a zero flux under any steady state.
The main reason for the presence of blocked reactions in various reconstructed metabolic
networks is the presence of dead end internal metabolites.
A dead end internal metabolite is one that either does not have a reaction consuming it
or does not have a reaction producing it in the network.
Since in any steady state the metabolite concentrations are constant, the sum of the
fluxes of reactions producing a metabolite is equal to the sum of the fluxes of reactions
consuming the metabolite.
As the dead end metabolites lack either a reaction producing it or a reaction consuming it,
the fluxes of reactions that involve such metabolites are guaranteed to have a zero value
under any steady state.
In FBA, we are interested in the steady state flux distribution for the metabolic network.
The blocked reactions can be removed from the reconstructed metabolic network for any
steady state analysis such as FBA.
The results such as optimal growth rate and set of essential reactions obtained by
implementing FBA on the original network and reduced network are the same.
In this thesis, we refer to the network obtained after removing blocked reactions from the
original metabolic network as the `reduced network' (see also section \ref{reducednetwork}
in chapter \ref{lowdeg}).


\subsection{Algorithm to determine blocked reactions}
\label{Burgard}

We will now describe an algorithm previously published by Burgard {\it et al} \cite{BNSM2004} to determine blocked reactions in the metabolic
network.
Burgard {\it et al} identify reactions in the metabolic network that have their maximum and minimum flux both equal to zero for any growth
medium.
Such reactions are blocked in the metabolic network for that growth medium.
To determine blocked reactions in the metabolic network, Burgard {\it et al} solved the following linear programming (LP) problem:\\
Maximize flux $v_k$ subject to stoichiometric constraints
\begin{equation}
\sum_{j=1}^{n} S_{ij}v_j = 0
\end{equation}
and flux capacity constraints
\begin{equation}
v_j \le \alpha_j,
\end{equation}
where $S_{ij}$ is the stoichiometric coefficient of metabolite $i$ in reaction $j$, $v_k$ is the flux of reaction $k$ that is being checked for
blocked status, $n$ is the number of reactions and $\alpha_j$ is the upper limit of reaction flux $v_j$ in the network.
A reaction $k$ is said to be always blocked if the maximum flux that is obtained by solving the above mentioned LP problem is equal to zero for
all possible media.

In FBA, we are interested in determining the maximum flux through the biomass reaction, given the stoichiometric constraints
(Eq. \ref{steadystate}) and flux capacity constraints (Eq. \ref{fluxcapacity}).
However, while determining blocked reactions, we are interested in determining the maximum flux through the reaction being checked for
blocked status, given the stoichiometric constraints and flux capacity constraints.


\section{Constrained FBA}
\label{constrainedfba}

The state of the genetic network at any time instant is given
by the configuration of its genes.
In the Boolean approach, any gene in the network at a given
time instant may be either active or inactive.
If a gene is active, the protein coded by the gene is produced
in the cell.
A reaction in the metabolic network can be assumed to be off if none
of the enzymes catalyzing it are being produced, or, equivalently,
if the genes coding for those enzymes are in the off state.
Thus, the state of the genetic network for any external environment
determines the subset of reactions in the metabolic network
that can happen under that condition.

In `pure FBA', we leave the flux through all internal reactions
unconstrained, i.e., all internal reactions are allowed to have a
nonzero flux value, to determine the optimal growth rate of the
cell for any given medium.
Using the state of the genetic network at any time for a given
medium, we can turn off all reactions whose corresponding genes
are in the off state in the genetic configuration in the input
to the FBA model \cite{CSP2001}.
Using FBA, we can then determine the optimal growth rate of the
cell for a given medium by constraining the maximum flux through
the reactions turned off by the genetic configuration to zero.
This variant of FBA technique where the constraints from the
state of the genetic network are incorporated to determine the
growth rate of the cell may be termed as `constrained FBA'.
Constrained FBA captures the effect of gene regulation on
metabolic function.
Using constrained FBA, one can track the optimal growth rate
as a function of time as the configuration of the genes changes
according to the dynamics of the genetic regulatory network,
as also discussed earlier by Covert {\it et al}
\cite{CSP2001,CKRHP2004}.
One can also calculate the growth rate in the attractor
configuration for any given medium.
The growth rate obtained from constrained FBA for any configuration
of the genes is, by definition, less than or equal to that obtained
from pure FBA (for the same medium).



\chapter{Computer Programs}
\label{program}

In this Appendix, we describe some of the computer programs used to
obtain results presented in this thesis.

\noindent These programs can be downloaded from the associated website:
\begin{center}
{\it \footnotesize \bf http://areejit.samal.googlepages.com/programs}
\end{center}
The website contains detailed instructions to run various codes.
These codes are free to use, modify and distribute for academic research.
For any further queries regarding the programs contact
{\bf Areejit Samal} at:
\begin{center}
{\it \footnotesize \bf areejit.samal@gmail.com}
\end{center}

\noindent  The codes mentioned below are meant for execution on any
Linux Operating System.
The programs have been mainly written using C++.
The GNU g++ compiler was used for compiling the C++ programs.
We have connected different C++ programs using shell scripts.
The bash shell must be used as the default environment to run these
shell scripts.


\section{Program to determine UP-UC clusters}

The C++ program UPUC.cpp can be used to determine UP-UC  metabolites,
UP/UC reactions and UP-UC clusters in the {\it E. coli} metabolic
network iJR904.
The program can be modified to determine UP-UC clusters in any
metabolic network.

\noindent There are two associated input files with this program:
Amatrix.txt and Metabolites.txt.
\begin{itemize}
\item The input file Amatrix.txt contains the bipartite matrix ${\bf A}$
of dimensions 618 $\times$ 1177 for the {\it E. coli} metabolic network
iJR904.
\item The input file Metabolites.txt contains the list of 618  internal
metabolites corresponding to rows of the bipartite matrix ${\bf A}$.
\end{itemize}

\noindent To run the program UPUC.cpp, download the three files: UPUC.cpp,
Amatrix.txt and Metabolites.txt from the associated website (mentioned at
the beginning of this Appendix) and put them in a single directory on a
Linux system.

\noindent Compile the code by invoking the command:
\begin{center}
{\footnotesize \bf  g++ UPUC.cpp}
\end{center}
followed by the instruction:
\begin{center}
{\footnotesize \bf  ./a.out}
\end{center}
to run the program.

\noindent The program generates the following output files:
UPUCmet.txt, UPUCreac.txt and UPUCcluster.txt.
\begin{itemize}
\item The output file UPUCmet.txt contains the UP(UC) status for each of the 618
internal metabolites in the {\it E. coli} metabolic network.
\item The output file UPUCreac.txt contains the UP/UC status for each of the 1176
reactions in the {\it E. coli} metabolic network.
\item The output file UPUCcluster.txt contains the UP-UC clusters in the {\it E. coli}
metabolic network.
Each line of this output file represents a single UP-UC cluster of reactions.
\end{itemize}
In addition to the three output files, the program UPUC.cpp prints out the number of
UP-UC metabolites, the number of UP/UC reactions and the size distribution of UP-UC
clusters in the {\it E. coli} metabolic network.


\section{Program to determine essential reactions}

The shell script Essential.sh can be used to determine essential reactions in the
{\it E. coli} metabolic network iJR904 using the technique of flux balance analysis
(FBA) for a specified growth medium.

\noindent The script requires the following files: glpsol, Essential.cpp, Sij.txt,
Rxnlist.txt and Media.txt associated with it.
\begin{itemize}
\item The executable file glpsol is the GNU linear programming kit solver that is used
to solve the linear programming problem.
\item The C++ program Essential.cpp generates the MPS file for the linear programming
problem to be solved.
\item The input file Sij.txt contains the stoichiometric matrix for the {\it E. coli}
metabolic network iJR904.
The dimensions of the matrix is 761 $\times$ 1463.
The rows of the matrix correspond to metabolites and the columns correspond to reactions.
\item The list of 1463 reactions in the {\it E. coli} metabolic network iJR904 corresponding
to the 1463 columns of the stoichiometric matrix contained in the file Sij.txt are listed
in the input file Rxnlist.txt.
\item The input file Media.txt contains the configuration of the growth medium.
At present, the growth medium is set to glucose aerobic minimal medium.
The file Media.txt consists of two columns.
The first column corresponds to external metabolites that the cell can uptake.
The second column of the input file Media.txt corresponds to the maximum uptake rate of
each external metabolite.
The second column should be suitably modified to represent different growth media.
\end{itemize}

\noindent The script and its associated files can be downloaded from the associated
website.

\noindent Invoke the shell script using the command:
\begin{center}
{\footnotesize \bf  sh Essential.sh}
\end{center}
The program prompts you to enter the reaction to be knocked out.
The program outputs the optimal growth rate of the organism for the truncated network
where the reaction knocked out is absent.
The steady state reaction fluxes for the truncated network is stored in the output file
Flux.txt for the simulated growth medium.


\section{Program to determine blocked reactions}

The shell script Blocked.sh can be used to determine blocked status of a reaction in the
{\it E. coli} metabolic network iJR904 using the algorithm by Burgard et al \cite{BNSM2004}
for a specified growth medium.

\noindent The script requires the following files: glpsol, Blocked.cpp, Sij.txt, Reversible.txt
and Media.txt associated with it.
\begin{itemize}
\item The executable file glpsol is the GNU linear programming kit solver that is used to solve
the linear programming problem.
\item The C++ program Blocked.cpp generates the MPS file for the linear programming problem to
be solved.
\item The input file Sij.txt contains the stoichiometric matrix for the {\it E. coli} metabolic network
iJR904.
The dimensions of the matrix is 761 $\times$ 1463.
The rows of the matrix correspond to metabolites and the columns correspond to reactions.
\item The list of 1463 reactions in the {\it E. coli} metabolic network iJR904 corresponding to the
1463 columns of the stoichiometric matrix contained in the file Sij.txt are listed in
the input file Rxnlist.txt.
\item The list of reversible reactions in the {\it E. coli} metabolic network is contained in the
input file Reversible.txt.
While determining the blocked status of a reaction, we need to set the maximum flux through reactions
that are exact reverse of the reaction being tested for the blocked status to zero, in order to avoid
trivial cycles in the network.
\item The input file Media.txt contains the configuration of the growth medium.
At present, the growth medium is set to be glucose aerobic minimal medium.
The file Media.txt consists of two columns.
The first column corresponds to external metabolites that the cell can uptake.
The second column of the input file Media.txt corresponds to the maximum uptake rate for each external
metabolite.
The second column should be suitably modified to represent different growth media.
\end{itemize}

\noindent The script and its associated files can be downloaded from the associated website.

\noindent Invoke the shell script using the command:
\begin{center}
{\footnotesize \bf  sh Blocked.sh}
\end{center}
The program prompts you to enter the reaction to be tested for blocked status.
The program then outputs the maximum flux possible for the reaction being tested for blocked status
under the given growth medium.
If the maximum flux possible is zero, then the reaction being tested is blocked under the simulated
growth medium.


\section{Program to simulate the genetic network controlling {\it E. coli} metabolism as a
Boolean dynamical system}

The programs associated with the simulation of the genetic network controlling {\it E. coli}
metabolism are contained in the following archive: SJ.tar.gz.
Download the archive from the associated website and extract it using the command:
\begin{center}
{\footnotesize \bf  tar zxvf SJ.tar.gz}
\end{center}
on a Linux operating system.
The successful execution of this command will create a directory named ``ECMC1010'' in the current directory.

\noindent The directory ECMC1010 has two sub-directories inside it: System1A and System1B
which contain the data and programs for the two dynamical systems A and B, respectively.

\subsection{Dynamical system A}

Inside the directory System1A, there are two sub-directories DATA and PROG.
\begin{itemize}
\item In the directory DATA, the file ENV.txt gives a list of variables whose values remain fixed at all times
for a given initial configuration of genes and minimal medium considered.
There are 128 such variables: 1 to 96 are the external metabolites, 97 to 117 are the 21 fluxes, then the two conditions
corresponding to pH and Growth,
followed by the 9 stimuli which are set to be always absent in our study.
We need to determine the column vector of 128 boolean variables (in the order mentioned in the file ENV.txt)
for a given minimal medium to determine the attractor of the dynamical system.
\item We study mostly the system under the 93 minimal media, which are listed in the file MinMedia contained in
directory DATA.
Each line of the file MinMedia corresponds to a different minimal medium.
\item We fix the 21 fluxes variables for a given medium based on their blocked reaction status for that medium.
For the 93 minimal media listed in the file MinMedia the information about blocked status for each of the 21 fluxes
is contained in the file FLUX1Aminmedia contained in directory DATA.
Here each column of the file FLUX1Aminmedia contains the blocked status information for the 21 fluxes for the minimal
medium mentioned at the corresponding line of the file MinMedia.
If a reaction flux is blocked for a given medium then it is set as 0 else 1.
The fluxes represented in the 21 rows of the file FLUX1Aminmedia are in the same order as in the file ENV.txt.
\item For example, the 23$^{rd}$ line of file MinMedia is glucose aerobic minimal medium (see file glcaermedia)
and hence the 23$^{rd}$ column of FLUX1Aminmedia file contains the blocked status information for the 21 fluxes
for the glucose aerobic minimal medium (see file glcaerFLUX).
\item The system contains 583 genes and we need to start with some initial condition for the 583 genes.
There are in principle $2^{583}$ possible initial conditions of genes.
One possible initial condition is contained in the file INITIAL in the directory DATA.
\item The genes corresponding to each line of the file INITIAL are contained in the file TRNgenes in the directory
DATA.
\end{itemize}

\noindent The program env.cpp in PROG directory is used to determine the vector of 128 fixed variables for a given
medium.
Compile the program using the command:
\begin{center}
{\footnotesize \bf  g++ PROG/env.cpp}
\end{center}
Run the program as follows:
\begin{center}
{\footnotesize \bf  ./a.out File1 File2 File3}
\end{center}
Here,
\begin{itemize}
\item File1 is a file containing the external metabolites in the minimal medium, i.e., one of the lines of the
file MinMedia.
\item File2 should contain the column corresponding to the minimal medium of File1 in the file FLUX1Aminmedia,
i.e., blocked reaction status of the 21 fluxes for the minimal medium studied.
\item File3 is the output file that will store the vector of 128 variables.
\end{itemize}
File1, File2 and File3 are command line arguments and can take any string as a name.

\noindent The program GP1A.cpp contained in the folder PROG is used to determine which proteins or transcription
factors (TF) are on/off given the configuration of genes at the previous time instant.

\noindent Compile the program using the command:
\begin{center}
{\footnotesize \bf  g++ PROG/GP1A.cpp}
\end{center}
Run the program as follows:
\begin{center}
{\footnotesize \bf  ./a.out File1 File2}
\end{center}
Here,
\begin{itemize}
\item File1 should contain the configuration of 583 genes at the previous time instant.
It can be DATA/INITIAL.
\item File2 is the output file containing the state of transcription factors at current time instant.
The lines in the output file File2 correspond to transcription factors listed in the file TF in
directory DATA.
\end{itemize}

\noindent Given the state of transcription factors at the current time instant (calculated using GP1A.cpp)
and the fixed values of 128 variables for the simulated minimal medium (calculated using env.cpp), we can use
TRN1A.cpp in the PROG directory to calculate the state of the 583 genes at the next time instant.

\noindent Compile the program using the command:
\begin{center}
{\footnotesize \bf  g++ PROG/TRN1A.cpp}
\end{center}
Run the program as follows:
\begin{center}
{\footnotesize \bf  ./a.out File1 File2 File3}
\end{center}
Here,
\begin{itemize}
\item File1 should contain the state of the transcription factors at the current time instant given the state of
the genes at previous time instant.
\item File2 should contain the values of the 128 fixed variables.
\item File3 is the output file in which state of the 583 genes at the current time instant is stored.
\end{itemize}

\noindent We have connected these programs using a shell script attractor.sh which can be used to compute
the attractor for a given minimal medium and initial configuration of genes.
The shell script attractor.sh is contained in the directory System1A.
The shell script first starts with the initial configuration of genes and a defined minimal medium and
computes the state of the 583 genes for the next 200 time steps.
Then a small perl script, attractor.pl in the PROG directory, is invoked to determine the attractor of the
system.
We are assuming here that the system reaches the attractor in 200 time steps.

\noindent Run the shell script using the command:
\begin{center}
{\footnotesize \bf  sh attractor.sh File1 File2 File3 File4}
\end{center}
where,
\begin{itemize}
\item File1 should contain the minimal medium.
\item File2 should contain the blocked status of 21 fluxes for the minimal medium in File1.
\item File3 should contain the initial configuration of genes.
\item File4 is the output file where the attractor is stored.
\end{itemize}

\noindent For example, run the script as follows:
\begin{center}
{\scriptsize \bf  sh attractor.sh DATA/glcaermedia DATA/glcaerFLUX DATA/INITIAL steady.txt}
\end{center}
\noindent The output is stored in the file steady.txt, i.e., it contains the attractor for
the glucose aerobic minimal medium for the initial configuration of genes specified in the
file DATA/INITIAL.

\noindent If the attractor is of period 1, i.e., a fixed point, then this vector is contained in a
line of the output file steady.txt.
If the attractor is of period 2, i.e., a two cycle, then the two vectors are contained in two lines
of the output file steady.txt.

\noindent There are some other files in the directory DATA:
\begin{itemize}
\item TRN1A.txt gives the rules for the update of each of the 583 genes.
\item GP1A.txt gives the Gene Protein association for transcription factors
in the model.
\item Condition1A.txt gives the rules for 17 out of the 19 conditions. The remaining two conditions, pH and Growth,
are set using the program env.cpp.
\end{itemize}
In these files, $!$ stands for the NOT operator, $\&\&$ for the AND operator
and $||$ for the OR operator.

\subsection{Dynamical system B}

Inside the directory System1B, there are two sub-directories DATA and PROG.
\begin{itemize}
\item In the directory DATA, the file ENV.txt gives a list of variables whose values remain fixed at all times
for a given initial configuration of genes and minimal medium considered.
There are 118 such variables: 1 to 96 are the external metabolites, 97 to 107 are the 11 fluxes, then the two
conditions corresponding to pH and Growth,
followed by the 9 stimuli which are set to be always absent in our study.
We need to determine the column vector of 118 boolean variables (in the order mentioned in the file ENV.txt)
for a given minimal medium to determine the attractor.
\item We study mostly the system under the 93 minimal media, which are listed in the file MinMedia contained in
directory DATA.
Each line of the file MinMedia corresponds to a different minimal medium.
\item We fix the 11 fluxes variables for a given medium based on their blocked reaction status for that medium.
For the 93 minimal media listed in the file MinMedia the information about blocked status for each of the 11 fluxes
is contained in the file FLUX1Bminmedia contained in directory DATA.
Here each column of the file FLUX1Bminmedia contains the blocked status information for the 11 fluxes for the minimal
medium mentioned at the corresponding line of the file MinMedia.
If a reaction flux is blocked for a given medium then it is set as 0 else 1.
The fluxes represented in the 11 rows of the file FLUX1Bminmedia are in the same order as in the file ENV.txt.
\item For example, the 23$^{rd}$ line of file MinMedia is glucose aerobic minimal medium (see file glcaermedia)
and hence the 23$^{rd}$ column of FLUX1Bminmedia file contains the blocked status information for the 11 fluxes
for the glucose aerobic minimal medium (see file glcaerFLUX).
\item The system contains 583 genes and we need to start with some initial condition for the 583 genes.
There are in principle $2^{583}$ possible initial conditions of genes.
One possible initial condition is contained in the file INITIAL in the directory DATA.
\item The genes corresponding to each line of the file INITIAL are contained in the file TRNgenes in the
directory DATA.
\end{itemize}

\noindent The program env.cpp in PROG directory is used to determine the vector of 118 fixed variables
for a given medium.
Compile the program using the command:
\begin{center}
{\footnotesize \bf  g++ PROG/env.cpp}
\end{center}
Run the program as follows:
\begin{center}
{\footnotesize \bf  ./a.out File1 File2 File3}
\end{center}
Here,
\begin{itemize}
\item File1 is a file containing the external metabolites in the minimal medium,
i.e., one of the lines of the file MinMedia.
\item File2 should contain the column corresponding to minimal medium of File1 in the file FLUX1Bminmedia,
i.e., blocked reaction status of the 11 fluxes for the minimal medium studied.
\item File3 is the output file that will store the vector of 118 variables.
\end{itemize}
File1, File2 and File3 are command line arguments and can take any string as a name.

\noindent The program GP1B.cpp contained in the folder PROG is used to determine which proteins or
transcription factors (TF) are on/off given the configuration of the genes at the previous time instant.

\noindent Compile the program using the command:
\begin{center}
{\footnotesize \bf  g++ PROG/GP1B.cpp}
\end{center}
Run the program as follows:
\begin{center}
{\footnotesize \bf  ./a.out File1 File2}
\end{center}
Here,
\begin{itemize}
\item File1 should contain the configuration of the 583 genes at the previous time instant.
It can be DATA/INITIAL.
\item File2 is the output file containing the state of the transcription factors at the current
time instant.
The lines in the output file File2 correspond to transcription factors listed in the file
TF in directory DATA.
\end{itemize}

\noindent Given the state of the transcription factors at the current time instant (calculated using GP1B.cpp)
and the fixed values of 118 variables for minimal medium (calculated using env.cpp), we can use TRN1B.cpp
in the PROG directory to calculate the state of the 583 genes at the next time instant.

\noindent Compile the program using the command:
\begin{center}
{\footnotesize \bf  g++ PROG/TRN1B.cpp}
\end{center}
Run the program as follows:
\begin{center}
{\footnotesize \bf  ./a.out File1 File2 File3}
\end{center}
Here,
\begin{itemize}
\item File1 should contain the state of the transcription factors at the current time instant given the state of
the genes at previous time instant.
\item File2 should contain the values of the 118 fixed variables.
\item File3 is the output file in which state of the 583 genes at the current time instant is stored.
\end{itemize}

\noindent We have connected these programs using a shell script attractor.sh which can be used to compute the
attractor of the system for a given minimal medium and initial configuration of genes.
The shell script attractor.sh is contained in the directory System1B.
The shell script first starts with a initial configuration of genes and a defined minimal medium and computes
the state of the 583 genes for the next 200 time steps.
Then a small perl script, attractor.pl in the PROG directory, is invoked to determine the attractor.
We are assuming here that the system reaches the attractor in 200 time steps.

\noindent Run the shell script using the command:
\begin{center}
{\footnotesize \bf  sh attractor.sh File1 File2 File3 File4}
\end{center}
where,
\begin{itemize}
\item File1 should contain the minimal medium.
\item File2 should contain the blocked status of 11 fluxes for the minimal medium in File1.
\item File3 should contain the initial configuration of genes.
\item File4 is the output file where the attractor is stored.
\end{itemize}

\noindent For example, run the script as follows:
\begin{center}
{\scriptsize \bf  sh attractor.sh DATA/glcaermedia DATA/glcaerFLUX DATA/INITIAL steady.txt}
\end{center}

\noindent The output is stored in the file steady.txt, i.e., it contains the attractor for the
glucose aerobic minimal medium for the initial configuration of genes specified in the file
DATA/INITIAL.

\noindent If the attractor is of period 1, i.e., a fixed point, then this vector is contained in a line
of the output file steady.txt.
If the attractor is of period 2, i.e., a two cycle, then the two vectors are contained in two lines of
the output file steady.txt.

\noindent There are some other files in the directory DATA:
\begin{itemize}
\item TRN1B.txt gives the rules for the update of each of the 583 genes.
\item GP1B.txt gives the Gene Protein association for transcription factors and enzymes in the
model.
\item Condition1B.txt gives the rules for 17 out of the 19 conditions. The remaining two conditions, pH and Growth,
are set using the program env.cpp.
\end{itemize}
In these files, $!$ stands for the NOT operator, $\&\&$ for the AND operator and
$||$ for the OR operator.



\addcontentsline{toc}{chapter}{References} 

\listoffigures
\addcontentsline{toc}{chapter}{List of Figures}

\listoftables
\addcontentsline{toc}{chapter}{List of Tables}


\chapter*{Acknowledgements}
\addcontentsline{toc}{chapter}{Acknowledgements}

\thispagestyle{empty}

I would like to thank Prof. Sanjay Jain and Prof. Shobhit Mahajan
under whose guidance the work reported in this thesis was
completed.

I thank Prof. Sanjay Jain for suggesting the questions addressed
in this thesis.
The work reported in this thesis has been done in close
collaboration with him.
He introduced me to the area of complex networks and systems
biology.
I have learned the subject from him through numerous scientific
discussions we had in the last five years.
He has through his actions influenced my personality and outlook.

I would like to thank Prof. Shobhit Mahajan for always being there
for me.
He has been very patient to take care of my problems and encouraged
me.
He has also shown a lot of faith and confidence in me.
I am indebted to him for his help during M.Sc. Physics.
His critical comments and practical advice have helped me immensely.
He has been a pillar of strength in my life.

I thank Shalini Singh, Varun Giri, Dr. Sandeep Krishna and
Dr. N. Raghuram for discussions and collaborations.
The work reported in Chapter 2 has been done in collaboration
with them.
I would also like to thank Dr. Sandeep Krishna, Shalini Singh
and Varun Giri for developing the flux balance analysis (FBA)
program used by our group.
The program has been used for obtaining various results in this
thesis.

I thank Shalini and Varun for their friendship and help.
We went together to many conferences in India.
I enjoyed their company during these trips.
I would like to thank Varun for spending a lot of his time 
everyday speculating on various matters with me and keeping 
our discussions lively.  

I would like to thank Aditya, Amit, Anubhav, Gagandeep, Utkarsh
and Vinay for discussions and their questions.
It was a learning experience for me while they were doing their
projects with Prof. Sanjay Jain.

I thank Prof. Amitabha Mukherjee for being part of my Ph.D.
advisory committee and for discussions.

I would like to thank all the members of Department of Physics
and Astrophysics, University of Delhi, who have helped me during
M.Sc. and Ph.D. programme.
In this regard, I would like to thank Prof. S. Annapoorni,
Prof. Debajyoti Choudhury, Prof. Patrick Das Gupta, Dr. Nivedita
Deo, Dr. R. S. Kaushal, Dr. Daksh Lohiya, Dr. Awadhesh Prasad,
Dr. T. R. Seshadri and Dr. Poonam Silotia for their help.
I would like to thank Mrs. M. Dawar for facilitating various
administrative matters.

I would like to especially thank Prof. D. S. Kulshreshtha for
his affection, encouragement and help during the past six years.

I would like to thank Council of Scientific and Industrial
Research (CSIR), Government of India for Junior Research
Fellowship (JRF) (2003 - 2005) and Senior Research Fellowship
(SRF) (2005 -).
In this regard, I would like to thank my senior and dear
friend Md. Naimuddin for his tips that were instrumental in
cracking the CSIR UGC JRF(NET) examination the very first time
during my Masters.

I would like to thank Prof. Ram Ramaswamy for chairing my CSIR JRF
to SRF evaluation.
I also thank him for various discussions and encouragement.
I would like to thank Dr. Ravi Mehrotra for chairing my CSIR SRF
extension evaluation and discussions.

I would like to thank Prof. Bernhard Palsson and Dr. Jennifer Reed
for kindly providing the GPR association of the {\it E. coli}
metabolic network iJR904.
I also thank all past and present members of the Palsson group for
making their reconstructions of metabolic and regulatory networks
freely available to the scientific community through their website
{\it http://gcrg.ucsd.edu/}.
Without the availability of these reconstructed networks, the work
reported in this thesis would not be possible.

I would like to thank Prof. Uri Alon for encouraging me during a
conference held at NCBS, Bangalore in December 2003.
The discussions I had with him helped me in deciding the problems
to pursue.

In the past five years, I have visited various universities and
research institutions in India.
I thank Himachal Pradesh University, IISc Bangalore, IIT Kanpur,
NCBS Bangalore, SNBNCBS Kolkata, and University of Pune for their
hospitality and infrastructural support.
I would like to also thank Dr. Kanury Rao, ICGEB, Delhi and all
his group members for the opportunity to spend two months in
their lab.

During my Ph.D., I have also been very fortunate to visit various
research groups outside India working in the area of complex networks.
In this regard, I would like to thank
Dr. Johannes Berg (University of Koln),
Dr. Oliver Ebenhoh (Max Planck Institute for Molecular Plant Physiology, Potsdam),
Dr. Alessandro Giuliani (ISS, Rome),
Prof. Hanspeter Herzel (Humboldt University, Berlin),
Prof. Herrmann-Georg Holzhutter (Charite, Humboldt University, Berlin),
Prof. Juergen Jost (Max Planck Institute for Mathematics in the Sciences, Leipzig),
Prof. Francois Kepes (Genopole, Evry),
Dr. Konstantin Klemm (University of Leipzig),
Prof. Matteo Marsili (ICTP, Trieste),
Prof. Olivier Martin (LPTMS, Orsay),
Prof. Hildegard Meyer-Ortmanns (Jacobs University, Bremen),
Dr. Jari Saramaki (HUT, Helsinki),
Prof. Stefan Schuster (University of Jena),
Prof. Frank Schweitzer (ETH, Zurich),
Prof. Roberto Serra (European Center for Living Technology, Venice)
Prof. Andreas Wagner (University of Zurich)
and
Dr. Martin Weigt (ISI, Torino)
for their kind support.
I would like to also thank all the group members of the above mentioned
scientists for their hospitality during my visits.

I also had a chance to attend a lot of conferences and
schools where I was unable to follow most of the talks.
But this time was the most fruitful as I made a lot of
friends in various scientific institutions.
I would like to thank Akhilesh, Amitabha, Arnab, Bhaswar,
Bino, Brojen, Dharmendra, Dhiraj, Dipanjan, Florian, Jeremiah,
Johannes, Julien, Moin, Niko, Rajat, Rajesh, Ramachandran,
Shilpi, Sonika, Srikanth, Sudhanshu, Vaibhao, Viren, Vishal
and Vivek for all the good times.

In the last ten years, I have spent most of my time in the
University of Delhi, North Campus.
I wish to thank Amit, Gaurav, Geetanjali, Harish, Jaswant,
Jiten, Keshwarjit, Kopal, Manish, Pooja, Pranav, Sandip,
Sanil, Sunil, Sushil, Uddipan, Umeshkanta, Varun, Vikrant,
Vikas, Wasim and Yogesh for their company.

I wish to especially thank Jiten for his friendship, help
and encouragement during the past seven years.
I am grateful to him for getting me to focus in life during
M.Sc. Physics.
He has always inspired me by his intelligence, modesty and
generosity.

During the past five years, I have realized that Ph.D. time
is full of confusions and tensions.
It is very easy to make mistakes and give up under certain
situations.
In such situations, you need a patient, mature and
understanding friend.
During the last four years, I have found such a friend in
Pranav.

Due to interdisciplinary nature of my work, it becomes very
essential that you have friends from other fields of science
with whom you can discuss your research, get suggestions and
criticisms.
In the last two years, I had many such stimulating discussions
with Dhiraj.

In the end, I want to mention that in spite of all my
shortcomings, my parents still continue to sacrifice their life
for me and love me more and more with each passing day.

\par
\vspace*{5ex} \noindent{\bf Areejit Samal} 


\begin{thebibliography}{100}
\providecommand{\url}[1]{[#1]}
\providecommand{\urlprefix}{}

\bibitem{WS1998}
Watts DJ, Strogatz SH: \textbf{Collective dynamics of `small-world' networks}.
  \emph{Nature} 1998, \textbf{393}:440--442.

\bibitem{JK1998}
Jain S, Krishna S: \textbf{Autocatalytic sets and the growth of complexity in
  an evolutionary model}. \emph{Phys Rev Lett} 1998, \textbf{81}:5684--5687.

\bibitem{BA1999}
Barabasi AL, Albert R: \textbf{Emergence of scaling in random networks}.
  \emph{Science} 1999, \textbf{286}:509--512.

\bibitem{S2001}
Strogatz SH: \textbf{Exploring complex networks}. \emph{Nature} 2001,
  \textbf{411}:268--276.

\bibitem{AB2002}
Albert R, Barabasi AL: \textbf{Statistical mechanics of complex networks}.
  \emph{Rev Mod Phys} 2002, \textbf{74}:47--97.

\bibitem{Handbookofgraphs}
Bornholdt S, Schuster HG: \emph{Handbook of Graphs and Networks: from the
  Genome to the Internet}. Wiley-VCH 2003.

\bibitem{HHLM1999}
Hartwell LH, Hopfield JJ, Leibler S, Murray AW: \textbf{From molecular to
  modular cell biology}. \emph{Nature} 1999, \textbf{402}:C47--C52.

\bibitem{K2002}
Kitano H: \textbf{Computational systems biology}. \emph{Nature} 2002,
  \textbf{420}:206--210.

\bibitem{A2003}
Alon U: \textbf{Biological Networks: The Tinkerer as an Engineer}.
  \emph{Science} 2003, \textbf{301}:1866--1867.

\bibitem{BO2004}
Barabasi AL, Oltvai ZN: \textbf{Network biology: Understanding the cell's
  functional organization}. \emph{Nat Rev Genet} 2004, \textbf{5}:101--113.

\bibitem{Palssonbook}
Palsson BO: \emph{Systems Biology: Properties of Reconstructed Networks}.
  Cambridge University Press 2006.

\bibitem{Alonbook}
Alon U: \emph{An Introduction to Systems Biology: Design Principles of
  Biological Circuits}. Chapman \& Hall 2007.

\bibitem{AA2003}
Alm E, Arkin AP: \textbf{Biological networks}. \emph{Curr Opin Struct Biol}
  2003, \textbf{13}:193--202.

\bibitem{B2003}
Bray D: \textbf{Molecular networks: the top-down view}. \emph{Science} 2003,
  \textbf{301}:1864--1865.

\bibitem{BLGT2004}
Babu MM, Lucsombe N, Gerstein M, Teichmann S: \textbf{Structure and evolution
  of gene regulatory networks}. \emph{Curr Opin Struc Biol} 2004,
  \textbf{14}:283--291.

\bibitem{MA2003}
Mangan S, Alon U: \textbf{Structure and function of the feed-forward loop
  network motif}. \emph{Proc Natl Acad Sci USA} 2003,
  \textbf{100}:11980--11985.

\bibitem{GEHL2002}
Guet C, Elowitz M, Hsing W, Leibler S: \textbf{Combinatorial synthesis of
  genetic networks}. \emph{Science} 2002, \textbf{296}:1466--1470.

\bibitem{BL1997}
Barkai N, Leibler S: \textbf{Robustness in simple biochemical networks}.
  \emph{Nature} 1997, \textbf{387}:855--857.

\bibitem{BI1999}
Bhalla US, Iyenger R: \textbf{Emergent properties of networks of biological
  signaling pathways}. \emph{Science} 1999, \textbf{283}:339--340.

\bibitem{VMMO2000}
von Dassow G, Meir E, Munro EM, Odell GM: \textbf{The segment polarity network
  is a robust developmental module}. \emph{Nature} 2000, \textbf{406}:188--192.

\bibitem{KB1973}
Kacser H, Burns JA: \textbf{The control of flux}. \emph{Symp Soc Exp Biol}
  1973, \textbf{25}:65--104.

\bibitem{GCRG}
\textbf{UCSD Systems Biology Research Group: {\it http://gcrg.ucsd.edu/}}.

\bibitem{KEGG}
\textbf{KEGG Database: {\it http://www.genome.jp/kegg/}}.

\bibitem{Ecocyc}
\textbf{Ecocyc Database: {\it http://ecocyc.org/}}.

\bibitem{JTAOB2000}
Jeong H, Tombor B, Albert R, Oltvai ZN, Barabasi AL: \textbf{The large-scale
  organization of metabolic networks}. \emph{Nature} 2000,
  \textbf{407}:651--654.

\bibitem{Hararybook}
Harary F: \emph{Graph Theory}. Addison-Wesley Publishing Company 1969.

\bibitem{Bollobasbook}
Bollobas B: \emph{Random Graphs}. Academic Press 1985.

\bibitem{WF2001}
Wagner A, Fell DA: \textbf{The small world inside large metabolic networks}.
  \emph{Proc R Soc Lond B} 2001, \textbf{268}:1803--1810.

\bibitem{MZ2003}
Ma HW, Zeng AP: \textbf{The connectivity structure, giant strong component and
  centrality of metabolic networks}. \emph{Bioinformatics} 2003,
  \textbf{19}:1423--1430.

\bibitem{BKMRRSTW2000}
Broder A, Kumar R, Maghoul F, Raghavan P, Rajagopalan S, Stata R, Tomkins A,
  Wiener J: \textbf{Graph structure in the web}. \emph{Computer Networks} 2000,
  \textbf{33}:309--320.

\bibitem{CD2004}
Csete M, Doyle J: \textbf{Bow ties, metabolism and disease}. \emph{Trends
  Biotechnol} 2004, \textbf{22}:446--450.

\bibitem{A2004}
Arita M: \textbf{The metabolic world of {\it Escherichia coli} is not small}.
  \emph{Proc Natl Acad Sci USA} 2004, \textbf{101}:1543--1547.

\bibitem{AJB2000}
Albert R, Jeong H, Barabasi AL: \textbf{Error and attack tolerance of complex
  networks}. \emph{Nature} 2000, \textbf{406}:378--382.

\bibitem{JMBO2001}
Jeong H, Mason SP, Barabasi AL, Oltvai ZN: \textbf{Lethality and centrality in
  protein networks}. \emph{Nature} 2001, \textbf{411}:41--42.

\bibitem{SSGKRJ2006}
Samal A, Singh S, Giri V, Krishna S, Raghuram N, Jain S: \textbf{Low degree
  metabolites explain essential reactions and enhance modularity in biological
  networks}. \emph{BMC Bioinformatics} 2006, \textbf{7}:118.

\bibitem{RegulonDB}
\textbf{RegulonDB Database: {\it http://regulondb.ccg.unam.mx/}}.

\bibitem{SMMA2002}
Shen-Orr S, Milo R, Mangan S, Alon U: \textbf{Network motifs in the
  transcriptional regulation network of {\it Escherichia coli}}. \emph{Nat
  Genet} 2002, \textbf{31}:64--68.

\bibitem{LRROBGHHTSZJMGRWTVFGY2002}
Lee TI, Rinaldi NJ, Robert F, Odom DT, Bar-Joseph Z, Gerber GK, Hannett NM,
  Harbison CT, Thompson CM, Simon I, Zeitlinger J, Jennings EG, Murray HL,
  Gordon DB, Ren B, Wyrick JJ, Tagne JB, Volkert TL, Fraenkel E, Gifford DK,
  Young RA: \textbf{Transcriptional regulatory networks in {\it Saccharomyces
  cerevisiae}}. \emph{Science} 2002, \textbf{298}:799--804.

\bibitem{GBBK2002}
Guelzim N, Bottani S, Bourgine P, Kepes F: \textbf{Topological and causal
  structure of the yeast transcriptional regulatory network}. \emph{Nat Genet}
  2002, \textbf{31}:60--63.

\bibitem{LBYSTG2004}
Lucsombe N, Babu MM, Yu H, Snyder M, Teichmann S, Gerstein M: \textbf{Genome
  analysis of regulatory network dynamics reveals large topological changes}.
  \emph{Nature} 2004, \textbf{431}:308--312.

\bibitem{CKRHP2004}
Covert MW, Knight EM, Reed JL, Herrgard MJ, Palsson BO: \textbf{Integrating
  high-throughput and computational data elcidates bacterial networks}.
  \emph{Nature} 2004, \textbf{429}:92--96.

\bibitem{MSIKCA2002}
Milo R, Shen-Orr S, Itzkovitz S, Kashtan N, Chklovskii D, Alon U:
  \textbf{Network Motifs: Simple Building Blocks of Complex Networks}.
  \emph{Science} 2002, \textbf{298}:824--827.

\bibitem{RRSA2002}
Ronen M, Rosenberg R, Shraiman BI, Alon U: \textbf{Assigning numbers to the
  arrows: Parameterizing a gene regulation network by using accurate expression
  kinetics}. \emph{Proc Natl Acad Sci USA} 2002, \textbf{99}:10555--10560.

\bibitem{CW2003}
Conant GC, Wagner A: \textbf{Convergent evolution of gene circuits}. \emph{Nat
  Genet} 2003, \textbf{34}:264--266.

\bibitem{W2003b}
Wagner A: \textbf{Does Selection Mold Molecular Networks?} \emph{Science
  Signaling} 2003, \textbf{pe}:41--43.

\bibitem{MBZ2004}
Ma HW, Buer J, Zeng AP: \textbf{Hierarchical structure and modules in the {\it
  Escherichia coli} transcriptional regulatory network revealed by a new
  top-down approach}. \emph{BMC Bioinformatics} 2004, \textbf{5}:199.

\bibitem{BBO2005}
Balazsi G, Barabasi AL, Oltvai ZN: \textbf{Topological units of environmental
  signal processing in the transcriptional regulatory network of {\it
  Escherichia coli}}. \emph{Proc Natl Acad Sci USA} 2005,
  \textbf{102}:7841--7846.

\bibitem{SJ2008}
Samal A, Jain S: \textbf{The regulatory network of {\it E. coli} metabolism as
  a Boolean dynamical system exhibits both homeostasis and flexibility of
  response}. \emph{BMC Systems Biology} 2008, \textbf{2}:21.

\bibitem{ASBL1999}
Alon U, Surette MG, Barkai N, Leibler S: \textbf{Robustness in bacterial
  chemotaxis}. \emph{Nature} 1999, \textbf{397}:168--171.

\bibitem{YHSD2000}
Yi TM, Huang Y, Simon MI, Doyle J: \textbf{Robust perfect adaptation in
  bacterial chemotaxis through integral feedback control}. \emph{Proc Natl Acad
  Sci USA} 2000, \textbf{97}:4649--4653.

\bibitem{I2004}
Ingolia NT: \textbf{Topology and robustness in the {\it Drosophila} segment
  polarity network}. \emph{PLoS Biol} 2004, \textbf{2}:e123.

\bibitem{B2005}
Bornholdt S: \textbf{Less Is More in Modeling Large Genetic Networks}.
  \emph{Science} 2005, \textbf{310}:449--450.

\bibitem{VP1994a}
Varma A, Palsson BO: \textbf{Stoichiometric flux balance models quantitatively
  predict growth and metabolic by-product secretion in wild-type {\it
  Escherichia coli} W3110}. \emph{Appl Environ Microbiol} 1994,
  \textbf{60}:3724--3731.

\bibitem{EP2000}
Edwards JS, Palsson BO: \textbf{The {\it {E}scherichia coli} MG1655 in silico
  metabolic genotype: its definition, characteristics, and capabilities}.
  \emph{Proc Natl Acad Sci USA} 2000, \textbf{97}:5528--5533.

\bibitem{EIP2001}
Edwards JS, Ibarra RU, Palsson BO: \textbf{In silico predictions of {\it
  Escherichi coli} metabolic capabilities are consistent with experimental
  data}. \emph{Nat Biotechnol} 2001, \textbf{19}:125--130.

\bibitem{IEP2002}
Ibarra RU, Edwards JS, Palsson BO: \textbf{{\it Escherichi coli} K-12 undergoes
  adaptive evolution to achieve in silico predicted optimal growth}.
  \emph{Nature} 2002, \textbf{420}:186--189.

\bibitem{K1969a}
Kauffman SA: \textbf{Metabolic stability and epigenesis in randomly constructed
  genetic nets}. \emph{J Theor Biol} 1969, \textbf{22}:432--467.

\bibitem{K1969b}
Kauffman SA: \textbf{Homeostasis and differentiation in random genetic control
  networks}. \emph{Nature} 1969, \textbf{224}:177--178.

\bibitem{Kauffmanbook}
Kauffman SA: \emph{Origins of Order: Self-Organization and Selection in
  Evolution}. Oxford University Press 1993.

\bibitem{ST2001}
Sanchez L, Thieffry D: \textbf{A logical analysis of the Drosophila gap-gene
  system}. \emph{J Theor Biol} 2001, \textbf{211}:115--141.

\bibitem{AO2003}
Albert R, Othmer HG: \textbf{The topology of the regulatory interactions
  predicts the expression pattern of the segment polarity genes in {\it
  Drosophila melanogaster}}. \emph{J Theor Biol} 2003, \textbf{223}:1--18.

\bibitem{LLLOT2004}
Li F, Long T, Lu Y, Ouyang Q, Tang C: \textbf{The yeast cell-cycle network is
  robustly designed}. \emph{Proc Natl Acad Sci USA} 2004,
  \textbf{101}:4781--4786.

\bibitem{EPA2004}
Espinosa-Soto C, Padilla-Longoria P, Alvarez-Buylla ER: \textbf{A gene
  regulatory network model for cell-fate determination during Arabidopsis
  thaliana flower development that is robust and recovers experimental gene
  expression profiles}. \emph{Plant Cell} 2004, \textbf{16}:2923--2939.

\bibitem{LAA2006}
Li S, Assmann SM, Albert R: \textbf{Predicting essential components of signal
  transduction networks: A dynamic model of guard cell abscisic acid
  signaling}. \emph{PLoS Biol} 2006, \textbf{4(10)}:e312.

\bibitem{HLPP2006}
Herrgard MJ, Lee BS, Portnoy V, Palsson BO: \textbf{Integrated analysis of
  regulatory and metabolic networks reveals novel regulatory mechanisms in {\it
  Saccharomyces cerevisiae}}. \emph{Genome Res} 2006, \textbf{16}:627--635.

\bibitem{RVSP2003}
Reed JL, Vo TD, Schilling CH, Palsson BO: \textbf{An expanded genome-scale
  model of {\it {E}scherichia coli} K-12 (i{JR904 GSM/GPR})}. \emph{Genome
  Biol} 2003, \textbf{4}:R54.1--R54.12.

\bibitem{DHP2004}
Duarte NC, Herrgard MJ, Palsson BO: \textbf{Reconstruction and validation of
  {\it {S}accharomyces cerevisiae} i{ND750}, a fully compartmentalized
  genome-scale metabolic model}. \emph{Genome Res} 2004,
  \textbf{14}:1298--1309.

\bibitem{BP2005}
Becker SA, Palsson BO: \textbf{Genome-scale reconstruction {\it Staphylococcus
  aureus} N315: an initial draft to the two-dimensional annotation}. \emph{BMC
  Microbiol} 2005, \textbf{5}:8.

\bibitem{Graphviz}
\textbf{Graphviz: {\it http://www.graphviz.org/}}.

\bibitem{PSNMS1999}
Pfeiffer T, Sanchez-Valdenebro I, Nuno JC, Montero F, Schuster S:
  \textbf{METATOOL: for studying metabolic networks}. \emph{Bioinformatics}
  1999, \textbf{15}:251--257.

\bibitem{SKWMP2002}
Schuster S, Klamt S, Weckwerth W, Moldenhauer F, Pfeiffer T: \textbf{Use of
  network analysis of metabolic systems in bioengineering}. \emph{Bioprocess
  Biosyst Eng} 2002, \textbf{24}:363--372.

\bibitem{SKBSG2002}
Stelling J, Klamt S, Bettenbrock K, Schuster S, Gilles ED: \textbf{Metabolic
  network structure determines key aspects of functionality and regulation}.
  \emph{Nature} 2002, \textbf{420}:190--193.

\bibitem{PPP2002}
Papin JA, Price ND, Palsson BO: \textbf{Extreme pathway lengths and reaction
  participation in genome-scale metabolic networks}. \emph{Genome Res} 2002,
  \textbf{12}:1889--1900.

\bibitem{RP2004}
Reed JL, Palsson BO: \textbf{Genome-scale in silico models of {\it E. coli}
  have multiple equivalent phenotypic states: assessment of correlated reaction
  subsets that comprise network states}. \emph{Genome Res} 2004,
  \textbf{14}:1797--1805.

\bibitem{BNSM2004}
Burgard AP, Nikolaev EV, Schilling CH, Maranas CD: \textbf{Flux coupling
  analysis of genome-scale metabolic network reconstructions}. \emph{Genome
  Res} 2004, \textbf{14}:301--312.

\bibitem{KTV1999}
Kannan R, Tetali P, Vempala S: \textbf{Simple Markov-chain algorithms for
  generating bipartite graphs and tournaments}. \emph{Random Structures and
  Algorithms} 1999, \textbf{14}:293--308.

\bibitem{MS2002}
Maslov S, Sneppen K: \textbf{Specificity and stability in topology of protein
  networks}. \emph{Science} 2002, \textbf{296}:910--913.

\bibitem{MIKLSASA2004}
Milo R, Itzkovitz S, Kashtan N, Levitt R, Shen-Orr S, Ayzenshtat I, Sheffer M,
  Alon U: \textbf{Superfamilies of Evolved and Designed Networks}.
  \emph{Science} 2004, \textbf{303}:1538--1542.

\bibitem{SP1992a}
Savinell JM, Palsson BO: \textbf{Optimal selection of metabolic fluxes for in
  vivo measurement. I. Development of mathematical methods}. \emph{J Theor
  Biol} 1992, \textbf{155}:201--214.

\bibitem{SP1992b}
Savinell JM, Palsson BO: \textbf{Optimal selection of metabolic fluxes for in
  vivo measurement. I. Application to {\it Escherichia coli} and hybridoma cell
  metabolism}. \emph{J Theor Biol} 1992, \textbf{155}:215--242.

\bibitem{VP1993a}
Varma A, Palsson BO: \textbf{Metabolic capabilities of {\it Escherichia coli}:
  I. Synthesis of biosynthetic precursors}. \emph{J Theor Biol} 1993,
  \textbf{165}:477--502.

\bibitem{VP1993b}
Varma A, Palsson BO: \textbf{Metabolic capabilities of {\it Escherichia coli}:
  II. Optimal growth patterns}. \emph{J Theor Biol} 1993,
  \textbf{165}:503--522.

\bibitem{VBP1993a}
Varma A, Boesch BW, Palsson BO: \textbf{Biochemical production capabilities of
  {\it Escherichia coli}}. \emph{Biotechnol Bioeng} 1993, \textbf{42}:59--73.

\bibitem{VBP1993b}
Varma A, Boesch BW, Palsson BO: \textbf{Stoichiometric interpretation of {\it
  Escherichia coli} glucose catabolism under various oxygenation rates}.
  \emph{Appl Environ Microbiol} 1993, \textbf{59}:2465--2473.

\bibitem{VP1994b}
Varma A, Palsson BO: \textbf{Metabolic flux balancing: Basic concepts,
  scientific and practical use}. \emph{Bio/Technology} 1994,
  \textbf{12}:994--998.

\bibitem{PK1997}
Pramanik J, Keasling JD: \textbf{Stoichimetric model of {\it Escherichia coli}
  metabolism: incorporation of growth-rate dependent biomass compostion and
  mechanistic energy requirements}. \emph{Biotechnol Bioeng} 1997,
  \textbf{56}:398--421.

\bibitem{PK1998}
Pramanik J, Keasling JD: \textbf{Effect of {\it Escherichia coli} biomass
  composition on central metabolic fluxes predicted by a stoichiometric model}.
  \emph{Biotechnol Bioeng} 1998, \textbf{60}:230--238.

\bibitem{SCFCEP2002}
Schilling CH, Covert MW, Famili I, Church GM, Edwards JS, Palsson BO:
  \textbf{Genome-scale metabolic model of {\it Helicobacter pylori} 26695}.
  \emph{J Bacteriol} 2002, \textbf{184}:4582--4593.

\bibitem{SVC2002}
Segre D, Vitkup D, Church GM: \textbf{Analysis of optimality in natural and
  perturbed metabolic networks}. \emph{Proc Natl Acad Sci USA} 2002,
  \textbf{99}:15112--15117.

\bibitem{FFFPN2003}
Forster J, Famili I, Fu P, Palsson B, Nielsen J: \textbf{Genome-scale
  reconstruction of the {\it Saccharomyces cerevisiae} metabolic network}.
  \emph{Genome Res} 2003, \textbf{13(2)}:244--253.

\bibitem{KPE2003}
Kauffman KJ, Prakash P, Edwards JS: \textbf{Advances in flux balance analysis}.
  \emph{Curr Opin Biotechnol} 2003, \textbf{14}:491--496.

\bibitem{AKVOB2004}
Almaas E, Kovacs B, Vicsek T, Oltvai ZN, Barabasi AL: \textbf{Global
  organization of metabolic fluxes in the bacterium {\it Escherichia coli}}.
  \emph{Nature} 2004, \textbf{427}:839--843.

\bibitem{SJunpublished}
Singh S, Jain S: \textbf{unpublished}.

\bibitem{MP2005}
Mahadevan R, Palsson BO: \textbf{Properties of Metabolic Networks: Structure
  vs. Function}. \emph{Biophys J} 2005, \textbf{88}:L7--L9.

\bibitem{SS1991}
Schuster S, Schuster R: \textbf{Detecting strictly detailed balanced
  subnetworks in open chemical reaction networks}. \emph{J Math Chem} 1991,
  \textbf{6}:17--40.

\bibitem{Heinrichbook}
Heinrich R, Schuster S: \emph{The Regulation of Cellular Systems}. Chapman \&
  Hall 1996.

\bibitem{GSCBRDSKAGBKDBGMFOBOO2003}
Gerdes SY, Scholle MD, Campbell JW, Balazsi G, Ravasz E, Daugherty MD, Somera
  AL, Kyrpides NC, Anderson I, Gelfand MS, Bhattacharya A, Kapatral V, D'Souza
  M, Baev MV, Grechkin Y, Mseeh F, Fonstein MY, Overbeek R, Barabasi AL, Oltvai
  ZN, Osterman AL: \textbf{Experimental determination and system level analysis
  of essential genes in {\it Escherichia coli} MG1655}. \emph{J Bacteriol}
  2003, \textbf{184}:152--164.

\bibitem{Cytoscape}
\textbf{Cytoscape: {\it http://www.cytoscape.org/}}.

\bibitem{T1973}
Thomas R: \textbf{Boolean formalisation of genetic control circuits}. \emph{J
  Theor Biol} 1973, \textbf{42}:565--583.

\bibitem{DW1986}
Derrida B, Weisbuch G: \textbf{Evolution of overlaps between configurations in
  random Boolean networks}. \emph{J Physique} 1986, \textbf{47}:1297--1303.

\bibitem{BS1998}
Bornholdt S, Sneppen K: \textbf{Neutral mutations and punctuated equilibrium in
  evolving genetic networks}. \emph{Phys Rev Lett} 1998, \textbf{81}:236--239.

\bibitem{HSWK2002}
Harris SE, Sawhill BK, Wuensche A, Kauffman SA: \textbf{A model of
  transcriptional regulatory networks based on biases in the observed
  regulation rules}. \emph{Complexity} 2002, \textbf{7}:23--40.

\bibitem{SDKZ2002}
Shmulevich I, Dougherty ER, Kim S, Zhang W: \textbf{Probabilistic Boolean
  Networks: A Rule-based Uncertainty Model for Gene Regulatory Networks}.
  \emph{Bioinformatics} 2002, \textbf{18(2)}:261--274.

\bibitem{ACK2003}
Aldana-Gonzalez M, Coppersmith S, Kadanoff LP: \emph{Perspectives and problems
  in nonlinear science, A celebratory volume in honor of Lawrence Sirovich},
  Springer, New York 2003 chap. Boolean Dynamics with Random Couplings.

\bibitem{KPST2003}
Kauffman SA, Peterson C, Samuelsson B, Troein C: \textbf{Random Boolean network
  models and the yeast transcriptional network}. \emph{Proc Natl Acad Sci USA}
  2003, \textbf{100}:14796--14799.

\bibitem{KPST2004}
Kauffman SA, Peterson C, Samuelsson B, Troein C: \textbf{Genetic networks with
  canalyzing Boolean rules are always stable}. \emph{Proc Natl Acad Sci USA}
  2004, \textbf{101}:17102--17107.

\bibitem{SVA2004}
Serra R, Villani M, Agostini L: \textbf{On the dynamics of random Boolean
  networks with scale-free outgoing connections}. \emph{Physica A} 2004,
  \textbf{339}:665--673.

\bibitem{KB2005a}
Klemm K, Bornholdt S: \textbf{Stable and unstable attractors in Boolean
  networks}. \emph{Phys Rev E} 2005, \textbf{72}:055101.

\bibitem{GD2005}
Greil F, Drossel B: \textbf{Dynamics of Critical Kauffman Networks under
  Asynchronous Stochastic Update}. \emph{Phys Rev Lett} 2005,
  \textbf{95}:048701.

\bibitem{BB2007}
Braunewell S, Bornholdt S: \textbf{Superstability of the yeast cell-cycle
  dynamics: Ensuring causality in the presence of biochemical stochasticity}.
  \emph{J Theor Biol} 2007, \textbf{245}:638--643.

\bibitem{BHRP2005}
Barrett CB, Herring CD, Reed JL, Palsson BO: \textbf{The Global Transcriptional
  Regulatory Network for Metabolism in {\it Escherichia coli} Exhibits Few
  Dominant Functional States}. \emph{Proc. Natl. Acad. Sci. USA} 2005,
  \textbf{102(52)}:19103--19108.

\bibitem{CSP2001}
Covert MW, Schilling CH, Palsson BO: \textbf{Regulation of Gene Expression in
  Flux Balance Models of Metabolism}. \emph{J Theor Biol} 2001,
  \textbf{213}:73--78.

\bibitem{AF2000}
Ancel LW, Fontana W: \textbf{Plasticity, Evolvability, and Modularity in RNA}.
  \emph{J Exp Zool} 2000, \textbf{288}:242--283.

\bibitem{Kirschnerbook}
Kirschner MW, Gerhart JC: \emph{The Plausibility of Life: Resolving Darwin's
  Dilemma}. Yale University Press 2005.

\bibitem{JK2002a}
Jain S, Krishna S: \textbf{Crashes, recoveries and core shifts in a model of
  evolving networks}. \emph{Phys Rev E} 2002, \textbf{65}:026103.

\bibitem{JK2002b}
Jain S, Krishna S: \textbf{Large extinctions in an evolutionary model: The role
  of innovation and keystone species}. \emph{Proc Natl Acad Sci USA} 2002,
  \textbf{99}:2055--2060.

\bibitem{PPH2004}
Papp B, Pal C, Hurst LD: \textbf{Metabolic network analysis of the causes and
  evolution of enzyme dispensability in yeast}. \emph{Nature} 2004,
  \textbf{439}:661--664.

\bibitem{M1999}
Morowitz HJ: \textbf{A theory of biochemical organization, metabolic pathways,
  and evolution}. \emph{Complexity} 1999, \textbf{4}:39--53.

\bibitem{SSSDD2004}
Stelling J, Sauer U, Szallasi Z, Doyle~III FJ, Doyle J: \textbf{Robustness of
  cellular functions}. \emph{Cell} 2004, \textbf{118}:675--685.

\bibitem{K2004}
Kitano H: \textbf{Biological robustness}. \emph{Nat Rev Genet} 2004,
  \textbf{5}:826--837.

\bibitem{Wagnerbook}
Wagner A: \emph{Robustness and Evolvability in Living Systems}. Princeton
  University Press 2005.

\bibitem{MED2002}
Mahadevan R, Edwards JS, Doyle FJ: \textbf{Dynamic Flux Balance Analysis of
  Diauxic Growth in {\it Escherichia coli}}. \emph{Biophys J} 2002,
  \textbf{83}:1331--1340.

\bibitem{LLTLZLL2006}
Luo RY, Liao S, Tao GY, Li YY, Zeng S, Li YX, Luo Q: \textbf{Dynamic analysis
  of optimality in myocardial energy metabolism under normal and ischemic
  conditions}. \emph{Mol Sys Biol} 2006, \textbf{2}:31.

\bibitem{SESR2007}
Shlomi T, Eisenberg Y, Sharan R, Ruppin E: \textbf{A genome-scale computational
  study of the interplay between transcriptional regulation and metabolism}.
  \emph{Mol Sys Biol} 2007, \textbf{3}:101.

\bibitem{P1984}
Papoutsakis ET: \textbf{Equations and calculations for fermentations of butyric
  acid bacteria}. \emph{Biotechnol Bioeng} 1984, \textbf{26}:174--187.

\bibitem{FS1986}
Fell DA, Small JA: \textbf{Fat synthesis in adipose tissue. An examination of
  stoichiometric constraints}. \emph{J Biochem} 1986, \textbf{238}:781--786.

\bibitem{MD1990}
Majewski RA, Domach MM: \textbf{Simple constrained optimization view of acetate
  overflow in {\it E. coli}}. \emph{Biotechnol Bioeng} 1990,
  \textbf{35}:731--738.

\bibitem{LPDG2000}
Lee S, Phalakornkule C, Domach MM, Grossmann IE: \textbf{Recursive MILP model
  for finding all alternate optima in LP models for metabolic networks}.
  \emph{Comp Chem Eng} 2000, \textbf{24}:711--716.

\bibitem{PLZKAGD2001}
Phalakornkule C, Lee S, Zhu T, Koepsel R, Ataai MM, Grossmann IE, Domach MM:
  \textbf{A MILP based flux alternative generation and NMR experimental design
  strategy for metabolic engineering}. \emph{Metab Eng} 2001,
  \textbf{3}:124--137.

\bibitem{KS2002}
Klamt S, Schuster S: \textbf{Calculating as many fluxes as possible in
  underdetermined metabolic networks}. \emph{Mol Biol Rep} 2002,
  \textbf{29}:243--248.

\end{thebibliography}
\end{document}